\documentclass[reqno,12pt]{article}
\usepackage{amsfonts,amssymb,
amsmath,amscd, bm, mathrsfs, 
mathrsfs,delarray,subfigure}
\usepackage{hyperref,cite}
\usepackage[all]{xy}
\usepackage[dvips]{color}
\usepackage[dvips]{graphicx}
\usepackage{pstricks}
\usepackage{pstricks-add}
\usepackage{pst-solides3d}

\DeclareMathOperator{\Ad}{Ad}

\DeclareMathOperator{\id}{id}

\DeclareMathOperator{\Map}{Map}
\DeclareMathOperator{\Knot}{Knot}

\DeclareMathOperator{\Gau}{Gau}

\DeclareMathOperator{\Aut}{Aut}

\DeclareMathOperator{\Diff}{Diff}

\DeclareMathOperator{\rank}{rank}

\DeclareMathOperator{\tr}{tr}

\DeclareMathOperator{\MCG}{MCG}

\DeclareMathOperator{\Cyc}{Cyc}

\numberwithin{equation}{subsection} 
\numberwithin{subsection}{section} 

\font\sansserif=cmss12
\font\scriptsansserif=cmss12 at 7 truept
\font\scriptscriptsansserif=cmss10 at 5 truept
\font\euler=eusm10 at 12.8 truept
\font\scripteuler=eusm7
\font\scriptscripteuler=eusm5 
\newcommand{\ul}[1]{{\underline{#1}}{}}

\CompileMatrices

\newtheorem{defi}{
Definition}[subsection]

\newtheorem{prop}{
Proposition}[subsection]
\newtheorem{lemma}{
Lemma}[subsection]

\begin{document}

\hrule\vskip.5cm
\hbox to 14.5 truecm{May 2015 \hfil DIFA 15}
\vskip.5cm\hrule
\vskip.7cm
\centerline{\textcolor{blue}{\bf ON HIGHER HOLONOMY INVARIANTS}}   
\centerline{\textcolor{blue}{\bf IN HIGHER GAUGE THEORY I}}   
\vskip.2cm
\centerline{by}
\vskip.2cm
\centerline{\bf Roberto Zucchini}
\centerline{\it Dipartimento di Fisica ed Astronomia, Universit\`a di Bologna}
\centerline{\it V. Irnerio 46, I-40126 Bologna, Italy}
\centerline{\it I.N.F.N., sezione di Bologna, Italy}
\centerline{\it E--mail: emanuele.soncini@studio.unibo.it, zucchinir@bo.infn.it}
\vskip.7cm
\hrule
\vskip.6cm
\centerline{\bf Abstract} 
\par\noindent
This is the first of a series of two technical papers devoted to the analysis of holonomy 
invariants in strict higher gauge theory with end applications in higher 
Chern--Simons theory. For a flat $2$--connection, we define the $2$-holonomy 
of surface knots of arbitrary genus and determine its covariance properties 
under $1$--gauge transformation and change of base data. 

 \vfil\eject

\tableofcontents

\vfil\eject

\section{\normalsize \textcolor{blue}{Introduction}}\label{sec:intro}

\vfil
\hspace{.5cm} One of the basic problems in $3$--dimensional topology is the classification 
of topologically distinct knots \cite{Kauffman:1991ds,Kauffman:1995hc}.
Knots which can be smoothly deformed into one other are topologically indistinguishable 
though they may appear completely unlike upon superficial inspection. Knot invariants take 
the same value for all topologically identical knots. Computing knot invariants, therefore, 
allows one to ascertain whether two given knots are topologically distinct or not. 
If they have different knot invariants, they are.

\vfil
Ordinary Chern--Simons theory allows in principle the computation of knot invariants
using quantum field theory \cite{Witten:1988hf,Witten:1992fb}. 
In this respect, it has been extraordinarily successful. A variety of methods 
have been devised, perturbative and non perturbative as well, including large level asymptotics,
canonical quantization and reduction to random matrix integrals \cite{Marino:2005sj}. 

\vfil
In topology, the problem of the classification of knots 
in $3$ dimensions has higher dimensional analogs. What is a higher knot?
As ordinary knots are embeddings of the circle $S^1$ into the $3$--sphere $S^3$, say, 
$n$--dimensional knots may be defined as embeddings of the $n$--sphere $S^n$ into 
the $n+2$--sphere $S^{n+2}$. Actually this is only one of 
the possible generalization of the notion of knot. $S^1$ is not only the lowest 
dimensional non trivial sphere, but it is also the only compact $1$--dimensional manifold
without boundary. Thus, one may also define $n$--dimensional knots as embeddings 
of a compact boundaryless $n$--fold $X$ into $S^{n+1}$. For $X=S^n$, the former 
more restrictive definition is recovered. 

\vfil
The problem is already quite involved for $2$--dimensional knots \cite{Carter:1998cas,Kamada:2002ska}. 
We distinguish $2$--knots, defined as embeddings of $S^2$ into $S^4$,
and surface knots, defined as embeddings of a compact surface $S$ with no boundary into $S^4$.
Surface knots can be classified according to the genus $\ell_S$ of $S$. $2$--knots are just 
genus $0$ surface knots. In this paper, we shall concentrate on surface knots.

Given the success of ordinary Chern--Simons theory as a quantum field theoretic framework
for the computation of ordinary knot invariants, it is conceivable that higher dimensional 
versions of Chern--Simons theory may compute higher knot invariants. In particular, 
a $4$--dimensional Chern--Simons theory would be required to deal with surface knot
invariants. However, this already poses a problem. The higher dimensional analogs of plane 
Chern--Simons theory exist only in odd dimensional spaces and are not guaranteed
to do the job. In fact, alternative approaches to the problem relying on $BF$ theory
instead have been developed \cite{Cattaneo:2002tk}. 
There is by now a growing evidence that the incarnation of Chern--Simons 
theory appropriate for higher knot theory may belong to the realm of higher gauge theory
\cite{Kotov:2007nr,Fiorenza:2011jr}.
See refs. \cite{Baez:2002jn,Baez:2010ya} for a readable introduction to the subject
and refs. \cite{Schreiber2011,Gruetzmann:2014ica} for a more comprehensive in depth treatment. 

In refs. \cite{Zucchini:2011aa,Soncini:2014ara}, a higher gauge theoretic $4$--dimensional Chern--Simons model 
whose symmetry is encoded in a balanced cyclic $2$--term $L_\infty$ algebra has been constructed.
The model resembles in many ways the usual Chern--Simons model, but at the moment 
it is not clear whether it exists as a quantum field theory, though there are some
indications in this direction. It is a promising candidate deserving further investigation. 

In ordinary Chern--Simons theory, the field theoretic computation of knot invariants 
involves the evaluation of traces of Wilson loops of the gauge field, mathematically holonomies 
of the gauge connection, along knots in representations of the gauge group. 
In the present endeavour, we consider the corresponding issue for surface knots in 
a strict higher gauge theory. 
The problem has two parts: $(a)$ the definition of surface holonomies and the study
of the way they depend on the choice of gauge and base data; $(b)$
the definition of the appropriate notion of trace for the gauge crossed module
such to give surface knot invariants upon application to surface holonomies. 
The first part is treated in the present work, the second one in a companion paper
\cite{SZ:2015}.  

\vfil

\subsection{\normalsize \textcolor{blue}{Scope and plan of the paper}}\label{sec:scope}

\vfil
\hspace{.5cm} The goal of the present paper is working out a rigorous definition of surface
knot holonomy in strict higher gauge theory that is suitable for a gauge theoretic 
computation of surface knot invariants. In this subsection, we outline our results 
with no pretence of mathematical rigour.

\vfil
Early investigations on surface holonomy include refs. \cite{Cattaneo:2002tk,Chepelev:2001mg}.
Higher parallel transport was studied in depth in refs. \cite{Baez:2004in,Baez:2005qu,Schrei:2009,Schrei:2011,
Schrei:2008,Martins:2007,Martins:2008,Chatterjee:2009ne,Chatterjee:2014pna,Soncini:2014zra}.
The problem of defining surface holonomy was tackled already in refs. \cite{Martins:2007,Martins:2008}
and, more recently, it has been analyzed in refs. \cite{Chatterjee:2010xa,Abbaspour:2012a,Abad:2014a,Abad:2014b}
from a variety of points of view.

\vfil
To make the reader appreciate the import of our results, it is useful to review first 
briefly knot holonomy in ordinary gauge theory. 
Consider a gauge theory with gauge group $G$ on a $3$--fold $M$ with 
a trivial principal $G$--bundle as gauge background, the situation normally envisaged in 
Chern--Simons theory. A connection, physically a gauge field, is then a 
$\mathfrak{g}$--valued $1$--form $\theta$. 

\vfil
Knots are closed embedded curves in $M$. 
As it turns out, knot holonomy is defined for based knots, that is knots which
begin and end at some point of $M$. For this reason, at an initial level of analysis, 
it is natural to assume that $M$ is pointed by a distinguished point 
$p_M$ and restrict the analysis to knots based at $p_M$.  

\vfil
Given a {\it flat} connection $\theta$, with each knot $\xi$ in $M$ 
there is associated its holonomy $F_\theta(\xi)\in G$
by a classical construction. The flatness of $\theta$ ensures that 
$F_\theta(\xi)$ is invariant under smooth deformations of $\xi$ which 
leave $p_M$ fixed. 

\vfil
A gauge transformation is a $G$--valued map $g$. Gauge transformations act on connections. 
Under a gauge transformation $g$, the flat connection $\theta$ turns into another one 
${}^g\theta$. Correspondingly, the holonomy of a knot $\xi$ transforms as
\begin{equation}
F_{{}^g\theta}(\xi)=g(p_M)F_\theta(\xi)g(p_M)^{-1}.
\vphantom{\ul{\ul{\ul{\ul{\ul{\ul{g}}}}}}}
\vphantom{\Big]}
\label{ihiholo2}
\end{equation}

Of course, we do not want to commit ourselves with a final choice of pointing of $M$. 
Thus, at the next level of analysis, it is necessary to find out how holonomy depends on the latter. 
Suppose that $p_{M0}$, $p_{M1}$ are two distinct pointings of $M$ such that 
there exists a curve $\gamma_1$ connecting $p_{M0}$ to $p_{M1}$. 
If $\xi_0$, $\xi_1$ are two knots based at $p_{M0}$, $p_{M1}$, respectively, which can be 
smoothly deformed into one another, then the holonomies $F_\theta(\xi_0)$, $F_\theta(\xi_1)$ satisfy
\begin{equation}
F_{\theta}(\xi_1)=F_\theta(\gamma_1)F_\theta(\xi_0)F_\theta(\gamma_1)^{-1},
\vphantom{\Big]}
\label{ihiholo3}
\end{equation}
where $F_\theta(\gamma_1)\in G$ is the customary parallel transport along $\gamma_1$.

In this paper, we propose a natural and physically intuitive extension of the above simple holonomic framework 
to strict higher gauge theory capable of computing holonomies of surface knots.

Consider a strict higher gauge theory with gauge crossed module $(G,H)$ on a $4$--fold $M$ with 
a trivial principal $(G,H)$--$2$--bundle as gauge background, the situation presumably relevant in the 
appropriate higher gauge theoretic version of $4$--dimensional Chern--Simons theory. A $2$--connection
is now a doublet of a $\mathfrak{g}$--valued $1$--form $\theta$ and a $\mathfrak{h}$--valued $2$--form 
$\varUpsilon$ satisfying a certain compatibility requirement known as zero fake curvature condition. 

Analogously to common knots, surface knots are closed embedded surfaces in $M$. 
Surface knot holonomy is defined for based surface knots of $M$, that is knots
whose independent non contractible loops, called characteristic knots, 
are specified closed curves. 
For this reason, much as in the ordinary case, at an initial level of analysis, 
it is natural to assume that $M$ is marked by a distinguished point 
$p_M$ and a special set of closed curves $\zeta_{Mi}$ based at $p_M$
and restrict to surface knots 
whose characteristic knots equal the $\zeta_{Mi}$. 
There are conditions on the number and type of the $\zeta_{Mi}$
which have to be met. A 
non trivial feature of surface knot theory is its requirement of a reference
``trivial'' surface knot.
A comprehensive treatment of surface knot theory 
is provided in subsects. \ref{sec:path}--\ref{sec:diffeo}.   

$2$--Connection doublets can be flat as ordinary connections. In subsect. \ref{sec:twopara}, we show that,
given a {\it flat} $2$--connection doublet $(\theta,\varUpsilon)$, with each surface knot $\varXi$ in $M$ 
of genus $\ell_\varXi$ there are associated its surface holonomy $F_{\theta,\varUpsilon}(\varXi)\in H$
and the holonomies $F_\theta(\zeta_{Mi})\in G$ of its $2\ell_\varXi$ characteristic knots $\zeta_{Mi}$. 
The flatness of $(\theta,\varUpsilon)$ ensures that $F_{\theta,\varUpsilon}(\varXi)$ and the 
$F_\theta(\zeta_{Mi})$ are invariant under smooth deformations of $\varXi_i$ which 
leave $p_M$ and the $\zeta_{Mi}$ fixed. 

The surface holonomy of a surface knot $\varXi$ belongs to the kernel of the target morphism $t$
of the crossed module $(G,H)$
\begin{equation}
t(F_{\theta,\varUpsilon}(\varXi))=1_H.
\label{icholo1}
\end{equation}
Non trivial surface holonomy is thus possible only if the kernel of $t$ is non trivial.

A $1$--gauge transformation doublet consists of $G$--valued map $g$ and an $\mathfrak{h}$--valued $1$--form $J$.
$1$--gauge transformation doublets act on $2$--connection doublets. 
Under a $1$--gauge transformation $(g,J)$, the flat $2$--connection
doublet $(\theta,\varUpsilon)$ turns into another one 
$({}^{g,J}\theta,{}^{g,J}\varUpsilon)$. In subsect. \ref{sec:gaupara}, we show that 
the surface holonomy of a surface knot $\varXi$ and the holonomies of its 
characteristic knots $\zeta_{Mi}$ correspondingly transform as \hphantom{xxxxxxxxxxx}
\begin{equation}
F_{{}^{g,J}\theta,{}^{g,J}\varUpsilon}(\varXi)=m(g(p_M))(F_{\theta,\varUpsilon}(\varXi)), 
\vphantom{\Big]}
\label{ihiholo6}
\end{equation}
where $m$ denotes the $G$--action morphism of the crossed module $(G,H)$, and
\begin{equation}
F_{{}^{g,J}\theta}(\zeta_{Mi})=g(p_M)t(G_{g,J;\theta}(\zeta_{Mi}))F_\theta(\zeta_{Mi})g(p_M)^{-1},
\vphantom{\Big]}
\label{ihiholo6/0}
\end{equation}
where $G_{g,J;\theta}(\zeta_{Mi})\in H$ is an appropriately defined gauge holonomy of $\zeta_{Mi}$. 

In subsect. \ref{sec:twoholo}, moving to the next level of analysis, 
we find out how surface holonomy depends on the choice of the marking of $M$.
Suppose that $p_{M0},\zeta_{M0i}$, $p_{M1},\zeta_{M1i}$ 
are two markings that can be smoothly deformed into one another, so that 
there exist a curve $\gamma_1$ joining $p_{M0}$ to $p_{M1}$ and, for each $i$, 
a surface $\varSigma_i$ stretching from $\zeta_{M0i}$ to $\zeta_{M1i}$. 
If $\varXi_0$, $\varXi_1$ are two surface knots based at the $\zeta_{M0i}$, $\zeta_{M1i}$, respectively, 
related by the same deformation, then  the surface holonomies 
$F_{\theta,\varUpsilon}(\varXi_0)$, $F_{\theta,\varUpsilon}(\varXi_1)$ are related as 
\begin{equation}
F_{{\theta,\varUpsilon}}(\varXi_1)=m(F_\theta(\gamma_1))(F_{\theta,\varUpsilon}(\varXi_0)),
\vphantom{\Big]}
\label{ihiholo8}
\end{equation}
whilst the line holonomies $F_\theta(\zeta_{M0i})$, $F_\theta(\zeta_{M1i})$ accord as
\begin{equation}
F_\theta(\zeta_{M1i})
=F_\theta(\gamma_1)t(F_{\theta,\varUpsilon}(\varSigma_i))
F_\theta(\zeta_{M0i})F_\theta(\gamma_1)^{-1},
\vphantom{\Big]}
\label{ihiholo7}
\end{equation}
where $F_\theta(\gamma_1)\in G$ is the line parallel transport
along $\gamma_1$ and, for each $i$,  
$F_{\theta,\varUpsilon}(\varSigma_i)\in H$ is the surface parallel transport along $\varSigma_i$,
all defined according to an appropriate higher gauge theoretic prescription. 

\subsection{\normalsize \textcolor{blue}{Outlook}}\label{sec:outlook}

The results we have succinctly expounded above suggest a possible 
avenue for the definition of surface knot invariants. 
In ordinary gauge theory, relations \eqref{ihiholo2} and \eqref{ihiholo3}
indicate that any construction of knot invariants out of knot holonomies 
that is gauge and pointing choice independent can be achieved only
by conjugation invariant mappings $\chi:G\rightarrow T$, where $T$ is some set.
An obvious choice of $\chi$ are the characters of the representations $R$ of $G$,
$\tr_R$. In strict higher gauge theory,  \eqref{ihiholo2} and \eqref{ihiholo3}
are replaced by relations \eqref{ihiholo6/0}, \eqref{ihiholo6} and \eqref{ihiholo7}, \eqref{ihiholo8},
respectively.
These show that, in order to obtain gauge and marking choice independent
surface knot invariants out of surface knot holonomies, one needs an extension of the notion 
of conjugation invariant mappings from groups to crossed modules. This problem will 
be analyzed in the companion paper \cite{SZ:2015} using the theory of higher quandles.

\vspace{.5cm}

\noindent
\textcolor{blue}{Acknowledgements.}
We thank E. Soncini for participating in the early stages of this project and 
Dan Margalit for sharing with us his insight about mapping class groups.  
We acknowledge financial support from INFN 
Research Agency. 

\vfil\eject

\section{\normalsize \textcolor{blue}{Surface knot theory}}\label{sec:hiknot}

\hspace{.5cm} In this section, the theory of knots is reviewed and that of 
surface knots is thoroughly expounded.
With the term ``knot theory'', we do not mean the topological theory of knots as detailed for instance 
in refs. \cite{Kauffman:1991ds,Kauffman:1995hc} and, in the higher case, in refs. 
\cite{Carter:1998cas,Kamada:2002ska}. 
What we are to delineate is a formal framework capable of associating with any knot a curve with sitting 
instants and, by extension, with any surface knot a surface with sitting instants with natural properties
\cite{Caetano:1993zf}. The definition and analysis of holonomy presented in the 
subsequent section, in fact, work only for curves and surfaces with sitting instants. 
Only very basic notions of the topological theory such as those of smooth embedding, smooth 
homotopy and ambient isotopy will be invoked in the following inquiry.  

Because of the nature of our approach we have just recalled, it is not required anywhere that knots or 
surface knots have codimension $2$ in the ambient manifold. Therefore, we only assume that the dimension 
of the latter is sufficiently large to allow for embeddings of the relevant types. 
The interesting applications, of course, belong to the codimension $2$ case.


\subsection{\normalsize \textcolor{blue}{Curves, surfaces and homotopy}}\label{sec:path}

\hspace{.5cm}  In this subsection, we survey the basic notions 
of curves and surfaces and of thin homotopy and homotopy thereof
expressing them in a way that is suited for higher parallel transport and holonomy 
theory. We do not stress however the categorical aspects of the subject, though 
they are admittedly important. This material is not original and is presented mainly to make 
this paper self--contained. For this reason, we give no proof of the basic results.

As an introduction to the topic, we review first
the familiar definitions of curves and thin homotopy and homotopy 
thereof used in the formulation of parallel transport 
of ordinary gauge theory.


\begin{defi} \label{def:homotp1}
Let $p_0,p_1$ be points of $M$. A curve $\gamma:p_0\rightarrow p_1$ of $M$ with sitting instants is a
mapping $\gamma\in\Map(\mathbb{R},M)$ satisfying
\begin{subequations}
\label{path1,2}
\begin{align}
&\gamma(x)=p_0\qquad \text{for $x<\epsilon$},
\vphantom{\Big]}
\label{path1}
\\
&\gamma(x)=p_1\qquad \text{for $x>1-\epsilon$}
\vphantom{\Big]}
\label{path2}
\end{align}
\end{subequations}
for some $\epsilon>0$ with $\epsilon<1/2$ depending on $\gamma$. The curve $\gamma$ is called closed 
if $p_0=p_1$. In such a case, $\gamma$ is said based at $p_0$.  
\end{defi}
All curves considered in this paper will be assumed to have sitting instants unless otherwise stated.

\begin{defi} \label{def:homotp2}
Let $p$ be a point of $M$. The unit curve $\iota_p:p\rightarrow p$ of $p$ is 
the curve defined by \hphantom{xxxxxxxxxxxxxxxxxxx}
\begin{equation}
\iota_p(x)=p.
\label{path3}
\end{equation}
Let $p_0,p_1$ be points and $\gamma:p_0\rightarrow p_1$ be a curve of $M$. The inverse
curve of $\gamma$ is the curve $\gamma^{-1_\circ}:p_1\rightarrow p_0$ given by \hphantom{xxxxxxxxx}
\begin{equation}
\gamma^{-1_\circ}(x)=\gamma(1-x).
\label{path4}
\end{equation}
Let $p_0,p_1,p_2$ be points and $\gamma_1:p_0\rightarrow p_1$, $\gamma_2:p_1\rightarrow p_2$ 
be curves of $M$. The composition of $\gamma_1$, $\gamma_2$ is the curve 
$\gamma_2\circ\gamma_1:p_0\rightarrow p_2$ defined piecewise by 
\begin{subequations}
\label{path5,6}
\begin{align}
&\gamma_2\circ\gamma_1(x)=\gamma_1(2x) \qquad \text{for $x\leq 1/2$},
\vphantom{\Big]}
\label{path5}
\\
&\gamma_2\circ\gamma_1(x)=\gamma_2(2x-1) \qquad \text{for $x\geq 1/2$}.
\vphantom{\Big]}
\label{path6}
\end{align}
\end{subequations}
\end{defi}

In gauge theory, parallel transport along curves, especially in the form of 
holonomy along closed curves, enjoys properties of homotopy invariance 
which make it one of the basic elements of the gauge theoretic construction 
of knot invariants. Homotopy plays therefore a fundamental role. 
It comes in two forms which are defined next. 


\begin{defi} \label{def:homotp3}
Let $p_0,p_1$ be points and $\gamma_0,\gamma_1:p_0\rightarrow p_1$ be curves. 
A thin homotopy $h:\gamma_0\Rightarrow \gamma_1$ is a mapping $h\in\Map(\mathbb{R}^2,M)$
such that
\begin{subequations}
\label{path7,8,9,10}
\begin{align}
&h(x,y)=p_0\qquad \text{for $x<\epsilon$},
\vphantom{\Big]}
\label{path7}
\\
&h(x,y)=p_1\qquad \text{for $x>1-\epsilon$},
\vphantom{\Big]}
\label{path8}
\\
&h(x,y)=\gamma_0(x)\qquad \text{for $y<\epsilon$},
\vphantom{\Big]}
\label{path9}
\\
&h(x,y)=\gamma_1(x)\qquad \text{for $y>1-\epsilon$}
\vphantom{\Big]}
\label{path10}
\end{align}
\end{subequations}
for some $\epsilon>0$ with $\epsilon<1/2$ and that 
\begin{equation}
\rank(dh(x,y))\leq 1.
\label{path11} 
\end{equation}
$\gamma_0$, $\gamma_1$ are called thin homotopy equivalent if there is thin homotopy 
$h:\gamma_0\Rightarrow \gamma_1$. 
If condition \eqref{path11} 
is dropped, then $h:\gamma_0\Rightarrow \gamma_1$ is called a homotopy and $\gamma_0$, $\gamma_1$ 
are called homotopy equivalent. 
\end{defi}

Let $N$, $M$ be manifolds and let $f:N\rightarrow M$ be a map. We can left 
compose by $f$ curves $\gamma$ and (thin) homotopies of curves $h$ in $N$ 
regarded as maps with target $N$ and obtain curves $f\circ\gamma$ and (thin) 
homotopies of curves $f\circ h$ in $M$ as established by the following 
proposition. 

 
\begin{prop} \label{prop:pushfwd1}
Let $p_0,p_1$ be points and $\gamma:p_0\rightarrow p_1$ be a curve of $N$. Then,
$f\circ\gamma:f(p_0)\rightarrow f(p_1)$ is a curve of $M$, called the push--forward 
of $\gamma$ by $f$. 

\noindent 
Let $p_0,p_1$ be points and $\gamma_0,\gamma_1:p_0\rightarrow p_1$ be curves of $N$.
If $h:\gamma_0\Rightarrow \gamma_1$ is a (thin) homotopy in $N$, then 
$f\circ h:f\circ \gamma_0\Rightarrow f\circ \gamma_1$ is a (thin) homotopy in $M$,
called the push--forward of $h$ by $f$.
\end{prop}

Let $\Pi_1M$ be the set of all curves of a manifold $M$. Def. \ref{def:homotp2} endows $(M,\Pi_1M)$ 
with a pregroupoid structure, that is a set of operations of the kind that would be required for 
$(M,\Pi_1M)$ to be a groupoid. As is well--known, however, $(M,\Pi_1M)$ is not one, as invertibility 
and associativity fail to hold. 

\vspace{1truemm}
\noindent
It is readily verified that 
both thin homotopy and homotopy are equivalence relations
on $\Pi_1M$ compatible with the pregroupoid structure of $(M,\Pi_1M)$. Upon replacing $\Pi_1M$ with the sets 
$P_1M$ and $P^0{}_1M$ of all thin homotopy and homotopy classes of curves of $M$, respectively, 
and equipping $(M,P_1M)$ and $(M,P^0{}_1M)$ with the pregroupoid structure induced by that of $(M,\Pi_1M)$, 
one obtains two genuine groupoids $(M,P_1M)$ and $(M,P^0{}_1M)$, called 
the path and the fundamental groupoids of $M$, respectively. The content of these groupoids 
has the intuitive diagrammatic representation \hphantom{xxxxxxxxxxxxxxxxxxxxxxxx}
\begin{equation}
\xymatrix{{\text{\footnotesize $p_1$}}&{\text{\footnotesize $p_0$}}\ar[l]_\gamma}\!,
\label{}
\end{equation}
where $\gamma$ is understood as a (thin) homotopy class of curves. 
$(M,P_1M)$, $(M,P^0{}_1M)$ enter in a basic way in the categorical theory
of parallel transport. A formulation along this lines has been developed in refs.
\cite{Schrei:2009,Schrei:2011,Schrei:2008}, 
but lies outside the scope of this paper. 

The above constructions have a higher extension. To curves 
connecting pairs of points, one adds surfaces stretching between pairs of curves with common endpoints. 
We now review the definitions of curves, surfaces and thin homotopy and homotopy 
thereof used in the formulation of higher parallel transport. 

\begin{defi} \label{def:homotp4}
For any two points $p_0,p_1$ of $M$, a curve $\gamma:p_0\rightarrow p_1$ with sitting instants, 
in particular a closed one,  is characterized as in def. \ref{def:homotp1}. 

\noindent
Let $p_0,p_1$ be points and $\gamma_0,\gamma_1:p_0\rightarrow p_1$ be curves of $M$. 
A surface $\varSigma:\gamma_0\Rightarrow\gamma_1$ of $M$ with sitting instants
is a map $\varSigma\in\Map(\mathbb{R}^2,M)$ such that  
\begin{subequations}
\begin{align}
&\varSigma(x,y)=p_0\qquad \text{for $x<\epsilon$},
\vphantom{\Big]}
\label{path12}
\\
&\varSigma(x,y)=p_1\qquad \text{for $x>1-\epsilon$},
\vphantom{\Big]}
\label{path13}
\\
&\varSigma(x,y)=\gamma_0(x)\qquad \text{for $y<\epsilon$},
\vphantom{\Big]}
\label{path14}
\\
&\varSigma(x,y)=\gamma_1(x)\qquad \text{for $y>1-\epsilon$}
\vphantom{\Big]}
\label{path15}
\end{align}
\end{subequations}
for some $\epsilon>0$ with $\epsilon<1/2$ depending on $\gamma_0$, $\gamma_1$, $\varSigma$. 
The surface $\varSigma$ is called closed 
if $\gamma_0=\gamma_1$. In such a case, $\varSigma$ is said to be based at $\gamma_0$.  
\end{defi}
As with curves, all surfaces considered in this paper will be assumed to have sitting instants unless otherwise stated.

\begin{defi} \label{def:homotp5}
For a point $p$ of $M$, the unit curve $\iota_p:p\rightarrow p$ of $p$ is defined as in \eqref{path3}.
For points $p_0,p_1$ and a curve $\gamma:p_0\rightarrow p_1$ of $M$, the inverse
curve $\gamma^{-1_\circ}:p_1\rightarrow p_0$ is defined as in \eqref{path4}.
For points $p_0,p_1,p_2$ and curves $\gamma_1:p_0\rightarrow p_1$, $\gamma_2:p_1\rightarrow p_2$ of $M$, 
the composed curve $\gamma_2\circ\gamma_1:p_0\rightarrow p_2$ is defined as in \eqref{path5,6}. 

\noindent
Let $p_0,p_1$ be points and $\gamma:p_0\rightarrow p_1$ be a curve of $M$. The unit surface 
$I_\gamma:\gamma\Rightarrow\gamma$ of $\gamma$ is the surface given by \hphantom{xxxxxxxxxx}
\begin{equation}
I_\gamma(x,y)=\gamma(x).
\label{path16}
\end{equation}
Let $p_0,p_1$ be points,  $\gamma_0,\gamma_1:p_0\rightarrow p_1$ be curves and 
$\varSigma:\gamma_0\Rightarrow\gamma_1$ be a surface of $M$. The vertical inverse of $\varSigma$
is the surface $\varSigma^{-1_\bullet}:\gamma_1\Rightarrow \gamma_0$ 
\begin{equation}
\varSigma^{-1_\bullet}(x,y)=\varSigma(x,1-y).
\label{path17}
\end{equation}
Let $p_0,p_1$ be points, 
$\gamma_0,\gamma_1,\gamma_2:p_0\rightarrow p_1$ be curves and 
$\varSigma_1:\gamma_0\Rightarrow\gamma_1$, $\varSigma_2:\gamma_1\Rightarrow\gamma_2$ be 
surfaces of $M$. The vertical composition of $\varSigma_1$, $\varSigma_2$
is the surface $\varSigma_2\bullet\varSigma_1:\gamma_0\Rightarrow \gamma_2$ defined piecewise by 
\begin{subequations}
\label{path18,19}
\begin{align}
&\varSigma_2\bullet\varSigma_1(x,y)=\varSigma_1(x,2y) \qquad \text{for $y\leq 1/2$},
\vphantom{\Big]}
\label{path18}
\\
&\varSigma_2\bullet\varSigma_1(x,y)=\varSigma_2(x,2y-1) \qquad \text{for $y\geq 1/2$}.
\vphantom{\Big]}
\label{path19}
\end{align}
\end{subequations}
Let $p_0,p_1$ be points, $\gamma_0,\gamma_1:p_0\rightarrow p_1$ be curves and 
$\varSigma:\gamma_0\Rightarrow\gamma_1$ be a surface of $M$. The horizontal inverse of $\varSigma$ 
is the surface $\varSigma^{-1_\circ}:\gamma_0{}^{-1_\circ}\Rightarrow \gamma_1{}^{-1_\circ}$ 
\begin{equation}
\varSigma^{-1_\circ}(x,y)=\varSigma(1-x,y).
\label{path20}
\end{equation}
Let $p_0,p_1,p_2$ be points, $\gamma_0,\gamma_1:p_0\rightarrow p_1$, $\gamma_2,\gamma_3:p_1\rightarrow p_2$ 
be curves and $\varSigma_1:\gamma_0\Rightarrow\gamma_1$, $\varSigma_2:\gamma_2\Rightarrow\gamma_3$ be 
surfaces of $M$. The horizontal composition of $\varSigma_1$, $\varSigma_2$
is the surface $\varSigma_2\circ\varSigma_1:\gamma_2\circ\gamma_0\Rightarrow \gamma_3\circ\gamma_1$ 
given piecewise by 
\begin{subequations}
\label{path21,22}
\begin{align}
&\varSigma_2\circ\varSigma_1(x,y)=\varSigma_1(2x,y) \qquad \text{for $x\leq 1/2$},
\vphantom{\Big]}
\label{path21}
\\
&\varSigma_2\circ\varSigma_1(x,y)=\varSigma_2(2x-1,y) \qquad \text{for $x\geq 1/2$}.
\vphantom{\Big]}
\label{path22}
\end{align}
\end{subequations}
\end{defi}

In higher gauge theory, analogously to the ordinary one, higher parallel transport along curves and surfaces, 
in particular holonomy along closed curves and surfaces, enjoys properties of homotopy invariance which 
make it a candidate as one of the basic elements of the higher gauge theoretic construction of surface knot 
invariants, which is the main goal of the present paper. Again, therefore, homotopy 
plays  a fundamental role. Again, it comes in two forms which are defined next.

\begin{defi} \label{def:homotp6} For any two points $p_0,p_1$ and curves $\gamma_0,\gamma_1:p_0\rightarrow p_1$ of $M$, 
a thin homotopy $h:\gamma_0\Rightarrow\gamma_1$ and the thin homotopy equivalence of 
$\gamma_0, \gamma_1$ are characterized as in def. \ref{def:homotp3}.

\noindent
Let $p_0,p_1$ be points, $\gamma_0,\gamma_1,\gamma_2,\gamma_3:p_0\rightarrow p_1$ be curves
and $\varSigma_0:\gamma_0\Rightarrow\gamma_1$, $\varSigma_1:\gamma_2\Rightarrow\gamma_3$
be surfaces of $M$. A thin homotopy of $H:\varSigma_0\Rrightarrow\varSigma_1$ is a 
mapping $H\in\Map(\mathbb{R}^3,M)$ such that 
\begin{subequations}
\label{path23,24,25,26,27,28}
\begin{align}
&H(x,y,z)=p_0 \qquad \text{for $x< \epsilon$}, \hspace{1.5cm}
\vphantom{\Big]}
\label{path23}
\\
&H(x,y,z)=p_1 \qquad \text{for $x>1-\epsilon$},
\vphantom{\Big]}
\label{path24}
\\
&H(x,y,z)=H(x,0,z) \qquad \text{for $y<\epsilon$},
\vphantom{\Big]}
\label{path25}
\\
&H(x,y,z)=H(x,1,z) \qquad \text{for $y>1-\epsilon$},
\vphantom{\Big]}
\label{path26}
\\
&H(x,y,z)=\varSigma_0(x,y) \qquad \text{for $z<\epsilon$},
\vphantom{\Big]}
\label{path27}
\\
&H(x,y,z)=\varSigma_1(x,y) \qquad \text{for $z>1-\epsilon$}
\vphantom{\Big]}
\label{path28}
\end{align}
\end{subequations}
for some $\epsilon>0$ with $\epsilon<1/2$ and that 
\begin{subequations}
\label{path29,30}
\begin{align}
&\rank(dH(x,0,z)),~\rank(dH(x,1,z))\leq 1,
\vphantom{\Big]}
\label{path29}
\\
&\rank(dH(x,y,z))\leq 2.
\vphantom{\Big]}
\label{path30} 
\end{align}
\end{subequations}
$\varSigma_0$, $\varSigma_1$ are called thin homotopy equivalent
if there is a thin homotopy $H:\varSigma_0\Rrightarrow\varSigma_1$. 
If condition \eqref{path30} 
is not imposed, then $H:\varSigma_0\Rrightarrow\varSigma_1$ is called a homotopy 
and $\varSigma_0$, $\varSigma_1$ are called homotopy equivalent. 
\end{defi}
We note that condition \eqref{path29} 
implies that the source curves $\gamma_0,\gamma_2$ of $\varSigma_0$, $\varSigma_1$ are thin homotopy equivalent, 
and similarly the target ones $\gamma_1,\gamma_3$. 

Let $N$, $M$ be manifolds and let $f:N\rightarrow M$ be a map. We can left compose by 
$f$ curves $\gamma$, surfaces $\varSigma$, thin homotopies of curves $h$ and (thin) homotopies 
of surfaces $H$ in $N$ regarded as maps with target $N$ and obtain curves $f\circ \gamma$, 
surfaces $f\circ \varSigma$, thin homotopies of curves $f\circ h$ and (thin) homotopies 
of surfaces $f\circ H$ in $M$. This is detailed in the following proposition.  


\begin{prop} \label{prop:pushfwd2}
The statements of prop. \ref{prop:pushfwd1} about curves and homotopies of curves 
continue to hold with the exclusion of those concerning non thin homotopies.

\noindent
Let $p_0,p_1$ be points, $\gamma_0,\gamma_1:p_0\rightarrow p_1$ be curves 
and $\varSigma:\gamma_0\Rightarrow\gamma_1$ be a surface of $N$. Then, 
$f\circ\varSigma:f\circ\gamma_0\Rightarrow f\circ \gamma_1$ is a surface of $M$, 
the push--forward of $\varSigma$ by $f$.

%

\noindent
Let $p_0,p_1$ be points, $\gamma_0,\gamma_1,\gamma_2,\gamma_3:p_0\rightarrow p_1$ be curves and
$\varSigma_0:\gamma_0\Rightarrow\gamma_1$, $\varSigma_1:\gamma_2\Rightarrow\gamma_3$ be surfaces 
of $N$. If $H:\varSigma_0\Rrightarrow\varSigma_1$ is a (thin) homotopy of surfaces in $N$, then
$f\circ H:f\circ\varSigma_0\Rrightarrow f\circ\varSigma_1$ is a (thin) homotopy in $M$, called
the push-forward of $H$ by $f$. 
\end{prop}



Let $\Pi_1M$, $\Pi_2M$ be the sets of all curves and surfaces of $M$, respectively.  
Def. \ref{def:homotp5} endows $(M,\Pi_1M,\Pi_2M)$ with a $2$--pregroupoid structure, 
that is a set of operations of the kind that would be required for 
$(M,\Pi_1M,\Pi_2M)$ to be a groupoid. Analogously to the ordinary case,  
$(M,\Pi_1M,\Pi_2M)$ is not one however, as invertibility 
and associativity fail to hold both for curves and 
surfaces in the vertical as well as the horizontal variants.

\noindent
Thin homotopy of curves and (thin) homotopy of surfaces are equivalence relations
on $\Pi_1M$, $\Pi_2M$, respectively, compatible with the $2$--pregroupoid structure 
of $(M,\Pi_1M,\Pi_2M)$. Upon replacing $\Pi_1M$ with the set $P_1M$ of all thin homotopy 
classes of curves and $\Pi_2M$ with the sets of thin homotopy and homotopy classes
of surfaces, respectively, and equipping $(M,P_1M,P_2M)$ and $(M,P_1M,P^0{}_2M)$ 
with the $2$--pregroupoid structure induced by that of $(M,\Pi_1M,\Pi_2M)$, 
one obtains two genuine $2$--groupoids $(M,P_1M,P_2M)$ and $(M,P_1M,P^0{}_2M)$, called 
the path and the fundamental $2$--groupoid of $M$, respectively. The content of these $2$--groupoids 
has the intuitive diagrammatic representation 
\begin{equation}
\xymatrix@C=3pc{
   {\text{\footnotesize $p_1$}}
& {\text{\footnotesize $p_0$}} \ar@/^1pc/[l]^{\gamma_1}="0"
           \ar@/_1pc/[l]_{\gamma_0}="1"           \ar@{=>}"1"+<0ex,-2.ex>;"0"+<0ex,2.ex>_{\varSigma\,}
}
\label{}
\end{equation}
where $\gamma_0$, $\gamma_1$ are understood as thin homotopy classes of curves
and $\varSigma$ as a (thin) homotopy class of surfaces. In \cite{Schrei:2009,Schrei:2011,Schrei:2008}, 
the categorical theory of higher parallel transport based on 
$(M,P_1M,P_2M)$, $(M,P_1M,P^0{}_2M)$ has been worked out, but we shall not rely on it 
in this paper.


\subsection{\normalsize \textcolor{blue}{$2$--knots and their associated surfaces}}\label{sec:knot}

\hspace{.5cm} In this subsection, we shall study how one can describe higher knots in a way suitable for 
higher gauge theory.

In what follows, the terms curves and surfaces will be always used with the meaning 
stated in defs. \ref{def:homotp1}, \ref{def:homotp4} unless otherwise stated.
Curves and surfaces understood as topological/differentiable $1$-- and $2$--dimensional manifolds will be called
topological/differentiable curves and surfaces, respectively.

Before proceeding to the study of knots, it is necessary to recall a few basic notions of 
differential topology. 

\begin{defi} \label{def:knot1}
Let $M$, $N$ be manifolds and let $f_0,~f_1:N\rightarrow M$ be maps. A homotopy $h$ of $f_0$, $f_1$ 
with sitting instants
is a smooth family $h_u:N\rightarrow M$, $u\in\mathbb{R}$, of maps such that 
\begin{subequations}
\label{zknot1,2}
\begin{align}
&h_u=f_0\qquad \text{for $u<\epsilon$},
\vphantom{\Big]}
\label{zknot1}
\\
&h_u=f_1\qquad ~~\text{for $u>1-\epsilon$},
\vphantom{\Big]}
\label{zknot2}
\end{align}
\end{subequations} 
for some number $\epsilon$ with $0<\epsilon<1/2$. The maps $f_0$, $f_1$ are called homotopic when they are 
related by a homotopy. If $f_0$, $f_1$ and the $h_u$ are all embeddings, then $h$ is called an isotopy 
and $f_0$, $f_1$ are called isotopic.
\end{defi} 

\noindent
Manifolds and maps and manifolds and maps up to homotopy form categories. 

There is a stronger notion of homotopy, called ambient isotopy. 

\begin{defi} \label{def:knot2}
Let $M$, $N$ be manifolds and let $f_0,~f_1:N\rightarrow M$ be maps. An ambient isotopy of  
$f_0$, $f_1$ with sitting instants is a smooth family $F_u$, $u\in\mathbb{R}$, of 
autodiffeomorphisms of $M$ with the property that 
\begin{subequations}
\label{knot7,8}
\begin{align}
&F_u=\id_M\qquad \text{for $u<\epsilon$},
\vphantom{\Big]}
\label{knot7}
\\
&F_u=F_1\qquad ~~\text{for $u>1-\epsilon$},
\vphantom{\Big]}
\label{knot8}
\end{align}
\end{subequations} 
for some number $\epsilon$ with $0<\epsilon<1/2$ and that 
\begin{equation}
f_1=F_{1}\circ f_0.
\label{knot9/0}
\end{equation}
The maps $f_0$, $f_1$ are called ambient isotopic when they are related by
an ambient isotopy. 
\end{defi}

The requirement that homotopies and isotopies have sitting instants is not 
necessary strictly speaking. We have introduced it for the sake of a more uniform formulation. 
\pagebreak All homotopies and isotopies considered in this paper will be assumed to have sitting 
instants unless otherwise stated. 

We provide next a formulation of ordinary knot theory
based on the smooth set--up of subsect. \ref{sec:path}.
Proceeding along the same lines, we shall then  
construct a full--fledged formulation of surface knot theory relying on the same framework. 


Let $C$ be an oriented differentiable closed curve, that is a circle. 

\begin{defi} \label{def:knot3} A pointing of $C$ consists in a distinguished point $p_C$ of $C$. 
\end{defi}

\noindent
Below, we shall assume that $C$ is equipped with a fixed pointing.

\begin{defi} \label{def:cfree}
A free $C$--knot $\xi$ of $M$ is an embedding $\xi:C\rightarrow M$.
\end{defi}

\noindent
The image $\xi(C)$ of a free $C$--knot $\xi$ is a differentiable closed curve in $M$ with the 
orientation induced by that of $C$ via the embedding. The image of $\xi$ should not be confused 
with $\xi$ itself.

Ambient isotopy of free $C$--knots is defined in the obvious fashion. 

\begin{defi} \label{def:cfreeamb}
Two free $C$--knot $\xi_0$, $\xi_1$ of $M$ are ambient isotopic 
if they are as embeddings of $C$ into $M$. 
\end{defi}

\noindent
So, if $F$ is the ambient isotopy transforming $\xi_0$ into $\xi_1$, then  
\begin{equation}
\xi_1=F_{1}\circ \xi_0.
\label{knot9/1}
\end{equation} 
Ambient isotopy of free $C$--knots is an equivalence relation.
We denote 
the set of all ambient isotopy classes of free $C$--knots by $\Knot_{\mathrm{F}C}(M)$.

The basic constructions with knots we are going to introduce work for based knots. 
This requires that a $C$--pointing on the target manifold $M$ be given. 

\begin{defi} \label{def:knot4} A manifold $M$ is called $C$--pointed, 
if a distinguished point $p_M$ of $M$ is specified. 
\end{defi}

\noindent
In common parlance, $M$ is called pointed. Clearly, $C$ itself is $C$--pointed. 

\begin{defi} \label{def:knot5} Let $M$, $N$ be $C$--pointed manifolds and let $f:N\rightarrow M$ be a map.
$f$ is said to be $C$--pointing preserving if 
\begin{equation}
f(p_N)=p_M.
\label{knot3/0}
\end{equation}
\end{defi}
In usual terms, $f$ is a pointed map.
One defines in analogous fashion  $C$--pointing preserving diffeomorphisms, autodiffeomorphisms etc.

\begin{defi} \label{def:knot6}
Let $M$, $N$ be $C$--pointed manifolds and let $f_0,~f_1:N\rightarrow M$ be $C$--pointing preserving maps.
A $C$--pointing preserving homotopy $h$ of $f_0$, $f_1$ is an ordinary homotopy of $f_0$, $f_1$ such that 
the maps $h_y:N\rightarrow M$, $y\in\mathbb{R}$, are %
all $C$--pointing preserving, 
\hphantom{xxxxxxxxxxxxxxxxxxxxxxx}
\begin{equation}
h_y(p_N)=p_M.
\label{zknot3/0}
\end{equation}
The $C$--pointing preserving maps $f_0$, $f_1$ are called  homotopic, if they are related by
a $C$--pointing preserving homotopy. If $f_0$, $f_1$ and the $h_y$ are 
all embeddings, $h$ is called a $C$--pointing preserving 
isotopy and $f_0$, $f_1$ are said isotopic.
\end{defi} 

\noindent
$C$--pointed manifolds and $C$--pointing preserving maps form a category. 
$C$--pointed manifolds and $C$--pointing preserving maps up to $C$--pointing preserving 
homotopy constitute also a category. 

\begin{defi} \label{def:knot7}
Let $M$, $N$ be $C$--pointed manifolds and let $f_0,~f_1:N\rightarrow M$ be $C$--pointing preserving maps.
A $C$--pointing preserving ambient isotopy of $f_0$, $f_1$ is an ordinary ambient isotopy of $f_0$, $f_1$ 
such that the autodiffeomorphisms $F_y$, $y\in\mathbb{R}$, are all $C$--pointing preserving,
\begin{equation}
F_y(p_M)=p_M.
\label{zknot10/0}
\end{equation}
The $C$--pointing preserving maps $f_0,~f_1$ are called ambient isotopic, if they are related by an 
$C$--pointing preserving ambient isotopy. 
\end{defi} 

Given a $C$--pointed manifold $M$, we can define $C$--knots. 

\begin{defi} \label{def:knot8}
A $C$--knot of $M$ 
in the $C$--pointed manifold $M$ is a $C$--pointing preserving embedding
$\xi:C\rightarrow M$.
\end{defi}
A $C$--knot $\xi$ is thus based in that \hphantom{xxxxxxxxxxx}
\begin{equation}
\xi(p_C)=p_M.
\label{knot5/0}
\end{equation}
A $C$--knot $\xi$ is also a free $C$--knot, if one forgets the $C$--pointing of $M$.
Conversely, a free $C$--knot $\xi$ is secretly a $C$--knot belonging to the $C$--pointing $p_M=\xi(p_C)$ of $M$.
We note that $\id_C$ is a $C$--knot of $C$, called tautological $C$--knot. 

We want to identify knots related by an appropriate form of ambient isotopy.

\begin{defi} \label{def:knot9}
Two $C$--knot $\xi_0$, $\xi_1$ in the $C$--pointed manifold $M$ are ambient isotopic 
if they are as embeddings of $C$ into $M$
via a $C$--pointing preserving ambient isotopy. 
\end{defi}

\noindent
If two $C$--knots are ambient isotopic, they also are as free $C$--knots.
The converse is clearly generally false.  
Ambient isotopy of $C$--knots is an equivalence relation.
We denote by $\Knot_C(M)$
the set of all ambient isotopy classes of $C$--knots.

We shall now describe knots in terms of curves and (thin) homotopy
thereof (cf. subsect. \ref{sec:path}). 

\begin{defi} \label{def:knot10} 
A curve $\gamma_C$ of $C$ is said to be compatible with the pointing of $C$, if 
$\gamma_C:p_C\rightarrow p_C$ and $I_C=\gamma_C{}^{-1}(C\setminus p_C)$ is an
open interval of $\mathbb{R}$ such that 
$\gamma_C\big|_{I_C}$ is an orientation preserving diffeomorphism 
onto $C\setminus p_C$.
\end{defi} 

\noindent 
It $\gamma_C$ is compatible, then 
\begin{subequations}
\label{knot16,17/0}
\begin{align}
&\gamma_C(x)=p_C\qquad \text{for $x<\epsilon$},
\vphantom{\Big]}
\label{knot16/0}
\\
&\gamma_C(x)=p_C\qquad \text{for $x>1-\epsilon$},
\vphantom{\Big]}
\label{knot17/0}
\end{align}
\end{subequations} 
for some number $\epsilon$ with $0<\epsilon<1/2$, by \eqref{path1,2}. 
Further, the interval $I_C$ is contained in the open interval $(\epsilon,1-\epsilon)$
and $\gamma_C\big|_{I_C}$ provides a one--to-one parametrization of $C\setminus p_C$
consistent with its orientation. 

Fix a curve $\gamma_C$ of $C$ compatible with the pointing of $C$.
Given a $C$--knot $\xi$, one can push--forward the curve $\gamma_C:p_C\rightarrow p_C$
of $C$ by the map $\xi:C\rightarrow M$ yielding a curve $\xi\circ \gamma_C:p_M\rightarrow p_M$ of $M$. 

\begin{defi} \label{def:knot11} For a $C$--knot $\xi$, let $\gamma_\xi:p_M\rightarrow p_M$
be the curve of $M$ 
\begin{equation}
\gamma_\xi=\xi\circ \gamma_C.
\label{knot21/0}
\end{equation}
\end{defi}



\noindent
By \eqref{knot16/0}, \eqref{knot17/0}, $\gamma_\xi$ satisfies 
\begin{subequations}
\label{knot23,24/0}
\begin{align}
&\gamma_\xi(x)=p_M\qquad \text{for $x<\epsilon$},
\vphantom{\Big]}
\label{knot23/0}
\\
&\gamma_\xi(x)=p_M\qquad \text{for $x>1-\epsilon$},
\vphantom{\Big]}
\label{knot24/0}
\end{align}
\end{subequations} 
for the same number $\epsilon$ as in \eqref{knot16,17/0}. Further, 
by the compatibility of $\gamma_C$, $\gamma_\xi\big|_{I_C}$
furnishes a one--to-one parametrization of $\xi(C)\setminus p_M$
consistent with its orientation. 
We notice also that the curve $\gamma_C$ is nothing but the curve
$\gamma_{\id_C}$ of $C$ of the tautological $C$--knot $\id_C$.  

The curve $\gamma_\xi$ associated with a $C$--knot $\xi$
enjoys several properties of invariance up to (thin) homotopy.

\begin{prop} \label{prop:knot2}
For any $C$--knot $\xi$ of $M$, the curve $\gamma_\xi$ is independent from 
the choice of the compatible curve $\gamma_C$ of $C$ up to thin homotopy.
\end{prop}

\noindent {\it Proof}. 
Fix an oriented differentiable universal covering $\tilde C$ of $C$
with orientation preserving covering map $\varpi_C:\tilde C\rightarrow C$. 

Let $\gamma_C$ be a curve of $C$ compatible with the pointing of $C$. 
Since $\gamma_C$ is a map $\gamma_C:\mathbb{R}\rightarrow C$ 
with a simply connected domain, there is a 
lifting map $\tilde\gamma_C:\mathbb{R}\rightarrow \tilde C$
such that $\varpi_C\circ\tilde\gamma_C=\gamma_C$ unique up to left composition 
by orientation preserving deck diffeomorphisms of $\tilde C$. 
The compatibility of $\gamma_C$ (cf. def. \ref{def:knot10}) implies 
further that there is an open subset $\tilde F_C$ of $\tilde C$ with the following 
properties. First, $\tilde F_C$ is the interior of a simply connected fundamental domain of 
the universal covering $\tilde C$. Second, $\tilde\gamma_C$ is a map of $\mathbb{R}$ onto 
$\overline{\tilde F}_C$ and $\tilde\gamma_C\big|_{I_C}$
is an orientation preserving diffeomorphism of $I_C$ 
onto $\tilde F_C$. Third, $\varpi_C\big|_{\tilde F_C}$is an orientation 
preserving diffeomorphism of $\tilde F_C$ onto the image of $\gamma_C\big|_{I_C}$.
The fact that $\gamma_C$ expresses a curve $\gamma_C:p_C\rightarrow p_C$ implies then that 
$\tilde \gamma_C$ similarly expresses a curve $\tilde\gamma_C:\tilde p_C\rightarrow \tilde p_C{}'$, where $\tilde p_C$, 
$\tilde p_C{}'$ are points of $\tilde C$ such that $\varpi(\tilde P_C)=\varpi_C(\tilde p_C{}')
=p_C$. As $\tilde F_C$ can be taken independent from $\gamma_C$ by the deck symmetry,
$\tilde p_C$, $\tilde p_C{}'$ also can. 

Concretely, modelling $C$ as $\mathbb{R}/\mathbb{Z}$,
$\tilde C$ as $\mathbb{R}$ and $\varpi_C$ as the quotient map of $\mathbb{R}$ modulo $\mathbb{Z}$
and taking $\tilde p_C=0$, $\tilde p_C{}'=1$,  
we can think of $\tilde \gamma_C$ as a 
surjective map $\tilde\gamma_C:\mathbb{R}\rightarrow [0,1]$
that is strictly increasing on $I_C$. 

Let $\gamma_{C0}$, $\gamma_{C1}$ be two curves of $C$ compatible with the pointing of $C$.
Let $\tilde \gamma_{C0}$, $\tilde \gamma_{C1}$ be the curves of $\tilde C$ lifting 
$\gamma_{C0}$, $\gamma_{C1}$, respectively. 
From the remarks of the previous paragraph, it follows that there is 
a map $\tilde h:\mathbb{R}^2\rightarrow\tilde C$ satisfying conditions \eqref{path7,8,9,10}
with $p_0=\tilde p_C$, $p_1=\tilde p_C{}'$ and $\gamma_0=\gamma_{C0}$, $\gamma_1=\gamma_{C1}$. 
By mere dimensional reasons, condition \eqref{path11} is fulfilled as well.
Hence, $\tilde h$ is a thin homotopy of $\tilde\gamma_{C0}$, $\tilde\gamma_{C1}$.
So, $h=\varpi_C\circ \tilde h:\mathbb{R}^2\rightarrow C$
is a thin homotopy of $\gamma_{C0}$, $\gamma_{C1}$. 

As $\gamma_{C0}$, $\gamma_{C1}$ are thin homotopic, for any $C$--knot $\xi$, 
$\xi\circ \gamma_{C0}$, $\xi\circ \gamma_{C1}$ also are. 
The statement follows. \hfill $\Box$  


Ambient isotopic $C$--knots yield homotopic curves, as it is natural.

\begin{prop} \label{prop:knot3}
If two $C$--knots $\xi_0$, $\xi_1$ of $M$ are ambient isotopic, then there exists
a homotopy $h:\gamma_{\xi_0}\Rightarrow \gamma_{\xi_1}$. 
\end{prop}

\noindent {\it Proof}. Let $F_y$, $y\in\mathbb{R}$, be a $C$--pointing preserving ambient isotopy relating
the knots $\xi_0$, $\xi_1$. Define a map $h:\mathbb{R}^2\rightarrow M$ by 
\begin{equation}
h(x,y)=F_{y}\circ\gamma_{\xi_0}(x).  \vphantom{\ul{\ul{\ul{\ul{\ul{x}}}}}}
\label{knotb1/0}
\end{equation}
Using relations \eqref{knot23,24/0} together with  \eqref{knot7,8} and \eqref{knot9/1}, 
it is readily checked that $h$ 
satisfies the relations \eqref{path7,8,9,10} with $p_0=p_1=p_M$ and $\gamma_0=\gamma_{\xi_0}$,
$\gamma_1=\gamma_{\xi_1}$. It follows that \eqref{knotb1/0} defines a homotopy  
$h:\gamma_{\xi_0}\Rightarrow \gamma_{\xi_1}$. \hfill $\Box$  

\noindent
As a consequence, the curve $\gamma_\xi$ depends only on the ambient isotopy class
of the knot $\xi$ up to homotopy and the map $\xi\rightarrow \gamma_\xi$ descends
to one from $\Knot_C(M)$ to set of the curves of $M$ mod homotopy. 


So far, we have considered $C$--knots belonging to a fixed $C$--pointing of $M$.
In view of topological applications, it is however necessary to compare
$C$--knots belonging to possibly distinct $C$--pointings. As before, 
these knots will be topologically equivalent if they are 
ambient isotopic. Since $C$--pointing preservation is now necessarily violated, 
the form of ambient isotopy involved here is more general than 
the one considered up to this point. We call it free for to comply 
with customary terminology. 
Prop. \ref{prop:knot3} extends also to freely ambient 
isotopic $C$--knots in a form which we shall make precise next. 

When several $C$--pointings $p_{M0},\, p_{M1},\,\ldots$ of $M$ are given, 
we shall denote the $C$--pointed manifold 
$M$ by $M_0,\,M_1,\,\ldots$ and shall call the $C$--knots $\xi$ of $M$ as 
$C$--knots of $M_0,\,M_1,\,\ldots$,
respectively.

Suppose we are given two possibly distinct $C$--pointings $p_{M0}$, $p_{M1}$ of $M$. 

\begin{defi} \label{def:cknotfreeamb}
Two $C$--knots $\xi_0$, $\xi_1$ of $M_0$, $M_1$, respectively, 
are called freely ambient isotopic if they are related by 
an ambient isotopy $W$.
\end {defi}
 
\noindent
We stress that, when the pointings are equal,
free ambient isotopy does not re\-duce to ambient isotopy, 
since it is not required to be $C$--pointing preserving.


We consider now two $C$--knots $\xi_0$, $\xi_1$ of $M_0$, $M_1$, respectively,
which are freely ambient isotopic. If $W$ is the underlying connecting ambient isotopy, 
$\xi_0$, $\xi_1$ are then related by definition as \hphantom{xxxxxxxxxxxxxxxxxx}
\begin{equation}
\xi_1=W_{1}\circ \xi_0. \vphantom{\ul{\ul{\ul{\ul{\ul{x}}}}}}
\label{knot38/0}
\end{equation}
In virtue of \eqref{knot5/0}, $p_{M0}$, $p_{M1}$ are related accordingly, 
\begin{equation}
p_{M1}=W_1(p_{M0}). 
\label{knot30/0}
\end{equation} 
These relations suggest using the isotopy $W$ to 
construct a smooth family $p_{My}$ of $C$--pointings interpolating $p_{M0}$, $p_{M1}$ and 
a smooth family $\xi_y$ of $C$--knots of $M_y$ interpolating $\xi_0$, $\xi_1$.
To carry this out, we pick a non decreasing smooth function $\alpha:\mathbb{R}\rightarrow[0,1]$ such that
$\alpha(u)=0$ for $u<\epsilon$ and $\alpha(u)=1$ for $u>1-\epsilon$, where  
$0<\epsilon<1/2$ such that the \eqref{knot16,17/0} hold. 

\begin{defi} \label{def:knot13} For $y\in\mathbb{R}$, let $p_{My}$ be the $C$--pointing of $M$ given by 
\begin{equation}
p_{My}=W_{\alpha(y)}(p_{M0}).
\label{knot32/0}
\end{equation} 
\end{defi}
We note that $p_{My}=p_{M0}$ for $y<\epsilon$ and $p_{My}=p_{M1}$ for $y>1-\epsilon$, where $p_{M0}$ and $p_{M1}$
are the two given $C$--pointings. 

\begin{defi} \label{def:knot14} For $y\in\mathbb{R}$, let $\xi_y$ be the $C$--knot  
of $M_y$ given by 
\begin{equation}
\xi_y=W_{\alpha(y)}\circ \xi_0.
\label{knot39/0}
\end{equation}
\end{defi}

\par\noindent
Property \eqref{knot5/0} is immediately checked. We note that $\xi_y=\xi_0$ for $y<\epsilon$ and 
$\xi_y=\xi_1$ for $y>1-\epsilon$, where here $\xi_0$ and $\xi_1$ stand for the two given knots.


Using the interpolating setup just introduced, one can prove the following proposition.

\begin{prop} \label{prop:knot4}
Let $\xi_0$, $\xi_1$ be $C$--knots of $M_0$, $M_1$, respectively. 
Assume further that $\xi_0$, $\xi_1$ are freely ambient isotopic. 
Then, there are a curve $\gamma_1:p_{M0}\rightarrow p_{M1}$ of $M$ and a homotopy 
$h:\gamma_{\xi_0}\Rightarrow \gamma_1{}^{-1_\circ}\circ\gamma_{\xi_1}\circ \gamma_1$. 
\end{prop}

\noindent
As composition of curves is non associative, in a multiple composition
of curves we have to fix the sequential order in which the single
compositions are performed. Here and in the following, we tacitly
apply the ``compose the rightmost first'' convention:  we compose 
first the rightmost and next to rightmost curve, then the result with the next 
to next to rightmost curve and so on. Any other choice of convention leads to 
a curve that is thin homotopic to the one obtained in this way. 

\noindent
{\it Proof}. As $y$ varies, the point $p_{My}$ traces a differentiable curve in $M$.
Conjoined with this is a family of curves 
$\gamma_y:p_{M0}\rightarrow p_{My}$ of $M$ given by 
\begin{equation}
\gamma_y(x)=W_{\alpha(x)\alpha(y)}(p_{M0}).
\label{knot35/0}
\end{equation}
This has the property that $\gamma_y=\iota_{p_{M0}}$ 
for $y<\epsilon$ and $\gamma_y=\gamma_1$ 
for $y>1-\epsilon$. 

With the $C$--knot $\xi_y$, there is associated the curve $\gamma_{\xi_y}$ according to \eqref{knot21/0}. 
By composing successively the curves $\gamma_y:p_{M0}\rightarrow p_{My}$, $\gamma_{\xi_y}:p_{My}\rightarrow p_{My}$
and $\gamma_y{}^{-1_\circ}:p_{My}\rightarrow p_{M0}$, we obtain a curve  
$\widehat{\gamma}_{\xi_y}:p_{M0}\rightarrow p_{M0}$, viz
\begin{equation}
\widehat{\gamma}_{\xi_y}=\gamma_y{}^{-1_\circ}\circ\gamma_{\xi_y}\circ \gamma_y.
\label{knot36/0}
\end{equation}

For continuously varying $y$, the curve $\widehat{\gamma}_{\xi_y}$ shifts from
the curve $\widehat{\gamma}_{\xi_0}$ to that $\widehat{\gamma}_{\xi_1}$ yielding 
a homotopy $\widehat{h}:\widehat{\gamma}_{\xi_0}\Rightarrow\widehat{\gamma}_{\xi_1}$. 
Indeed, as a map $\widehat{h}:\mathbb{R}^2\rightarrow M$, $\widehat{h}$ is given by 
\begin{equation}
\widehat{h}(x,y)=\widehat{\gamma}_{\xi_y}(x).
\label{knot37/0}
\end{equation}
By the properties of the curves $\gamma_{\xi_y}$ and $\gamma_y$ 
and \eqref{knot37/0}, it is evident that $\widehat{h}$ is smooth and 
it is readily verified that $\widehat{h}$ satisfies the \eqref{path7,8,9,10} with
$p_0=p_1=p_{M0}$, $\gamma_0=\widehat{\gamma}_{\xi_0}$, $\gamma_1=\widehat{\gamma}_{\xi_1}$
and perhaps a different value of $\epsilon$. 

By \eqref{knot36/0}, the source and target curves of $\widehat{h}$ explicitly are
$\widehat{\gamma}_{\xi_0}=\iota_{p_{M0}}{}^{-1_\circ}\circ\gamma_{\xi_0}\circ \iota_{p_{M0}}$ 
and $\widehat{\gamma}_{\xi_1}
=\gamma_1{}^{-1_\circ}\circ\gamma_{\xi_1}\circ \gamma_1$. Since 
$\iota_{p_{M0}}{}^{-1_\circ}\circ\gamma_{\xi_0}\circ \iota_{p_{M0}}$
is thin homotopic to $\gamma_{\xi_0}$ itself, $\widehat{h}$ induces a homotopy 
$h:\gamma_{\xi_0}\Rightarrow \gamma_1{}^{-1_\circ}\circ\gamma_{\xi_1}\circ \gamma_1$. \hfill $\Box$ 


Since a free $C$--knot $\xi$ is a $C$--knot belonging to
the $C$--pointing $\xi(p_C)$ of $M$, we can construct the associated curve 
$\gamma_\xi:\xi(p_C)\rightarrow \xi(p_C)$ as detailed above. 

The curve $\gamma_\xi$ of a free $C$--knot $\xi$ \pagebreak 
depends on the the pointing $p_C$ of $C$, which up to this point 
we have tacitly assumed to be fixed, while $\xi$ as a mere embedding
of $C$ into $M$ does not. 
Of course, there are infinitely many choices of $p_C$ and any particular one 
is merely conventional. To be natural, our construction should be independent from the 
choice made in a suitable sense. The following analysis is in order.

When several pointings $p_{C0},~p_{C1},~\ldots$ of $C$ are envisaged, we shall denote $C$ 
by $C_0,~C_1,~\ldots$, call a target $C$--pointed manifold $M$ a $C_0$--, $C_1$--, $\ldots$ pointed 
manifold and the $C$--knots $\xi$ it supports $C_0$--, $C_1$--, $\ldots$ knots, respectively, 
to indicate which pointing of $C$ is referred to. 

Suppose that two pointings $p_{C0}$ and $p_{C1}$ of $C$ are given and that the manifold $M$ is 
$C_0$-- and $C_1$-- pointed by $p_{M0}$, $p_{M1}$, respectively.
Suppose further that $\xi$ is simultaneously a $C_0$--knot of $M_0$ and a $C_1$--knot of $M_1$. 
The natural question arises
about the relation between the curves $\gamma_{0\xi}$, $\gamma_{1\xi}$ associated with $\xi$ 
in the two pointings. The following proposition provides a simple answer to it. 

\begin{prop} \label{prop:c0knot}
If $\xi$ is at the same time a $C_0$--knot of $M_0$ and a $C_1$--knot of $M_1$, then there 
are a curve $\gamma_1:p_{M0}\rightarrow p_{M1}$ of $M$ and a thin homotopy 
$h:\gamma_{0\xi}\Rightarrow \gamma_1{}^{-1_\circ}\circ\gamma_{1\xi}\circ \gamma_1$, 
both with image contained in that of $\xi$.
\end{prop}


\noindent
{\it Proof}. 
To build the curves $\gamma_{0\xi}$, $\gamma_{1\xi}$, 
compatible curves $\gamma_{C0}$, $\gamma_{C1}$ of $C_0$, $C_1$, respectively, must be picked. 
By prop. \ref{prop:knot2}, $\gamma_{0\xi}$, $\gamma_{1\xi}$ are independent 
from the choice of $\gamma_{C0}$, $\gamma_{C1}$ up to thin homotopy.
This allows us to assume a convenient relationship between $\gamma_{C0}$, $\gamma_{C1}$.
We choose  $\gamma_{C0}$, $\gamma_{C1}$ satisfying  
$\gamma_{C1}=f\circ \gamma_{C0}$, where $f$ is an orientation preserving autodiffeomorphism 
of $C$ such that $p_{C1}$ $=f(p_{C0})$.

The composition $\xi\circ f$ is a $C_0$--knot of $M_1$. 
Since $C$ is the circle manifold which has no isotopically non trivial 
orientation preserving autodiffeomorphisms, the autodiffeomorphisms $f$ is necessarily 
isotopic to $\id_C$. There is thus an orientation preserving
isotopy $k_y$, $y\in\mathbb{R}$, deforming $\id_C$ into $f$.
The composition $\xi\circ k_y$ defines then an isotopy of the $C_0$--knots $\xi$ 
and $\xi\circ f$. By the isotopy extension theorem, there is an ambient isotopy 
$W_y$, $y\in\mathbb{R}$, of $\xi$ and $\xi\circ f$ such that $W_y\circ \xi=\xi\circ k_y$. 
The $C_0$--knots $\xi$, $\xi\circ f$ are thus freely ambient isotopic.

By prop. \ref{prop:knot4} with $\xi_0=\xi$, $\xi_1=\xi\circ f$, there is then a homotopy 
$h:\gamma_{0\xi}\Rightarrow \gamma_1{}^{-1_\circ}\circ\gamma_{0\xi\circ f}\circ \gamma_1$, 
where $\gamma_1:p_{M0}\rightarrow p_{M1}$ is a curve of $M$. By \eqref{knot21/0},
as $\gamma_{C1}=f\circ \gamma_{C0}$, $\gamma_{1\xi}=\gamma_{0\xi\circ f}$. The above is thus also a homotopy
$h:\gamma_{0\xi}\Rightarrow \gamma_1{}^{-1_\circ}\circ\gamma_{1\xi}\circ \gamma_1$. 
As explained in the proof of prop. \ref{prop:knot4}, 
the homotopy $h$ is the composition of the standard thin homotopy $\gamma_{0\xi}\Rightarrow
\iota_{p_{M0}}{}^{-1_\circ}$ $\circ\,\gamma_{0\xi}\circ\iota_{p_{M0}}$ and the homotopy 
$\widehat h:\iota_{p_{M0}}{}^{-1_\circ}\circ\gamma_{0\xi}\circ\iota_{p_{M0}}
\Rightarrow \gamma_1{}^{-1_\circ}\circ\gamma_{1\xi}\circ \gamma_1$ defined 
by \eqref{knot36/0}, \eqref{knot37/0}, here with $\gamma_{\xi_y}$ replaced by $\gamma_{0\xi_y}$.
The curves $\gamma_y$ and $\gamma_{0\xi_y}$ entering in \eqref{knot36/0} 
are given by \eqref{knot39/0}, \eqref{knot35/0}, 
where $W$ is the free ambient isotopy introduced in the previous paragraph. 
Since $W_y\circ \xi=\xi\circ k_y$, we have $\gamma_y(x) 
=W_{\alpha(x)\alpha(y)}(\xi(p_{C0}))=\xi(k_{\alpha(x)\alpha(y)}(p_{C0}))$
and $\gamma_{0\xi_y}(x)=W_{\alpha(y)}(\xi(\gamma_{C0}(x)))=\xi(k_{\alpha(y)}(\gamma_{C0}(x)))$.
Thus, the images of $\gamma_y$ and $\gamma_{0\xi_y}$ are contained in that of $\xi$. As this latter is 
a differentiable curve in $M$, $\widehat{h}$ satisfies \eqref{path11} for dimensional reasons. 
$\widehat{h}$ is hence a thin homotopy. It follows that $h$ is a thin homotopy.
By the above property of the curves $\gamma_y$, $\gamma_{0\xi_y}$, 
the images of the curve $\gamma_1$ and homotopy $h$ lie in that of the knot $\xi$.  \hfill $\Box$ 



Inspired by the theory of ordinary knots expounded above, we are now ready 
to formulate its extension to surface knots. 

To construct surface knot theory, 
we need a slight generalization of the notion of (free) $C$--knot given above which does not require 
full smoothness. 

Assume that $C$ is endowed with a pointing $p_C$. 

\begin{defi} \label{def:spiknot1}
A spiky free $C$--knot of $M$ is a topological embedding 
$\xi:C\rightarrow M$ that is smooth on $C\setminus p_C$ and has the following properties.
\begin{enumerate}

\item The jets of any order of $\xi$ have finite limits at both ends of $C\setminus p_C$.

\item The jet of order $1$ of $\xi$ has non zero limits at both ends of $C\setminus p_C$.

\end{enumerate}
\end{defi}

\noindent
The regularity conditions at $p_C$ ensure that \pagebreak $\xi$ has a well defined tangent vector at each 
end of $C\setminus p_C$ and that with $\xi$ there can be associated a curve $\gamma_\xi$ of $M$ 
as for a $C$--knot.
Every free $C$--knot is also a spiky free $C$--knot, while not every spiky free $C$--knot 
is a free $C$--knot. 

Suppose now that the manifold $M$ is $C$--pointed by $p_M$. 

\begin{defi} \label{def:spiknot1y}
A spiky $C$--knot of $M$ is a $C$--pointing preservation topological embedding 
$\xi:C\rightarrow M$ satisfying the same regularity conditions as 
a spiky free $C$--knot.
\end{defi}

\noindent
The property of $C$ pointing preservation has again the statement 
\eqref{knot5/0}. Compare with def. \ref{def:knot8}. 
Every  $C$--knot is also a spiky  $C$--knot, while not every spiky  $C$--knot 
is a  $C$--knot.

Let $\gamma_C$ be a curve of $C$ compatible with the pointing of $C$ (cf. def. \ref{def:knot10}). 

\begin{defi} \label{def:spiknot2}
For a spiky $C$--knot $\xi$ of $M$, let $\gamma_\xi:p_M\rightarrow p_M$
be the curve of $M$ defined by \eqref{knot21/0}.
\end{defi}

\noindent
Compare with def. \ref{def:knot11}. The curve $\gamma_\xi$ of a spiky $C$--knot enjoys the same 
properties as that of a $C$--knot. In particular, the \eqref{knot23,24/0} hold. 

Let $S$ be an oriented differentiable closed surface of genus $\ell_S$, 
that is a sphere with $\ell_S$ handles. 

\begin{defi} \label{def:knot17} A marking of $S$ consists of:
\begin{enumerate}

\item a $C$--pointing $p_S$ of $S$;

\item $2\ell_S$ spiky $C$--knots $\zeta_{Si}$ of $S$ with the following properties.

\item The sets $\zeta_{Si}(C)$ pairwise intersect only at $p_S$.

\item The $\zeta_{Si}$ represent the $2\ell_S$ homotopy classes of the standard 
oriented $a$-- and $b$--cycles of $S$.

\end{enumerate}
\end{defi}


\vspace{1.5truemm} \eject
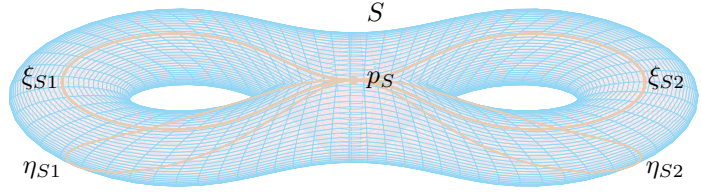
\begin{figure}[!t]
\begin{pspicture}(-7,-1.8)(5,2.5)
\psset{viewpoint= 20 90 75 rtp2xyz,Decran=28}    

\defFunction[algebraic]%
{genus}(u,v)
{cos(2*\psPi*u)*(1.2*cos(2*\psPi*u)^2 +1.2*2^(-2*sin(2*\psPi*u)^2)+1.2*2^(-2*cos(2*\psPi*u)^2)
+.82*(1-.25*2^(-2*cos(\psPi*u)^2)-.25*2^(-2*sin(\psPi*u)^2))*cos(2*\psPi*v))}  
{sin(2*\psPi*u)*(1.2*cos(2*\psPi*u)^2+.82*(1-.25*2^(-2*cos(\psPi*u)^2)-.25*2^(-2*sin(\psPi*u)^2))*cos(2*\psPi*v))}
{.82*(1-.25*2^(-2*cos(\psPi*u)^2)-.25*2^(-2*sin(\psPi*u)^2))*sin(2*\psPi*v)}

\psSolid[
object=surfaceparametree,
base=.5 neg .5  0 1,        
function=genus, 
fillcolor=red!10, incolor=cyan!10, opacity=.75,
linewidth=0.3\pslinewidth,
linecolor=cyan!40,
ngrid=75 75 
]%

\defFunction[algebraic]%
{circle1}(u)
{cos(2*\psPi*u)*(1.2*cos(2*\psPi*u)^2 +1.2*2^(-2*sin(2*\psPi*u)^2)+1.2*2^(-2*cos(2*\psPi*u)^2))}
{sin(2*\psPi*u)*1.2*cos(2*\psPi*u)^2}
{.82*(1-.25*2^(-2*cos(\psPi*u)^2)-.25*2^(-2*sin(\psPi*u)^2))}

\psSolid[
object=courbe, r=0.002,
range= .5 neg 5,    
function=circle1, 
linewidth=0.5pt, 
linecolor=brown!40,plotpoints=50 
]%

\defFunction[algebraic]%
{circle2}(v) 
{cos(2*\psPi*(.25-.210*sin(\psPi*v)^(.55)))*(1.2*cos(2*\psPi*(.25-.210*sin(\psPi*v)^(.55)))^2 
+1.2*2^(-2*sin(2*\psPi*(.25-.210*sin(\psPi*v)^(.55)))^2)+1.2*2^(-2*cos(2*\psPi*(.25-.210*sin(\psPi*v)^(.55)))^2)
+.82*(1-.25*2^(-2*cos(\psPi*(.25-.210*sin(\psPi*v)^(.55)))^2)
-.25*2^(-2*sin(\psPi*(.25-.210*sin(\psPi*v)^(.55)))^2))*sin(2*\psPi*v))}  
{sin(2*\psPi*(.25-.210*sin(\psPi*v)^(.55)))*(1.2*cos(2*\psPi*(.25-.210*sin(\psPi*v)^(.55)))^2
+.82*(1-.25*2^(-2*cos(\psPi*(.25-.210*sin(\psPi*v)^(.55)))^2)
-.25*2^(-2*sin(\psPi*(.25-.210*sin(\psPi*v)^(.55)))^2))*sin(2*\psPi*v))}
{.82*(1-.25*2^(-2*cos(\psPi*(.25-.210*sin(\psPi*v)^(.55)))^2)
-.25*2^(-2*sin(\psPi*(.25-.210*sin(\psPi*v)^(.55)))^2))*cos(2*\psPi*v)}


\psSolid[
object=courbe, r=0.002,
range= 0.001 .999, 
function=circle2, 
linewidth=0.5pt, 
linecolor=brown!40,plotpoints=50 
]%

\defFunction[algebraic]%
{circle3}(v) 
{cos(2*\psPi*(.25+.210*sin(\psPi*v)^(.55)))*(1.2*cos(2*\psPi*(.25+.210*sin(\psPi*v)^(.55)))^2 
+1.2*2^(-2*sin(2*\psPi*(.25+.210*sin(\psPi*v)^(.55)))^2)+1.2*2^(-2*cos(2*\psPi*(.25+.210*sin(\psPi*v)^(.55)))^2)
+.82*(1-.25*2^(-2*cos(\psPi*(.25+.210*sin(\psPi*v)^(.55)))^2)
-.25*2^(-2*sin(\psPi*(.25+.210*sin(\psPi*v)^(.55)))^2))*sin(2*\psPi*v))}  
{sin(2*\psPi*(.25+.210*sin(\psPi*v)^(.55)))*(1.2*cos(2*\psPi*(.25+.210*sin(\psPi*v)^(.55)))^2
+.82*(1-.25*2^(-2*cos(\psPi*(.25+.210*sin(\psPi*v)^(.55)))^2)
-.25*2^(-2*sin(\psPi*(.25+.210*sin(\psPi*v)^(.55)))^2))*sin(2*\psPi*v))}
{.82*(1-.25*2^(-2*cos(\psPi*(.25+.210*sin(\psPi*v)^(.55)))^2)
-.25*2^(-2*sin(\psPi*(.25+.210*sin(\psPi*v)^(.55)))^2))*cos(2*\psPi*v)}


\psSolid[
object=courbe, r=0.002,
range= 0.001 .999, 
function=circle3, 
linewidth=0.5pt, 
linecolor=brown!40,plotpoints=50 
]%

\psPoint(0,0,.82 .75 mul){I}

\psdot[linecolor=brown!40](I)

\composeSolid

\psPoint(0,0,.82 .75 mul){PP}

\psPoint(1.2 2.25 mul,0,.82 .875 mul){A1}
\psPoint(1.2 2.25 mul neg,0,.82 .875 mul){A2}
\psPoint(1.2 2.25 mul,0,.82 .875 mul neg 4.25 mul){B1}
\psPoint(1.2 2.25 mul neg,0,.82 .875 mul neg 4.25 mul){B2}
\psPoint(0,1.2 .55 mul neg,.82 .75 mul){S}

\uput[r](PP){\footnotesize $p_S$}
\uput[l](A1){\footnotesize $\xi_{S1}\!\!\!$}
\uput[r](A2){\footnotesize $\!\!\!\xi_{S2}$}
\uput[l](B1){\footnotesize $\eta_{S1}\,\,\,\,\,\,\,$}
\uput[r](B2){\footnotesize $\,\,\,\,\,\,\,\eta_{S2}$}
\uput[r](S){\footnotesize $S$}

\end{pspicture}
\caption{The surface $S$ with the $C$--knots $\xi_{Sr}$, $\eta_{Sr}$ shown. Here, $\ell_S=2$. 
\label{fig:genuscl}}
\end{figure}
\noindent
For a given index $i$, the knot $\zeta_{Si}$ will be denoted also as $\xi_{Sr}$ or $\eta_{Sr}$
according to whether it corresponds to the $r$--th $a$-- or $b$--cycle, respectively. 
See. fig. \ref{fig:genuscl}. 
For $\ell_S=0$, the $\zeta_{Si}$ are absent. 
Below, we assume that $S$ is equipped with a marking fixed once and for all.

\begin{defi} \label{def:sfree}
A free $S$--surface knot $\varXi$ of $M$ is an embedding $\varXi:S\rightarrow M$.
\end{defi}

\noindent
The image of $\varXi(S)$ of a free $S$--surface knot $\varXi$ is a differentiable closed 
surface of $M$ with the orientation induced by that of $S$ via the embedding. 
The image of $\varXi$ should not be confused with $\varXi$ itself.

Ambient isotopy of free $S$--surface knots is defined in the obvious fashion. 

\begin{defi} \label{def:sfreeamb}
Two free $S$--surface knot $\varXi_0$, $\varXi_1$ of $M$ are ambient isotopic 
if they are as embeddings of $S$ into $M$. 
\end{defi}

\noindent 
Hence, if $F$ is the ambient isotopy transforming $\varXi_0$ into $\varXi_1$, then
\begin{equation}
\varXi_1=F_1\circ \varXi_0.
\label{knot9}
\end{equation}
Ambient isotopy of free $S$--surface knots is an equivalence relation.
We denote 
the set of all ambient isotopy classes of free $S$--surface knots by $\Knot_{\mathrm{F}S}(M)$.

\begin{defi} \label{def:knot22/0}
For any free $S$--surface knot $\varXi$, the spiky free $C$--knots 
\begin{equation}
\zeta_{Si}{}^*\varXi=\varXi\circ\zeta_{Si}
\label{knot9y}
\end{equation}
are called the characteristic $C$--knots of $\varXi$. 
\end{defi}

\noindent

Ambient isotopic surface knots have ambient isotopic characteristic $C$--knots.

\begin{prop}
Let $\varXi_0$, $\varXi_1$ be two ambient isotopic free $S$--surface knots of $M$. 
Then, their characteristic $C$--knots $\zeta_{Si}{}^*\varXi_0$, $\zeta_{Si}{}^*\varXi_1$ are ambient isotopic.
\end{prop}

\noindent {\it Proof}.
This is an immediate consequence of \eqref{knot9} and \eqref{knot9y}. \hfill $\Box$

If $F$ is the ambient isotopy turning $\varXi_0$ into $\varXi_1$, then 
\begin{equation}
\zeta_{Si}{}^*\varXi_1=F_1\circ\zeta_{Si}{}^*\varXi_0.
\label{knot9/ck}
\end{equation}

As for ordinary knots, the basic constructions with surface knots we are going to introduce 
work for based surface knots.

\begin{defi} \label{def:knot18} A manifold $M$ is said $S$--marked when the following data are specified:
\begin{enumerate}

\item a $C$--pointing $p_M$ of $M$;

\item $2\ell_S$ spiky $C$--knots $\zeta_{Mi}$ of $M$ with the following property.

\item The sets $\zeta_{Mi}(C)$ pairwise intersect only at $p_M$.

\end{enumerate}
\end{defi}
For a given index $i$, the knot $\zeta_{Mi}$ will be denoted also as $\xi_{Mr}$ or $\eta_{Mr}$
according to whether the knot $\zeta_{Si}$ 
equals $\xi_{Sr}$ or $\eta_{Sr}$ in $S$ as detailed above, respectively. 
For $\ell_S=0$, the knots $\zeta_{Mi}$ are absent. 
Note that the differentiable surface $S$ itself is $S$--marked. 

\begin{defi} \label{def:knot19} Let $M$, $N$ be $S$--marked \pagebreak manifolds and let $f:N\rightarrow M$ be a map.
$f$ is said $S$--marking preserving if 
\begin{subequations}
\label{knot3,4}
\begin{align}
&f(p_N)=p_M,
\vphantom{\Big]}
\label{knot3}
\\
&f\circ \zeta_{Ni}=\zeta_{Mi}.
\vphantom{\Big]}
\label{knot4}
\end{align}
\end{subequations}
\end{defi}
One defines analogously $S$--marking preserving diffeomorphisms, autodiffeomorphisms etc. 
For $\ell_S>0$, \eqref{knot3} follows from \eqref{knot4} upon evaluation at $p_C$. 
By \eqref{knot3}, $f$ is $C$--pointing preserving. 

\begin{defi} \label{def:knot20}
Let $M$, $N$ be $S$--marked manifolds and let $f_0,~f_1:N\rightarrow M$ be $S$--marking preserving 
maps. An $S$--marking preserving homotopy $h$ of $f_0$, $f_1$ is an ordinary homotopy of $f_0$, $f_1$ such that 
the maps $h_z:N\rightarrow M$, $z\in\mathbb{R}$, are all $S$--marking preserving, 
\begin{subequations}
\label{zknot3,4}
\begin{align}
&h_z(p_N)=p_M,
\vphantom{\Big]}
\label{zknot3}
\\
&h_z\circ \zeta_{Ni}=\zeta_{Mi}.
\vphantom{\Big]}
\label{zknot4}
\end{align}
\end{subequations}
The $S$--marking preserving maps $f_0$, $f_1$ are said homotopic when they are related by
$S$--marking preserving homotopy. If $f_0$, $f_1$ and the $h_z$ are 
all embeddings, then $h$ is called an $S$--marking preserving 
isotopy and $f_0$, $f_1$ are said isotopic.
\end{defi}

\noindent
$S$--marked manifolds and $S$--marking preserving maps form a category. 
$S$--marked manifolds and $S$--marking preserving maps up to $S$--marking preserving 
homotopy constitute also a category. 

\begin{defi} \label{def:knot21}
Let $M$, $N$ be $S$--marked manifolds and let $f_0,~f_1:N\rightarrow M$ be $S$--marking preserving maps.
An $S$--marking preserving ambient isotopy of $f_0$, $f_1$ is an ordinary ambient isotopy of $f_0$, $f_1$ 
such that the autodiffeomorphisms $F_z$, $z\in\mathbb{R}$, are all $S$--marking preserving ,
\begin{subequations}
\label{knot10,11}
\begin{align}
&F_z(p_M)=p_M,
\vphantom{\Big]}
\label{knot10}
\\
&F_{z}\circ \zeta_{Mi}=\zeta_{Mi}.
\vphantom{\Big]}
\label{knot11}
\end{align}
\end{subequations}
The $S$--marking preserving maps $f_0,~f_1$ are said ambient isotopic when they are related by an 
$S$--marking preserving ambient isotopy. 
\end{defi}

Given 
an $S$--marked manifold $M$, we can define $S$--surface knots, which are a higher analogue of 
ordinary based knots.

\begin{defi} \label{def:knot22}
An $S$--surface knot in the $S$--marked manifold $M$ is an $S$--marking preserving embedding
$\varXi:S\rightarrow M$.
\end{defi}

\noindent
A $S$--surface knot $\varXi$ is based in that 
\begin{subequations}
\label{knot5,6}
\begin{align}
&\varXi(p_S)=p_M,
\vphantom{\Big]}
\label{knot5}
\\
&\varXi\circ \zeta_{Si}=\zeta_{Mi}.
\vphantom{\Big]}
\label{knot6}
\end{align}
\end{subequations}
An $S$--surface knot $\varXi$ is also a free $S$--surface knot, if one forgets the $S$--marking of $M$. Conversely,
a free $S$--surface knot $\varXi$ is secretly an $S$--surface knot belonging to the $S$--marking 
$\{p_M=\varXi(p_S),\zeta_{Mi}=\zeta_{Si}{}^*\varXi\}$ of $M$.
$\id_S$ is an $S$--surface knot of $S$, the tautological $S$--surface knot. 

Clearly, by \eqref{knot6}, the characteristic $C$--knots $\zeta_{Si}{}^*\varXi$ of an $S$--surface knots $\varXi$ belonging
an $S$--marking of $M$ $\{p_M,\zeta_{Mi}\}$ of $M$ belong to the $C$--pointing $p_M$
of $M$ and are equal to $\zeta_{Mi}$. 

It is important to stress that, while an $S$--marking of $M$ of genus $\ell_S=0$ always supports
$S$--surface knots, for a given $S$--marking of $M$ of genus $\ell_S>0$, there may exist
no $S$--surface knots at all, as in the latter case, for generic $C$--knots $\zeta_{Mi}$, 
\eqref{knot6} may be incompatible with the $\zeta_{Si}$ 
representing the $2\ell_S$ homotopy classes of the standard oriented $a$-- and
$b$--cycles of $S$ and $\varXi$ being an embedding of $S$ in $M$. 
By contrast, for every $C$--pointing, there always are $C$--knots
belonging to it. 

\begin{defi}
An $S$--marking of $M$ is called admissible if there exists at least
one $S$--surface knot.
\end{defi}

\noindent
All $S$--markings considered in the following will be tacitly assumed to be admissible
unless otherwise stated.

We want to identify surface knots related by an appropriate form of ambient isotopy, 
which we define next.

\begin{defi} \label{def:knot23}
Two $S$--surface knot $\varXi_0$, $\varXi_1$ in the $S$--marked manifold $M$ are ambient isotopic 
if they are as embeddings of $S$ into $M$ via an $S$--marking preserving ambient isotopy. 
\end{defi}

\noindent
Ambient isotopy of $S$--surface knots is an equivalence relation.
We denote by $\Knot_S(M)$
the set of all ambient isotopy classes of $S$--surface knots.

Let us now see whether it is possible to describe surface knots using curves, surfaces and (thin) homotopy
thereof (cf. subsect. \ref{sec:path}). 

To begin with, we fix a curve $\gamma_C$ of $C$ compatible with the pointing of $C$
to  construct the curves associated with spiky $C$--knots.

\begin{figure}[!b]

{\psset{xunit=1.153,yunit=1.153}

\begin{pspicture}(-6.3,-2.55)(1.7,2.85)

\psline*[linewidth=2pt,linecolor=red!10]{->}(2,0)(1.4142,1.4142)(0, 2)(-1.4142,1.4142)(-2,0)(-1.4142,-1.4142)(0,-2)
(1.4142,-1.4142)(2,0)

\psline[linewidth=2pt,linecolor=brown!40]{->}(2,0)(1.4142,1.4142)
\psline[linewidth=2pt,linecolor=brown!40]{->}(1.4142,1.4142)(0, 2)
\psline[linewidth=2pt,linecolor=brown!40]{->}(0,2)(-1.4142,1.4142)
\psline[linewidth=2pt,linecolor=brown!40]{->}(-1.4142,1.4142)(-2,0)
\psline[linewidth=2pt,linecolor=brown!40]{->}(-2,0)(-1.4142,-1.4142)
\psline[linewidth=2pt,linecolor=brown!40]{->}(-1.4142,-1.4142)(0,-2)
\psline[linewidth=2pt,linecolor=brown!40]{->}(0,-2)(1.4142,-1.4142)
\psline[linewidth=2pt,linecolor=brown!40]{->}(1.4142,-1.4142)(2,0)

\uput[r](1.84776, 0.765367){\footnotesize $\xi_{S1}$}
\uput[ur](0.765367, 1.84776){\footnotesize $\eta_{S1}$}
\uput[l](-1.84776, 0.765367){\footnotesize $\eta_{S1}{}^{-1}\!\!\!$}
\uput[ul](-0.765367, 1.84776){\footnotesize $\xi_{S1}{}^{-1}\!\!\!$}
\uput[dl](-0.765367, -1.84776){\footnotesize $\eta_{S2}$}
\uput[l](-1.84776, -0.765367){\footnotesize $\xi_{S2}$}
\uput[dr](0.765367, -1.84776){\footnotesize $\!\!\!\xi_{S2}{}^{-1}$}
\uput[r](1.84776, -0.765367){\footnotesize $\!\!\!\eta_{S2}{}^{-1}$}

\uput[r](2,0){\footnotesize $p_S$}
\uput[ur](1.4142,1.4142){\footnotesize $p_S$}
\uput[u](0, 2){\footnotesize $p_S$}
\uput[ul](-1.4142,1.4142){\footnotesize $p_S$}
\uput[l](-2,0){\footnotesize $p_S$}
\uput[dl](-1.4142,-1.4142){\footnotesize $p_S$}
\uput[d](0,-2){\footnotesize $p_S$}
\uput[dr](1.4142,-1.4142){\footnotesize $p_S$}


\end{pspicture}
}
\caption{The surface $S$ unfolds upon being cut along the $C$--knots
$\xi_{Sr}$, $\eta_{Sr}$. Here, $\ell_S=2$.
\label{fig:genusop}}
\end{figure}
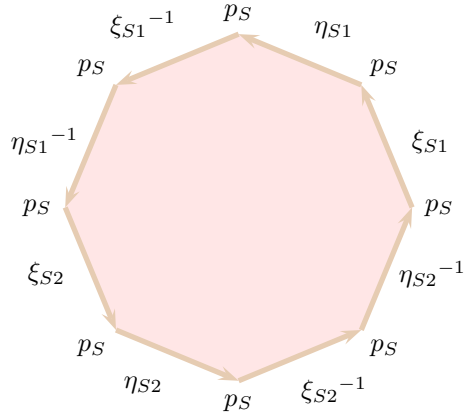

The curves of the spiky $C$--knots of the marking of $S$ are constructed 
according the prescription \eqref{knot21/0}. We have 
\begin{equation}
\gamma_{Si}=\gamma_{\zeta_{Si}}
\label{xknot1}
\end{equation}
Correspondingly, we set
\begin{equation}
\alpha_{Sr}=\gamma_{\xi_{Sr}}, \qquad \beta_{Sr}=\gamma_{\eta_{Sr}},
\label{xknot2}
\end{equation}
when we sort the knots $\zeta_{Si}$ according to their correspondence to the 
$a$-- and $b$--cycles of $S$. 
Next, we now define a curve $\tau_S:p_S\rightarrow p_S$ of $S$ by 
\begin{equation}
\tau_S=\beta_{S\ell_S}{}^{-1_\circ}\circ\alpha_{S\ell_S}{}^{-1_\circ}\circ\beta_{S\ell_S}\circ\alpha_{S\ell_S}\circ \cdots \circ
\beta_{S1}{}^{-1_\circ}\circ\alpha_{S1}{}^{-1_\circ}\circ\beta_{S1}\circ\alpha_{S1}
\vphantom{\Bigg]}
\label{knot15}
\end{equation}
(cf. fig. \ref{fig:genusop}).
The ``compose the rightmost first'' convention for multiple composition of curves is tacitly 
applied here and in the following. 
When $\ell_S=0$, we set conventionally that $\tau_S=\iota_{p_S}$. 

\begin{defi} \label{def:knot24}
A surface $\varSigma_S$ is said to be compatible with the marking of $S$, if 
$\varSigma_S:\iota_{p_S}\Rightarrow \tau_S$ and 
$D_S=\varSigma_S{}^{-1}(S\setminus\cup_i\zeta_{Si}(C))$ is an open simply connected domain
of $\mathbb{R}^2$ such that $\varSigma_S\big|_{D_S}$ is an orientation preserving diffeomorphism
onto $S\setminus\cup_i\zeta_{Si}(C)$. 
\end{defi}

\noindent
If $\varSigma_S$ is compatible, then 
\begin{subequations}
\label{knot16,17,18,19}
\begin{align}
&\varSigma_S(x,y)=p_S\qquad \text{for $x<\epsilon$},
\vphantom{\Big]}
\label{knot16}
\\
&\varSigma_S(x,y)=p_S\qquad \text{for $x>1-\epsilon$},
\vphantom{\Big]}
\label{knot17}
\\
&\varSigma_S(x,y)  
=p_S\qquad \text{for $y<\epsilon$},
\vphantom{\Big]}
\label{knot18}
\\
&\varSigma_S(x,y)=\tau_S(x)\qquad \text{for $y>1-\epsilon$}
\vphantom{\Big]}
\label{knot19}
\end{align}
\end{subequations} 
for some number $\epsilon$ with $0<\epsilon<1/2$, by \eqref{path12}--\eqref{path15},
where we used that $\iota_{p_S}(x)=p_S$. 
Further, the domain $D_S$ is contained in the open square $(\epsilon,1-\epsilon)^2$
and $\varSigma_S\big|_{D_S}$ provides a one--to-one parametrization of $S\setminus\cup_i\zeta_{Si}(C)$
consistent with its orientation. 

It is important to realize that, \pagebreak in high genus, the choice of 
a compatible surface $\varSigma_S$ of $S$ involves a prior choice of a compatible curve 
$\gamma_C$ of $C$. 

\begin{defi} \label{def:consch}
A choice of a surface $\varSigma_S$ of $S$ compatible with the marking of $S$ 
is said congruent with one of a curve $\gamma_C$ of $C$ compatible with the pointing 
of $C$ if either $\ell_S=0$ or $\ell_S>0$ and $\tau_S$ 
is expressed by \eqref{knot15} with the curves $\alpha_{Sr}$, $\beta_{Sr}$ 
defined through \eqref{knot21/0} and \eqref{xknot2} in terms of $\gamma_C$.
\end{defi}

Fix a surface $\varSigma_S$ of $S$ compatible with the marking of $S$ and congruent with 
the chosen compatible curve $\gamma_C$.
Below, we shall tacitly assume that the value of the number $\epsilon$ appearing in the 
\eqref{knot16,17/0} and \eqref{knot16,17,18,19} has been adjusted so that 
the \eqref{knot16,17/0} and \eqref{knot16,17,18,19} simultaneously hold. 

Next, we construct the curves of the spiky $C$--knots $\zeta_{Mi}$ 
of the $S$--marked manifold $M$ using again the prescription \eqref{knot21/0}. We set
\begin{equation}
\gamma_{Mi}=\gamma_{\zeta_{Mi}}
\label{xknot3}
\end{equation}
Correspondingly, we set
\begin{equation}
\alpha_{Mr}=\gamma_{\xi_{Mr}}, \qquad \beta_{Mr}=\gamma_{\eta_{Mr}},
\label{xknot4}
\end{equation} 
when we sort the knots $\zeta_{Mi}$ according to their correspondence to the 
$a$-- and $b$--cycles of $S$. 
We can define in this way a curve $\tau_M:p_M\rightarrow p_M$ of $M$ by 
\begin{equation}
\tau_M=\beta_{M\ell_S}{}^{-1_\circ}\circ\alpha_{M\ell_S}{}^{-1_\circ}\circ\beta_{M\ell_S}\circ\alpha_{M\ell_S}
\circ \cdots \circ\beta_{M1}{}^{-1_\circ}\circ\alpha_{M1}{}^{-1_\circ}\circ\beta_{M1}\circ\alpha_{M1}.
\vphantom{\Bigg]}
\label{knot22}
\end{equation} 
When $\ell_S=0$, $\tau_M=\iota_{p_M}$. $\tau_M$ clearly answers to $\tau_S$ and indeed
it reduces to this latter when $M$ is the $S$--marked manifold $S$.

Given an $S$--surface knot $\varXi$ of $M$, one can push--forward the surface 
$\varSigma_S:\iota_{p_S}$ $\Rightarrow \tau_S$ of $S$ by the map $\varXi:S\rightarrow M$.
Since
\begin{equation}
\tau_M=\varXi\circ \tau_S 
\label{knot20} 
\end{equation}
by \eqref{knot6}, we get a surface $\varXi\circ\varSigma_S:\iota_{p_M}\Rightarrow \tau_M$ of $M$.

\begin{defi} \label{def:knot25} For an $S$--surface knot $\varXi$, 
let $\varSigma_\varXi:\iota_{p_M}\Rightarrow\tau_M$ be the surface of $M$ 
defined as \hphantom{xxxxxxxxxxxxxxxxxxxxx}
\begin{equation}
\varSigma_\varXi=\varXi\circ \varSigma_S.
\label{knot21}
\end{equation}
\end{defi}



\noindent
By \eqref{path12}--\eqref{path15}, we have 
\begin{subequations}
\label{knot23,24,25,26}
\begin{align}
&\varSigma_\varXi(x,y)=p_M\qquad \text{for $x<\epsilon$},
\vphantom{\Big]}
\label{knot23}
\\
&\varSigma_\varXi(x,y)=p_M\qquad \text{for $x>1-\epsilon$},
\vphantom{\Big]}
\label{knot24}
\\
&\varSigma_\varXi(x,y)=p_M\qquad \text{for $y<\epsilon$},
\vphantom{\Big]}
\label{knot25}
\\
&\varSigma_\varXi(x,y)=\tau_M(x)\qquad \text{for $y>1-\epsilon$}
\vphantom{\Big]}
\label{knot26}
\end{align}
\end{subequations} 
for the same number $\epsilon$ as in \eqref{knot16,17,18,19}, where we used that $\iota_{p_M}(x)=p_M$. 
Further, by the compatibility of $\varSigma_S$, $\varSigma_\varXi\big|_{D_S}$
furnishes a one--to-one parametrization of $\varXi(S)\setminus\cup_i\zeta_{Mi}(C)$
consistent with its orientation. 
We note also that the surface $\varSigma_S$ is just the surface 
$\varSigma_{\id_S}$ of $S$ associated with the tautological $S$--knot $\id_S$.

At first glance, it would seem that the surface naturally associated
with an $S$--surface knot $\varXi$ should be $\varSigma_\varXi$, 
since this appears to be the obvious counterpart of the curve 
$\gamma_\xi$ associated with a $C$--knot $\xi$. Things are 
however a bit subtler and the reason why they are so is ultimately related
to the end goal of the present construction, the use of flat connection 
holonomy and its homotopy invariance to study knot topology. 

In ordinary knot theory, 
the curve $\gamma_\xi$ associated with a $C$--knot $\xi$ encodes 
the topological properties of $\xi$. 
It so provides also a measure of its non triviality. 
A generic $C$--knot $\xi$ is non trivial  
to the extent to which its curve $\gamma_\xi$ differs from
the curve $\gamma_M=\gamma_{\xi_M}$ of a reference knot $\xi_M$,
that is trivial according to topological knot theory. 
A measure of the triviality of $\xi$ is then provided by the curve ratio
$\gamma^\sharp{}_\xi=\gamma_M{}^{-1_\circ}\circ\gamma_\xi$.

Every $C$--pointing of $M$ supports a canonical 
topologically trivial $C$--knot unique up to $C$--pointing  preserving 
ambient isotopy, the
unknot $\xi_M$, which is characterized  
by its image being the boundary of an embedded $2$--dimensional disk. 
The curve $\gamma_M=\gamma_{\xi_M}$ associated with $\xi_M$ 
is so equipped with a trivializing $C$--pointing preserving homotopy 
$k_M:\gamma_M\Rightarrow \iota_{p_M}$.
For a $C$--knot $\xi$, the curve $\gamma^\sharp{}_\xi$ measuring its non triviality 
is thus homotopy equivalent to $\iota_{p_M}{}^{-1_\circ}\circ \gamma_\xi$ 
through $k_M$ and then to $\gamma_\xi$ by the standard 
thin homotopy $\iota_{p_M}{}^{-1_\circ}\circ\gamma_\xi\Rightarrow\gamma_\xi$. 
For the purpose of an evaluation of the non triviality of $\xi$, $\gamma_\xi$ is hence 
as good as $\gamma^\sharp{}_\xi$. 

Let us now examine to what extent the above consideration extend to higher knots.
In surface knot theory, analogously to ordinary knot theory,  
the surface $\varSigma_\varXi$ associated with an $S$--surface knot $\varXi$ encodes 
the topological properties of $\varXi$. 
It so provides also a measure of its non triviality. 
A generic $S$--surface knot $\varXi$ is non trivial  
inasmuch as its surface $\varSigma_\varXi$ differs from
the surface $\varSigma_M=\varSigma_{\varXi_M}$ of a reference knot $\varXi_M$,
that is trivial according to some topological knot theoretic principle. 
A measure of the triviality of $\varXi$ is then provided by the surface vertical ratio
$\varSigma^\sharp{}_\varXi=\varSigma_M{}^{-1_\bullet}\bullet\varSigma_\varXi$.

For genus $\ell_S=0$, every $S$--marking of $M$ supports a canonical 
topologically trivial $S$--surface knot unique up to $S$--marking preserving
ambient isotopy, the
unknot $\varXi_M$, which is characterized  
by its image being the boundary of an embedded $3$--dimensional disk. 
The surface $\varSigma_M=\varSigma_{\varXi_M}$ associated with $\varXi_M$ 
is so equipped with a trivializing $S$--marking preserving homotopy 
$K_M:\varSigma_M\Rrightarrow I_{\iota_{p_M}}$.
For an $S$--surface knot $\varXi$, the surface $\varSigma^\sharp{}_\varXi$ measuring its non triviality 
is thus homotopy equivalent to $I_{\iota_{p_M}}{}^{-1_\bullet}\bullet \varSigma_\varXi$ 
through $K_M$ and then to $\varSigma_\varXi$ by the standard 
thin homotopy $I_{\iota_{p_M}}{}^{-1_\bullet}\bullet\varSigma_\varXi\Rrightarrow\varSigma_\varXi$. 
For the purpose of an evaluation of the non triviality of $\varXi$, $\varSigma_\varXi$ is hence 
as good as $\varSigma^\sharp{}_\varXi$. For genus $\ell_S>0$, the above reasoning fails because the homotopy 
$K_M$ no longer exists in general, as it is 
topological obstructed. In that case, for a given $S$--marking of $M$, the explicit indication of a 
reference $S$--surface knot $\varXi_M$ is then an indispensable element of the construction.  
The surface $\varSigma^\sharp{}_\varXi$ rather than 
$\varSigma_\varXi$ then adequately describes the non triviality of $\varXi$. 
This is why we shall concentrate on $\varSigma^\sharp{}_{\varXi}$ in the following. 

The encoding the topology of the $S$--surface knots $\varXi$ in the surfaces
$\varSigma^\sharp{}_{\varXi}$  
will be natural only if there is a well defined topological prescription that 
associates with any $S$--marking of $M$ a reference topologically trivial $S$--surface knot $\varXi_M$ 
unique up to $S$--marking preserving ambient isotopy. As we cannot prove this in general,
we shall assume it as a working hypothesis. Many of the results reported below, however, do not 
hinge on it. 

\begin{defi} \label{def:knot26} For an $S$--surface knot $\varXi$, let 
$\varSigma^\sharp{}_{\varXi}:\iota_{p_M}\Rightarrow\iota_{p_M}$ be the surface of $M$ defined by 
\hphantom{xxxxxxxxxxxxxxxxxxxxx}
\begin{equation}
\varSigma^\sharp{}_{\varXi}=\varSigma_M{}^{-1_\bullet}\bullet\varSigma_{\varXi},
\label{knot40}
\end{equation}
where $\varSigma_M=\varSigma_{\varXi_M}$.  
\end{defi}

\noindent
Since $\varSigma_{\varXi},\varSigma_M:\iota_{p_M}\Rightarrow\tau_M$, the composition is well defined
and has $\iota_{p_M}$ as source and target curves. 
By \eqref{path12}--\eqref{path15}, we have 
\begin{subequations}
\label{knot23,24,25,26/*}
\begin{align}
&\varSigma^\sharp{}_\varXi(x,y)=p_M\qquad \text{for $x<\epsilon$},
\vphantom{\Big]}
\label{knot23/*}
\\
&\varSigma^\sharp{}_\varXi(x,y)=p_M\qquad \text{for $x>1-\epsilon$},
\vphantom{\Big]}
\label{knot24/*}
\\
&\varSigma^\sharp{}_\varXi(x,y)=p_M\qquad \text{for $y<\epsilon$},
\vphantom{\Big]}
\label{knot25/*}
\\
&\varSigma^\sharp{}_\varXi(x,y)=p_M\qquad \text{for $y>1-\epsilon$}
\vphantom{\Big]}
\label{knot26/*}
\end{align}
\end{subequations} 
for the same number $\epsilon$ as in \eqref{knot16,17,18,19}, where we used that $\iota_{p_M}(x)=p_M$. 
In the following, we shall call 
$\varSigma^\sharp{}_\varXi$ the normalized surface associated to $\varXi$ to distinguish
it from the unnormalized surface $\varSigma_\varXi$. 

On account of def. \ref{def:knot22/0},
the characteristic $C$--knots $\zeta_{Si}{}^*\varXi$ of an $S$--surface knot $\varXi$ have an obvious description
as curves of $M$. 

\begin{defi} \label{def:knot23/1}
Let $\varXi$ be an $S$--surface knot of $M$. Let $\gamma_{\zeta_{Si}{}^*\varXi}:p_M\rightarrow p_M$ be
the curves of $M$ defined as \hphantom{xxxxxxxxxxxxxxxxxxxxxx}
\begin{equation}
\gamma_{\zeta_{Si}{}^*\varXi}=\gamma_{Mi}.
\label{knot9/2}
\end{equation}
\end{defi}

The surface $\varSigma^\sharp{}_\varXi$ and the curves $\gamma_{\zeta_{Si}{}^*\varXi}$ 
associated with an $S$--surface knot $\varXi$ and its characteristic $C$--knots $\zeta_{Si}{}^*\varXi$ 
enjoy several properties of invariance up to (thin) homotopy. 


\begin{prop} \label{prop:knot6}
For any $S$--surface knot $\varXi$ of $M$, the surface $\varSigma^\sharp{}_\varXi$ is independent 
from the choice of the compatible curve $\gamma_C$ of $C$ 
and surface $\varSigma_S$ of $S$ congruent with it 
up to thin homotopy.
\end{prop}

\noindent {\it Proof}. 
Suppose first that the genus $\ell_S>0$. Fix an oriented differentiable 
universal covering $\tilde S$ of $S$ with orientation preserving covering map 
$\varpi_S:\tilde S\rightarrow S$. 

Let $\gamma_C$ be a curve of $C$ compatible with the pointing of $C$ 
and $\varSigma_S$ be a surface of $S$ compatible with the marking of $S$ 
congruent with $\gamma_C$ (cf. def. \ref{def:consch}). Since $\varSigma_S:\mathbb{R}^2\rightarrow S$ is a map 
with a simply connected domain, there is a lifting map $\tilde\varSigma_S:\mathbb{R}^2\rightarrow \tilde S$
such that $\varpi_S\circ\tilde\varSigma_S=\varSigma_S$ unique up to left composition 
by orientation preserving deck diffeomorphisms of $\tilde S$. The compatibility of $\varSigma_S$ 
(cf. def. \ref{def:knot24}) implies further that there is an open set $\tilde F_S$ of $\tilde S$ 
with the following properties.
First, $\tilde F_S$ is the interior of a simply connected fundamental domain of the covering $\tilde S$.
Second, $\tilde\varSigma_S$ is a map of $\mathbb{R}^2$ onto $\overline{\tilde F}_S$ and 
$\tilde\varSigma_S\big|_{D_S}$ is an orientation preserving diffeomorphism of $D_S$ onto 
$\tilde F_S$. Third, $\varpi_S\big|_{\tilde F_S}$ is an orientation preserving diffeomorphism 
of $\tilde F_S$ onto the image of $\varSigma_S\big|_{D_S}$.
The fact that $\varSigma_S$ expresses a surface $\varSigma_S:\iota_{p_S}\Rightarrow \tau_S$ of $S$
implies then that $\tilde\varSigma_S$ expresses similarly a surface 
$\tilde\varSigma_S:\iota_{\tilde p_S}\Rightarrow \tilde\tau_S$ of $\tilde S$, 
where $\tilde p_S$ and $\tilde\tau_S$ are respectively a point and a curve of $\tilde S$ 
such that $\varpi_S(\tilde p_S)=p_S$ and $\varpi_S\circ \tilde \tau_S=\tau_S$. 
Since $\tilde F_S$ can be considered to be independent from $\gamma_C$ and $\varSigma_S$
by the deck symmetry, $\tilde p_S$ and the image of $\tilde \tau_S$ also are. $\tilde\tau_S$ itself,
instead, is independent from $\gamma_C$ only up to thin homotopy. 
Indeed, reasoning as in the proof of prop. \ref{prop:knot2}, it is found that 
the curves $\alpha_{Sr}$, $\beta_{Sr}$ composing $\tau_S$
according to \eqref{knot15} and consequently $\tau_S$ itself
are independent from the choice of $\gamma_C$ only up to thin homotopy. 
As it is simple to see, also $\tilde\tau_S$ then is. 

Modelling $S$ as $\mathbb{A}/G$, $\tilde S$ as $\mathbb{A}$ and $\varpi_S$ as the quotient map of
$\mathbb{A}/G$, where $\mathbb{A}$ and $G$ are respectively $\mathbb{R}^2$ and a lattice $\varLambda$ of $\mathbb{R}^2$ 
for $\ell_S=1$ and the unit disk $\mathbb{D}_2$ and a Fuchsian group $\varGamma$ for $\ell_S\geq 2$,
we can think of $\tilde\varSigma_S$ as a surjective map $\tilde\varSigma_S:\mathbb{R}^2\rightarrow \mathbb{F}_2$
onto a closed set $\mathbb{F}_2$ of $\mathbb{R}^2$ mapping $D_S$ 
diffeomorphically and orientation preservingly onto $\mathbb{F}_2$. Relatedly, we can think of 
$\tilde\tau_S$ as a surjective map $\tilde\tau_S:\mathbb{R}\rightarrow \partial \mathbb{F}_2$.

Let $\gamma_{C0}$, $\gamma_{C1}$ be two curves of $C$ compatible with the pointing of $C$
and $\varSigma_{S0}$, $\varSigma_{S1}$ be two surfaces of $S$ compatible with the marking of $S$
congruent with $\gamma_{C0}$, $\gamma_{C1}$, respectively. Let $\tilde \varSigma_{S0}$, $\tilde \varSigma_{S1}$ 
be the surfaces of $\tilde S$ lifting respectively $\varSigma_{S0}$, $\varSigma_{S1}$. 
From the remarks of the previous paragraph, on account of 
the thin homotopy of $\tilde\tau_{S0}$, $\tilde\tau_{S1}$, 
there is a map $\tilde h: \mathbb{R}^2\rightarrow \tilde S$ satisfying conditions 
\eqref{path7,8,9,10}
with $p_0=\tilde p_S$, $p_1=\tilde p_S$ and $\gamma_0=\tilde\tau_{S0}$, $\gamma_1=\tilde\tau_{S1}$
and \eqref{path11}. 
Further, there is a map a map $\tilde H:\mathbb{R}^3\rightarrow\tilde S$ fulfilling 
conditions \eqref{path23,24,25,26,27,28} with $p_0=p_1=\tilde p_S$, 
$\tilde H(x,0,z)=\tilde p_S$, $\tilde H(x,1,z)=\tilde h(x,z)$ 
and $\varSigma_0=\tilde\varSigma_{S0}$, $\varSigma_1=\tilde\varSigma_{S1}$. 
Since $\rank(d\tilde h(x,z))\leq 1$, conditions 
\eqref{path29} are also fulfilled. By mere dimensional reasons, condition \eqref{path30} 
is fulfilled as well.
Thus, $\tilde H$ is a thin homotopy of $\tilde \varSigma_{S0}$, $\tilde \varSigma_{S1}$. 
It follows immediately that $H=\varpi_S\circ \tilde H:\mathbb{R}^3\rightarrow S$
is a thin homotopy of $\varSigma_{S0}$, $\varSigma_{S1}$. 

As $\varSigma_{S0}$, $\varSigma_{S1}$ are thin homotopic, for any $S$--surface knot $\varXi$, 
$\varXi\circ \varSigma_{S0}$, $\varXi\circ \varSigma_{S1}$ also are. In particular,
$\varXi_M\circ \varSigma_{S0}$, $\varXi_M\circ \varSigma_{S1}$ are. Thus, both $\varSigma_\varXi$
and $\varSigma_M=\varSigma_{\varXi_M}$ are independent from the choice of $\varSigma_S$ up to thin homotopy. 
By \eqref{knot40} and the fact thin homotopy of surfaces is preserved by vertical composition,
we conclude that $\varSigma^\sharp{}_\varXi$ is independent from 
the choice of $\varSigma_S$ up to thin homotopy. 

The proof can be extended straightforwardly to the genus $\ell_S=0$ case as follows,
where consistency of the chosen compatible surface $\varSigma_S$ of $S$ with a given 
compatible curve $\gamma_C$ of $C$ is no longer an issue. For $\ell_S=0$, 
$S$ is the usual differentiable sphere, which 
be identified 
with the one point compactification $\mathbb{R}^2\cup\infty$ of 
$\mathbb{R}^2$ via the stereographic projection. 
Given any two surfaces $\varSigma_{S0}$, $\varSigma_{S1}$ of $S$ compatible with the marking of $S$, we can find
a thin homotopy $H$ of $\varSigma_{S0}$, $\varSigma_{S1}$ relying on the projection 
We conclude then again that $\varSigma^\sharp{}_\varXi$ is independent from 
the choice of $\varSigma_S$ up to thin homotopy for any $S$--surface knot $\varXi$. \hfill $\Box$  


Suppose that we are given two choices $\varXi_{M0}$, $\varXi_{M1}$ of the reference 
$S$--surface knot. For any $S$--surface knot $\varXi$, let $\varSigma^\sharp{}_{\varXi|0}$, $\varSigma^\sharp{}_{\varXi|1}$
be the normalized surfaces associated with $\varXi$ according to \eqref{knot40}
with $\varXi_M=\varXi_{M0},\varXi_{M1}$, respectively.

\begin{prop} \label{prop:refind}
If the reference $S$--surface knots $\varXi_{M0}$, $\varXi_{M1}$ are ambient isotopic, then
there is a homotopy $H_M{}^\sharp:\varSigma^\sharp{}_{\varXi|0}\Rrightarrow \varSigma^\sharp{}_{\varXi|1}$ 
for any $S$--surface knot $\varXi$. 
\end{prop}

\noindent {\it Proof}. Let $F_{Mz}$, $z\in\mathbb{R}$, be an $S$--marking preserving ambient isotopy relating
the knots $\varXi_{M0}$, $\varXi_{M1}$. Define a map $H_M:\mathbb{R}^3\rightarrow M$ by 
\begin{equation}
H_M(x,y,z)=F_{Mz}\circ \varSigma_{\varXi_{M0}}(x,y).
\label{knotb1/M}
\end{equation}
Using relations \eqref{knot23,24,25,26} together with \eqref{knot7,8}, \eqref{knot9} and \eqref{knot10,11}, 
it is readily  checked that $H_M$ 
satisfies the relations \eqref{path23,24,25,26,27,28} with $p_0=p_1=p_M$ and $\varSigma_0=\varSigma_{\varXi_{M0}}$,
$\varSigma_1=\varSigma_{\varXi_{M1}}$ for some value of $\epsilon$. From \eqref{knot10}, \eqref{knot11} 
and \eqref{knot22}, it is checked further that $H_M(x,0,z)=\iota_{p_M}(x)$,
$H_M(x,1,z)=\tau_M(x)$, so that conditions \eqref{path29} are satisfied too. It follows that 
\eqref{knotb1} defines a homotopy $H_M:\varSigma_{M0}\Rrightarrow \varSigma_{M1}$,
where $\varSigma_{M0}=\varSigma_{\varXi_{M0}}$, $\varSigma_{M1}=\varSigma_{\varXi_{M1}}$.
Given an $S$--surface knot $\varXi$,
this induces a homotopy $H_M{}^\sharp:\varSigma^\sharp{}_{\varXi|0}\Rrightarrow \varSigma^\sharp{}_{\varXi|1}$
on account of \eqref{knot40}. 
\hfill $\Box$  

The normalized surfaces of ambient isotopic $S$--surface knots are homotopic.

\begin{prop} \label{prop:knot7}
If two $S$--surface knots $\varXi_0$, $\varXi_1$ of $M$ are ambient isotopic, then 
there exists a homotopy $H^\sharp:\varSigma^\sharp{}_{\varXi_0}\Rrightarrow \varSigma^\sharp{}_{\varXi_1}$. 
\end{prop}

\noindent {\it Proof}. Let $F_z$, $z\in\mathbb{R}$, be an $S$--marking preserving ambient isotopy relating
the knots $\varXi_0$, $\varXi_1$. Define a map $H:\mathbb{R}^3\rightarrow M$ by 
\begin{equation}
H(x,y,z)=F_{z}\circ \varSigma_{\varXi_0}(x,y).
\label{knotb1}
\end{equation}
Reasoning in the same way as in the proof of prop. \ref{prop:refind}, 
one finds that \eqref{knotb1} defines a homotopy $H:\varSigma_{\varXi_0}\Rrightarrow \varSigma_{\varXi_1}$.
This induces a homotopy $H^\sharp:\varSigma^\sharp{}_{\varXi_0}\Rrightarrow \varSigma^\sharp{}_{\varXi_1}$
on account of \eqref{knot40}. The statement is so shown.
\hfill $\Box$  

\noindent
As a consequence, the surface $\varSigma^\sharp{}_\varXi$ depends only on the ambient isotopy class
of the knot $\varXi$ up to homotopy and the map $\varXi\rightarrow \varSigma^\sharp{}_\varXi$ descends
to one from $\Knot_S(M)$ to the set of surfaces of $M$ mod homotopy. 

We consider now the curves $\gamma_{\zeta_{Si}{}^*\varXi}$ associated with the characteristic $C$--knots 
$\zeta_{Si}{}^*\varXi$ of an $S$--surface knot $\varXi$. 

\begin{prop} \label{prop:cknot2}
For any $S$--surface knot $\varXi$ of $M$, the curves $\gamma_{\zeta_{Si}{}^*\varXi}$ are independent from 
the choice of the compatible curve $\gamma_C$ up to thin homotopy.
\end{prop}

\noindent See def \ref{def:knot11}, eq.  \eqref{knot21/0}.

\noindent {\it Proof}. The proof 
is essentially identical to that of 
prop. \ref{prop:knot2}.
\hfill $\Box$

The following proposition is an important though obvious remark. 

\begin{prop} \label{prop:knot4/1}
If two $S$--surface knots $\varXi_0$, $\varXi_1$ of $M$ are ambient isotopic,
then $\gamma_{\zeta_{Si}{}^*\varXi_1}=\gamma_{\zeta_{Si}{}^*\varXi_0}$
\end{prop}

\noindent
{\it Proof}.
This follows trivially from \eqref{knot9/2}. 
\hfill $\Box$


Until now, we have considered $S$--surface knots belonging to a fixed $S$--marking of $M$.
Analogously to $C$--knots, in topological applications it is necessary to compare
$S$--surface knots belonging to possibly distinct $S$--markings. Again, 
these knots will be topologically equivalent if they are 
ambient isotopic. Since $S$--marking preservation is now necessarily broken,  
the form of ambient isotopy involved here is more general than 
the one considered earlier. We call it free in keeping with 
ordinary knot terminology.
Prop. \ref{prop:knot7} extends also to freely ambient 
isotopic $S$--surface knots in a form which we shall make precise next.

When several $S$--markings $\{p_{M0},\zeta_{M0i}\}$, $\{p_{M1},\zeta_{M1i}\},\,\ldots$ 
of $M$ are given, we shall denote the $S$--marked manifold 
$M$ by $M_0,\,M_1,\,\ldots$ and shall call \pagebreak the $S$--surface knots $\varXi$ of $M$ as $S$--surface
knots of $M_0,\,M_1,\,\ldots$, respectively. 

Suppose that two possibly distinct $S$--markings $\{p_{M0},\zeta_{M0i}\}$, $\{p_{M1},\zeta_{M1i}\}$ of $M$
are given. 

\begin{defi} \label{def:sknotfreeamb}
Two $S$--surface knots $\varXi_0$, $\varXi_1$ of $M_0$, $M_1$, respectively, 
are called freely ambient isotopic if they are related by an ambient isotopy $W$. 
\end {defi}
 
\noindent

As for $C$--knots, when the markings are equal,
free ambient isotopy does not re\-duce to ambient isotopy, 
since $S$--marking preservation is not required.

We consider now two $S$--surface knots $\varXi_0$, $\varXi_1$
of $M_0$, $M_1$, respectively,  which are freely ambient isotopic. 
If $W$ is the underlying connecting ambient isotopy, 
$\varXi_0$, $\varXi_1$ are then related as \hphantom{xxxxxxxxxxxxxxxxxx}
\begin{equation}
\varXi_1=W_{1}\circ \varXi_0.
\label{knot38}
\end{equation}
In virtue of \eqref{knot5,6}, $\{p_{M0},\zeta_{M0i}\}$, $\{p_{M1},\zeta_{M1i}\}$ are connected accordingly,  
\begin{subequations}
\label{knot30,31}
\begin{align}
&p_{M1}=W_1(p_{M0}),
\vphantom{\Big]}
\label{knot30}
\\
&\zeta_{M1i}=W_1\circ \zeta_{M0i}.
\vphantom{\Big]}
\label{knot31}
\end{align}
\end{subequations} 
These relations suggest using the isotopy $W$ to construct a smooth family 
$\{p_{Mz}$, $\zeta_{Mzi}\}$ of $S$--markings interpolating $\{p_{M0},\zeta_{M0i}\}$, $\{p_{M1},\zeta_{M1i}\}$
and a smooth family $\varXi_z$ of $S$--surface knots of $M_z$ interpolating $\varXi_0$, $\varXi_1$. 
To this end, we pick again a non decreasing smooth function $\alpha:\mathbb{R}\rightarrow[0,1]$ 
with the property that 
$\alpha(u)=0$ for $u<\epsilon$ and $\alpha(u)=1$ for $u>1-\epsilon$, where $0<\epsilon<1/2$ such that the 
\eqref{knot16,17/0} and \eqref{knot16,17,18,19} simultaneously hold. 

\begin{defi} \label{def:knot28} For $z\in\mathbb{R}$, let $\{p_{Mz},\zeta_{Mzi}\}$ be the $S$--marking of $M$ given by 
\begin{subequations}
\label{knot32,33}
\begin{align}
&p_{Mz}=W_{\alpha(z)}(p_{M0}),
\vphantom{\Big]}
\label{knot32}
\\
&\zeta_{Mzi}=W_{\alpha(z)}\circ \zeta_{M0i}.
\vphantom{\Big]}
\label{knot33}
\end{align}
\end{subequations} 
\end{defi}

\par\noindent
We note that $p_{Mz}=p_{M0}$,
$\zeta_{Mzi}=\zeta_{M0i}$ for $z<\epsilon$ and $p_{Mz}=p_{M1}$, $\zeta_{Mzi}=\zeta_{M1i}$ 
for $z>1-\epsilon$, where $\{p_{M0},\zeta_{M0i}\}$ and $\{p_{M1},\zeta_{M1i}\}$
are the two given $S$--markings. 

\begin{defi} \label{def:knot31} For $z\in\mathbb{R}$, let $\varXi_z$ be the $S$--surface knot  
of $M_z$ given by 
\begin{equation}
\varXi_z=W_{\alpha(z)}\circ \varXi_0.
\label{knot39}
\end{equation}
\end{defi}

\par\noindent
Properties \eqref{knot5,6} are immediately checked. We note that $\varXi_z=\varXi_0$ for $z<\epsilon$ and 
$\varXi_z=\varXi_1$ for $z>1-\epsilon$, where $\varXi_0$ and $\varXi_1$ stand for the given 
surface knots.

On account of def. \ref{def:knot22/0} and \eqref{knot31}, the characteristic $C$--knots 
$\zeta_{Si}{}^*\varXi_0$, $\zeta_{Si}{}^*\varXi_1$ of $\varXi_0$, $\varXi_1$
are related as
\begin{equation}
\zeta_{Si}{}^*\varXi_1=W_1\circ\zeta_{Si}{}^*\varXi_0.
\label{ccknot1}
\end{equation}
Furthermore, by \eqref{knot33}, the characteristic $C$--knots 
$\zeta_{Si}{}^*\varXi_z$ of $\varXi_z$ smoothly interpolate between 
$\zeta_{Si}{}^*\varXi_0$, $\zeta_{Si}{}^*\varXi_1$, 
\begin{equation}
\zeta_{Si}{}^*\varXi_z=W_{\alpha(z)}\circ\zeta_{Si}{}^*\varXi_0.
\label{ccknot2}
\end{equation}
Clearly, $\zeta_{Si}{}^*\varXi_z=\zeta_{Si}{}^*\varXi_0$ for $z<\epsilon$ and 
$\zeta_{Si}{}^*\varXi_z=\zeta_{Si}{}^*\varXi_1$ for $z>1-\epsilon$, 
$\zeta_{Si}{}^*\varXi_0$, $\zeta_{Si}{}^*\varXi_1$ being precisely the characteristic $C$--knots 
of $\varXi_0$, $\varXi_1$.

In view of \eqref{knot40}, 
the comparison of the normalized surfaces $\varSigma^\sharp{}_{\varXi_0}$, $\varSigma^\sharp{}_{\varXi_1}$
of two $S$--surface knots $\varXi_0$, $\varXi_1$ of $M_0$, $M_1$, respectively, 
which are freely ambient isotopic is meaningful only provided the underlying 
reference $S$--surface knots $\varXi_{M0}$, $\varXi_{M1}$ of $M_0$, $M_1$ are also freely ambient 
isotopic and the ambient isotopy of $\varXi_0$, $\varXi_1$ 
is suitably compatible with that of $\varXi_{M0}$, $\varXi_{M1}$. 

To the above end, the notion of concordance is apt. 

\begin{defi} \label{def:concoknot}
Let $X$ be a manifold $S$--marked by $\{p_X,\zeta_{Xi}\}$.
Two smooth families $F_z$, $G_z$, $z\in\mathbb{R}$, of autodiffeomorphisms
of $X$ are called concordant on the $S$--marking 
if $F_z(p_X)=G_z(p_X)$ and $F_z\circ\zeta_{Xi}=G_z\circ \zeta_{Xi}$ for all $z$. 
\end{defi}

\vspace{-1.5truemm}\eject

\noindent
For genus $l_S>0$, the condition of concordance on $p_X$ actually follows from that on the $\zeta_{Xi}$ 
upon evaluation at $p_C$.
We assume now that the reference $S$--surface knots 
$\varXi_{M0}$, $\varXi_{M1}$ are freely ambient isotopic and that the $S$--surface 
knots $\varXi_0$, $\varXi_1$ are freely ambient isotopic 
with the underlying isotopy concordant with that of $\varXi_{M0}$, $\varXi_{M1}$ 
on $\{p_{M0},\zeta_{M0i}\}$.
%
Stated explicitly, $\varXi_{M0}$, $\varXi_{M1}$ are related by an ambient isotopy $W_M$ 
and $\varXi_0$, $\varXi_1$ are related by an ambient isotopy 
$W$ such that $W_z(p_{M0})=W_{Mz}(p_{M0})$ and 
$W_z\circ\zeta_{M0i}=W_{Mz}\circ \zeta_{M0i}$ for all $z$. 
In this way, for each $z$, the interpolating $S$--markings $\{p_{Mz},\zeta_{Mzi}\}$ of $M$ constructed 
according to \eqref{knot32,33} are the same for $W_M$ and $W$. Moreover, 
the reference surface knots $\varXi_{Mz}$ interpolating 
$\varXi_{M0}$, $\varXi_{M1}$ and the surface knots $\varXi_z$ interpolating 
$\varXi_0$, $\varXi_1$ built  via \eqref{knot39} using $W_M$ and $W$, respectively, 
are both $S$--surface knots of $M_z$.
Out of $\varXi_{Mz}$, $\varXi_z$, using \eqref{knot21} and \eqref{knot40}, we can then build a smooth 
family of normalized surfaces $\varSigma^\sharp{}_{\varXi_z}$ interpolating 
$\varSigma^\sharp{}_{\varXi_0}$, $\varSigma^\sharp{}_{\varXi_1}$.


Using the interpolating set–up constructed above, one can prove the following proposition.

\begin{prop} \label{prop:knot9} 
Assume that the reference $S$--surface knots $\varXi_{M0}$, $\varXi_{M1}$ 
of $M_0$. $M_1$, respectively, are freely ambient isotopic. 
Let $\varXi_0$, $\varXi_1$ be $S$--surface knots 
of $M_0$. $M_1$, respectively.  
Assume that $\varXi_0$, $\varXi_1$ are freely ambient isotopic through an isotopy 
concordant with that of $\varXi_{M0}$, $\varXi_{M1}$ on 
$\{p_{M0},\zeta_{M0i}\}$. Then, there are a curve $\gamma_1:p_{M0}\rightarrow p_{M1}$ of $M$ 
and a homotopy 
$H:\varSigma^\sharp{}_{\varXi_0}\Rrightarrow I_{\gamma_1}{}^{-1_\circ}\circ\varSigma^\sharp{}_{\varXi_1}\circ I_{\gamma_1}$.
\end{prop}

\noindent
Here and below, the ``compose the rightmost first'' convention is extended
to multiple horizontal composition of surfaces. 

\noindent {\it Proof}. 
As $z$ varies, the point $p_{Mz}$ traces a differentiable curve in $M$. 
Since $p_{Mz}$ is by itself a $C$ pointing, this is described by the family of curves 
$\gamma_z:p_{M0}\rightarrow p_{Mz}$ defined by \eqref{knot35/0} and enjoying the same properties.
The ambient isotopies $W_z$, $W_{Mz}$ connecting $\varXi_0$ to $\varXi_1$ and $\varXi_{M0}$ to $\varXi_{M1}$,
respectively, can be employed indifferently in 
\eqref{knot35/0}, as these are concordant on $\{p_{M0},\zeta_{M0i}\}$ by assumption. 

With the $S$--surface knot $\varXi_z$, there is associated the surface $\varSigma^\sharp{}_{\varXi_z}$ according to \eqref{knot40}.
By successively composing the curves $\gamma_z:p_{M0}\rightarrow p_{Mz}$, $\iota_{p_{Mz}},
:p_{Mz}\rightarrow p_{Mz}$ and $\gamma_z{}^{-1_\circ}:p_{Mz}\rightarrow p_{M0}$, 
we obtain a curve $\widehat{\iota}_{p_{Mz}}:p_{M0}\rightarrow p_{M0}$, 
\begin{equation}
\widehat{\iota}_{p_{Mz}}=\gamma_z{}^{-1_\circ}\circ\iota_{p_{Mz}}\circ\gamma_z.
\label{knot41}
\end{equation}
By successively horizontally composing the surfaces 
$I_{\gamma_z}:\gamma_z\Rightarrow \gamma_z$, $\varSigma^\sharp{}_{\varXi_z}:\iota_{p_{Mz}}\Rightarrow\iota_{p_{Mz}}$ and 
$I_{\gamma_z}{}^{-1_\circ}:\gamma_z{}^{-1_\circ}\Rightarrow \gamma_z{}^{-1_\circ}$, 
we obtain then a surface $\widehat{\varSigma}^\sharp{}_{\varXi_z}:\widehat{\iota}_{p_{Mz}}\Rightarrow\widehat{\iota}_{p_{Mz}}$, 
\begin{equation}
\widehat{\varSigma}^\sharp{}_{\varXi_z}=I_{\gamma_z}{}^{-1_\circ}\circ\varSigma^\sharp{}_{\varXi_z}\circ I_{\gamma_z}.
\label{knot42}
\end{equation}
%
%

For continuously varying $z$, the surface $\widehat{\varSigma}^\sharp{}_{\varXi_z}$ 
shifts from the surface $\widehat{\varSigma}^\sharp{}_{\varXi_0}$ to that 
$\widehat{\varSigma}^\sharp{}_{\varXi_1}$ yielding a homotopy 
$\widehat{H}:\widehat{\varSigma}^\sharp{}_{\varXi_0}\Rrightarrow \widehat{\varSigma}^\sharp{}_{\varXi_1}$. 
Indeed, viewed as a map $\widehat{H}:\mathbb{R}^3\rightarrow M$, $\widehat{H}$ is given by \hphantom{xxxxxxxxxx}
\begin{equation}
\widehat{H}(x,y,z)=\widehat{\varSigma}^\sharp{}_{\varXi_z}(x,y).
\label{knot43}
\end{equation}
By the properties of the curves $\iota_{p_{Mz}}$ and $\gamma_z$ 
and the surfaces $\varSigma_{\varXi_z}$, $\varSigma_{Mz}$ 
and \eqref{knot43}, it is evident that $\widehat{H}$ is smooth.  
It is readily verified that $\widehat{H}$ satisfies the \eqref{path23,24,25,26,27,28} with
$p_0=p_1=p_{M0}$, $\varSigma_0=\widehat{\varSigma}^\sharp{}_{\varXi_0}$, $\varSigma_1=\widehat{\varSigma}^\sharp{}_{\varXi_1}$
and perhaps a different value of $\epsilon$.
We notice next that $\widehat{H}(x,0,z)=\widehat{H}(x,1,z)=\widehat{\iota}_{p_{Mz}}(x)$. Here, $\widehat{\iota}_{p_{Mz}}$
is given by the composition \eqref{knot41}, whose factors $\gamma_z$, $\iota_{p_{Mz}}$ have the property that 
$d_x\gamma_z(x)$, $d_z\gamma_z(x)$
are mutually proportional, by \eqref{knot35/0}, and that $d_x\iota_{p_{Mz}}(x)=0$, by \eqref{path3}. 
It follows that  \eqref{path29} also holds. 

By \eqref{knot42}, the source and target surfaces of $\widehat{H}$ explicitly are
$\widehat{\varSigma}^\sharp{}_{\varXi_0}=I_{\iota_{p_{M0}}}{}^{-1_\circ}\circ\varSigma^\sharp{}_{\varXi_0}\circ I_{\iota_{p_{M0}}}$
and $\widehat{\varSigma}^\sharp{}_{\varXi_1}=
I_{\gamma_1}{}^{-1_\circ}\circ\varSigma^\sharp{}_{\varXi_1}\circ I_{\gamma_1}$.
Since $I_{\iota_{p_{M0}}}{}^{-1_\circ}\circ\varSigma^\sharp{}_{\varXi_0}\circ I_{\iota_{p_{M0}}}$
is thin homotopic to $\varSigma^\sharp{}_{\varXi_0}$, 
$\widehat{H}$ induces a homotopy 
$H:\varSigma^\sharp{}_{\varXi_0}\Rrightarrow I_{\gamma_1}{}^{-1_\circ}\circ\varSigma^\sharp{}_{\varXi_1}\circ I_{\gamma_1}$. \hfill $\Box$   

Our last task is the comparison of the curves $\gamma_{\zeta_{Si}{}^*\varXi_0}$, $\gamma_{\zeta_{Si}{}^*\varXi_1}$
of the characteristic $C$--knots $\zeta_{Si}{}^*\varXi_0$, $\zeta_{Si}{}^*\varXi_1$ 
of two freely ambient isotopic $S$--surface knots $\varXi_0$, $\varXi_1$.

\vspace{1.5truemm}\eject

%
%
%
%

\begin{prop} \label{prop:knot9/1} 
Let $\varXi_0$, $\varXi_1$ be  $S$--surface knots of $M_0$, $M_1$, respectively. Assume further that 
$\varXi_0$, $\varXi_1$ are freely ambient isotopic. Then, there are a curve $\gamma_1:p_{M0}\rightarrow p_{M1}$ 
and $2\ell_S$ homotopies $h_i:\gamma_{\zeta_{Si}{}^*\varXi_0}\Rightarrow \gamma_1{}^{-1_\circ}\circ\gamma_{\zeta_{Si}{}^*\varXi_1}\circ \gamma_1$.
Further, $\gamma_1$ can be taken to be be the same curve as that
indicated in prop. \ref{prop:knot9} under the hypotheses of this latter. 
\end{prop}

\noindent {\it Proof}. Fix $i=1,\ldots,2\ell_S$. 
As $z$ varies, the knots $\zeta_{Mzi}$ trace $2\ell_S$ 
differentiable surfaces in $M$. These are described by the families of curves $\gamma_{Mzi}:p_{Mz}\rightarrow p_{Mz}$
associated to the $S$--markings $\{p_{Mz},\zeta_{Mzi}\}$ according to \eqref{xknot3}. 
The expected properties are verified: $\gamma_{Mzi}=\gamma_{M0i}$ for $z<\epsilon$, 
$\gamma_{Mzi}=\gamma_{M1i}$ for $z>1-\epsilon$.

With the ambient isotopy $W$ connecting $\varXi_0$ to $\varXi_1$, there is attached
again the family of curves $\gamma_z$ defined by expression \eqref{knot35/0}
as in the proof of prop. \ref{prop:knot9}. 
By composing successively the curves $\gamma_z:p_{M0}\rightarrow p_{Mz}$, $\gamma_{Mzi}:p_{Mz}\rightarrow p_{Mz}$
and $\gamma_z{}^{-1_\circ}:p_{Mz}\rightarrow p_{M0}$, we obtain the curves 
$\widehat{\gamma}_{Mzi}:p_{M0}\rightarrow p_{M0}$ 
\begin{equation}
\widehat{\gamma}_{Mzi}=\gamma_z{}^{-1_\circ}\circ\gamma_{Mzi}\circ \gamma_z
\label{knot36}
\end{equation}

For continuously varying $z$, the curve $\widehat{\gamma}_{Mzi}$ 
shifts from the curve $\widehat{\gamma}_{M0i}$  to that $\widehat{\gamma}_{M1i}$ 
giving a homotopy of these latter. 
To see this, define $\widehat{h}_i:\mathbb{R}^2\rightarrow M$ by 
\begin{equation}
\widehat{h}_i(x,y)=\widehat{\gamma}_{Myi}(x).
\label{knot37}
\end{equation}
By the properties of the curves $\gamma_{Mzi}$ and $\gamma_z$ 
and \eqref{knot37}, it is evident that $\widehat{h}_i$ is smooth and 
it is readily verified that $\widehat{h}_i$ satisfies the \eqref{path7,8,9,10} with
$p_0=p_1=p_{M0}$, $\gamma_0=\widehat{\gamma}_{M0i}$, $\gamma_1=\widehat{\gamma}_{M1i}$
for some value of $\epsilon$.
So, \eqref{knot37} defines a homotopy $\widehat{h}_i:\widehat{\gamma}_{M0i}\Rightarrow\widehat{\gamma}_{M1i}$. %

By \eqref{knot36}, the source and target curves of $\widehat{h}_i$ are  
$\widehat{\gamma}_{M0i}=\iota_{p_{M0}}{}^{-1_\circ}\circ\gamma_{M0i}\circ \,\iota_{p_{M0}}$ 
and $\widehat{\gamma}_{M1i}=\gamma_1{}^{-1_\circ}\circ\gamma_{M1i}\circ \gamma_1$. As 
$\iota_{p_{M0}}{}^{-1_\circ}\circ\gamma_{M0i}\circ \iota_{p_{M0}}$ is thin homotopic
$\gamma_{M0i}$, the homotopy $\widehat{h}_i$ 
induces a homotopy $h_i:\gamma_{M0i}\Rightarrow \gamma_1{}^{-1_\circ}\circ\gamma_{M1i}\circ \gamma_1$.
Recalling that according to \eqref{knot9/2} $\gamma_{\zeta_{Si}{}^*\varXi_z}=\gamma_{Mzi}$, we conclude that 
the statement holds. \hfill $\Box$

As a free $S$--surface knot $\varXi$ is an $S$--surface knot belonging to
the $S$--marking $\{\varXi(p_S),\zeta_{Si}{}^*\varXi\}$ of $M$, we can construct the associated 
normalized surface
$\varSigma^\sharp{}_\varXi:\iota_{\varXi(p_S)}\Rightarrow \iota_{\varXi(p_S)}$ and $2\ell_S$ curves 
$\gamma_{\zeta_{Si}{}^*\varXi}:\varXi(p_S)\rightarrow \varXi(p_S)$ as detailed above.
The definition of $\varSigma^\sharp{}_\varXi$ involves a reference $S$--surface knot of
$\{\varXi(p_S),\zeta_{Si}{}^*\varXi\}$. 

\begin{defi} \label{defi:concom}
For a free $S$--surface knot $\varXi$, the reference $S$--surface knot $\varXi_{M\varXi}$ 
of the $S$--marking $\{\varXi(p_S),\zeta_{Si}{}^*\varXi\}$ is said concomitant to $\varXi$. 
\end{defi}

\noindent
Unlike for an $S$--surface knot $\varXi$, for which the reference $S$--surface knot $\varXi_M$ is a datum, 
for a free $S$--surface knot $\varXi$, the concomitant reference $S$--surface knot $\varXi_{M\varXi}$ is 
dependent on $\varXi$. Under our working hypothesis, it is determined by $\varXi$
up to $S$--marking preserving ambient isotopy. 

The surface $\varSigma^\sharp{}_\varXi$ and the curves $\gamma_{\zeta_{Si}{}^*\varXi}$
of a free $S$--surface knot $\varXi$ depend on the the marking $\{p_S,\zeta_{Si}\}$ of $S$, 
which up to now we have tacitly assumed to be fixed, 
while $\varXi$ as a mere embedding of $S$ into $M$ does not. 
Of course, there are infinitely many choices of $\{p_S,\zeta_{Si}\}$ and any particular one 
is merely conventional. To be natural, our construction should be independent from the 
choice made in a suitable sense. Let us see how. 

When several markings $\{p_{S0},\zeta_{S0i}\},~\{p_{S1},\zeta_{S1i}\},~\ldots$ 
of $S$ are envisaged, we shall denote $S$ 
by $S_0,~S_1,~\ldots$, call a target $S$--marked manifold $M$ an $S_0$--, $S_1$--, $\ldots$ marked 
manifold and the $S$--surface knots $\varXi$ it supports $S_0$--, $S_1$--, $\ldots$ surface knots, respectively, 
to indicate which marking of $S$ is referred to. 

Assume that two markings $\{p_{S0},\zeta_{S0i}\}$ and $\{p_{S1},\zeta_{S1i}\}$ of $S$ 
with the same underlying $C$--pointing $p_C$ 
are given and that the manifold $M$ is 
$S_0$-- and $S_1$-- marked by $\{p_{M0},\zeta_{M0i}\}$, $\{p_{M1},\zeta_{M1i}\}$, respectively. 
Suppose further that $\varXi$ is simultaneously an $S_0$--surface knot of $M_0$ and an $S_1$--surface 
knot of $M_1$. The natural question arises about the 
relation between the normalized surfaces $\varSigma^\sharp{}_{0\varXi}$, $\varSigma^\sharp{}_{1\varXi}$ 
associated with $\varXi$ in the two markings. 
Since the definition of such surfaces according to \eqref{knot40} involves the 
reference $S_0$--surface knot $\varXi_{M 0}$ of $M_0$ and $S_1$--surface knots $\varXi_{M 1}$
of $M_1$, this relation will not be simple unless $\varXi_{M 0}$, $\varXi_{M 1}$
themselves are simply related. In what follows, we assume that 
$\varXi_{M 0}=\varXi_{M 1}=\varXi_{M}$ perhaps up to $S_0$-- and 
$S_1$--marking preserving ambient isotopy, respectively, 
where $\varXi_{M}$ is simultaneously an $S_0$--surface knot of $M_0$ and an $S_1$--surface knot of $M_1$. 

\begin{lemma} \label{lemma:s0knot}
There is an orientation preserving isotopy $k_z$, $z\in\mathbb{R}$, relating the 
markings $\{p_{S0},\zeta_{S0i}\}$ and $\{p_{S1},\zeta_{S1i}\}$. 
\end{lemma}

\noindent
{\it Proof}. The markings $\{p_{S0},\zeta_{S0i}\}$ and $\{p_{S1},\zeta_{S1i}\}$ of $S$ are 
homotopically equivalent in the sense of algebraic topology, since
both the $\zeta_{S0i}$ and the $\zeta_{S1i}$ represent the $2\ell_S$ homotopy classes 
of the standard oriented $a$-- and $b$--cycles of $S$. There is thus an orientation preserving
isotopy $k_z$, $z\in\mathbb{R}$, deforming $\{p_{S0},\zeta_{S0i}\}$ into
$\{p_{S1},\zeta_{S1i}\}$, that is such that $k_1(p_{S0})=p_{S1}$, $k_1\circ \zeta_{S0i}=\zeta_{S1i}$.
\hfill $\Box$

We now have the following proposition. 

\begin{prop} \label{prop:s0knot1}
Suppose that the reference $S_0$--surface knot $\varXi_{M0}$ of $M_0$ and $S_1$--surface knot $\varXi_{M1}$
of $M_1$ are both equal to a simultaneous $S_0$-- and $S_1$ surface knot $\varXi_M$  of $M_0$ 
and $M_1$ perhaps up to $S_0$-- and $S_1$--marking preserving ambient isotopy, respectively.
Let $\varXi$ be a simultaneous $S_0$-- and $S_1$--surface knot of $M_0$ and $M_1$, respectively, 
having the following properties: $\varXi\circ k_z(p_{S0})=\varXi_M\circ k_z(p_{S0})$ and 
$\varXi\circ k_z\circ \zeta_{S0i}=\varXi_M\circ k_z\circ \zeta_{S0i}$ for all 
$z$, where $k_z$ is an orientation preserving isotopy with the properties stated 
in lemma \ref{lemma:s0knot}. 
Then, there are a curve $\gamma_1:p_{M0}\rightarrow p_{M1}$ of $M$ and a thin homotopy 
$H:\varSigma^\sharp{}_{0\varXi}\Rrightarrow I_{\gamma_1}{}^{-1_\circ}\circ\varSigma^\sharp{}_{1\varXi}\circ I_{\gamma_1}$,
both with image contained in that of $\varXi$.
\end{prop}

\noindent
{\it Proof}. 
Preliminarily, we examine the unnormalized surfaces $\varSigma_{0\varXi}$, $\varSigma_{1\varXi}$
of a simultaneous $S_0$-- and $S_1$--surface knot $\varXi$ of $M_0$ and $M_1$. To construct them, 
compatible surfaces $\varSigma_{S0}$, $\varSigma_{S1}$ of $S_0$, $S_1$, respectively,
must be picked. From the proof of prop. \ref{prop:knot6}, $\varSigma_{0\varXi}$, $\varSigma_{1\varXi}$ 
are independent from the choice of $\varSigma_{S0}$, $\varSigma_{S1}$ up to thin homotopy.
This allows us to enforce a convenient relationship between $\varSigma_{S0}$, $\varSigma_{S1}$.
We select $\varSigma_{S0}$, $\varSigma_{S1}$ related as $\varSigma_{S1}=f\circ \varSigma_{S0}$ with $f=k_1$, 
where $k_z$ is an orientation preserving isotopy with the properties of lemma \ref{lemma:s0knot}. 

%

The composition $\varXi\circ f$ is an $S_0$--surface of $M_1$. As $f=k_1$,
the composition $\varXi\circ k_z$ defines an isotopy of the $S_0$--surface knots $\varXi$ 
and $\varXi\circ f$. By the isotopy extension theorem, there is an ambient isotopy 
$W_z$, $z\in\mathbb{R}$, of $\varXi$ and $\varXi\circ f$ with $W_z\circ \varXi=\varXi\circ k_z$. 
The same holds true 
for the reference $S_0$--and $S_1$--surface knot $\varXi_M$
of $M_0$ and $M_1$, with a generally different ambient isotopy $W_{Mz}$. The $S_0$--surface 
knots $\varXi$, $\varXi\circ f$ are thus freely ambient isotopic and similarly 
$\varXi_M$, $\varXi_M\circ f$.

Because of the assumptions made, we have 
$W_z(p_{M0})=W_z\circ\varXi(p_{S0})=\varXi\circ k_z(p_{S0})=\varXi_M\circ k_z(p_{S0})
=W_{Mz}\circ\varXi_M(p_{S0})=W_{Mz}(p_{M0})$ and 
$W_z\circ \zeta_{M0i}=W_z\circ \varXi\circ\zeta_{S0i}=\varXi\circ k_z\circ\zeta_{S0i}=
\varXi_M\circ k_z\circ\zeta_{S0i}=W_{Mz}\circ \varXi_M\circ\zeta_{S0i}=W_{Mz}\circ \zeta_{M0i}$
for all real $z$.
The ambient isotopies $W_z$, $W_{Mz}$ are therefore concordant on 
$\{p_{M0},\zeta_{M0i}\}$ (cf. def. \ref{def:concoknot}). 


By prop. \ref{prop:knot9} 
with $\varXi_{M0}=\varXi_M$, $\varXi_{M1}=\varXi_M\circ f$
and $\varXi_0=\varXi$, $\varXi_1=\varXi\circ f$, there is a homotopy 
$H:\varSigma^\sharp{}_{0\varXi}\Rrightarrow I_{\gamma_1}{}^{-1_\circ}\circ\varSigma^\sharp{}_{0\varXi\circ f}\circ I_{\gamma_1}$,
where $\gamma_1:p_{M0}\rightarrow p_{M1}$ is a curve of $M$.
Now, as $\varSigma_{S1}=f\circ \varSigma_{S0}$, $\varSigma_{1\varXi}=\varSigma_{0\varXi\circ f}$ and 
$\varSigma_{1\varXi_{M1}}=\varSigma_{0\varXi_{M1}\circ f}$ by \eqref{knot21}. 
Taking \eqref{knot40} into account, we find that 
the above is also a homotopy
$H:\varSigma^\sharp{}_{0\varXi}\Rrightarrow I_{\gamma_1}{}^{-1_\circ}\circ\varSigma^\sharp{}_{1\varXi}\circ I_{\gamma_1}$.

As explained in the proof of prop. \ref{prop:knot9}, 
the homotopy $H$ is the composition of the standard thin homotopy 
$\varSigma^\sharp{}_{0\varXi}\Rrightarrow I_{\iota_{p_{M0}}}{}^{-1_\circ}\circ\varSigma^\sharp{}_{0\varXi}\circ I_{\iota_{p_{M0}}}$
and the homotopy $\widehat{H}:I_{\iota_{p_{M0}}}{}^{-1_\circ}\circ\varSigma^\sharp{}_{0\varXi}\circ I_{\iota_{p_{M0}}}
\Rrightarrow I_{\gamma_1}{}^{-1_\circ}\circ\varSigma^\sharp{}_{1\varXi}\circ I_{\gamma_1}$
defined by \eqref{knot41}--\eqref{knot43}, here with $\varSigma^\sharp{}_{\varXi_z}$ replaced by $\varSigma^\sharp{}_{0\varXi_z}$.
The curve $\gamma_z$ and the surface $\varSigma^\sharp{}_{0\varXi_z}$ entering in \eqref{knot42} 
are built from \eqref{knot39} and its reference analog, \eqref{knot40} and \eqref{knot35/0},
with $W$, $W_M$ the free ambient isotopies introduced above. 
Since $W_z\circ \varXi=\varXi\circ k_z$, $W_{Mz}\circ \varXi_M=\varXi_M\circ k_y$, we have $\gamma_y(x) 
=W_{\alpha(x)\alpha(y)}(\varXi(p_{S0}))=\varXi(k_{\alpha(x)\alpha(y)}(p_{S0}))$
and $\varSigma_{0\varXi_z}(x,y)=W_{\alpha(z)}(\varXi(\varSigma_{S0}(x,y)))=\varXi(k_{\alpha(z)}(\varSigma_{S0}(x,y)))$,
$\varSigma_{0Mz}(x,y)=W_{M\alpha(z)}(\varXi_M(\varSigma_{S0}(x,y)))=\varXi_M(k_{\alpha(z)}(\varSigma_{S0}(x,y)))$.
Hence, the images of $\gamma_z$ and $\varSigma^\sharp{}_{0\varXi_z}$ are contained in that of $\varXi$. As this latter is 
a differentiable surface in $M$, $\widehat{H}$ satisfies \eqref{path30} for dimensional reasons. 
$\widehat{H}$ is therefore a thin homotopy. It follows that $H$ is a thin homotopy.
By the property of the curve $\gamma_y$ just found, 
the image of the curve $\gamma_1$ as well as the homotopy $H$ are contained in that of the knot $\xi$.
\hfill $\Box$
 


The characteristic $C$--knots $\zeta_{S0i}{}^*\varXi$, $\zeta_{S1i}{}^*\varXi$ of 
an $S_0$-- and $S_1$--knot $\varXi$ of $M_0$ and $M_1$ depend on which 
marking one assumes. However, they are independent form the marking 
up to homotopy. 

\begin{prop} \label{prop:s0knot2}
Let $\varXi$ be a simultaneous S$_0$-- and $S_1$--surface knot of $M_0$ and $M_1$,
respectively. Then, there are a curve $\gamma_1:p_{M0}\rightarrow p_{M1}$ of $M$ and 
$2\ell_S$ homotopies $h_i:\gamma_{0\zeta_{S0i}{}^*\varXi}\Rightarrow 
\gamma_1{}^{-1_\circ}\circ\gamma_{1\zeta_{S1i}{}^*\varXi}\circ \gamma_1$ all
with image contained in that of $\varXi$. 
Further, $\gamma_1$ can be taken to be the same curve as that 
indicated in prop. \ref{prop:s0knot1} under the hypotheses of this latter. 
\end{prop}

\noindent {\it Proof}. Let again $f=k_1$, where $k_z$ is the orientation preserving isotopy 
of lemma \ref{lemma:s0knot}. We showed in the proof of prop. \ref{prop:s0knot1} that the 
$S_0$--surface knots $\varXi$, $\varXi\circ f$ are freely ambient isotopic
with the underlying ambient isotopy $W_z$ satisfying $W_z\circ\varXi=\varXi\circ k_z$.   
By prop. \ref{prop:knot9/1} with $\varXi_0=\varXi$, $\varXi_1=\varXi\circ f$
there is then a homotopy $h_i:\gamma_{0\zeta_{S0i}{}^*\varXi}\Rightarrow 
\gamma_1{}^{-1_\circ}\circ\gamma_{0\zeta_{S0i}{}^*(\varXi\circ f)}\circ \gamma_1$,
where the curve $\gamma_1:p_{M0}\rightarrow p_{M1}$ is defined as in the proof of prop. 
\ref{prop:s0knot1} and has hence the properties shown there.  
Since $\gamma_{0\zeta_{S0i}{}^*(\varXi\circ f)}=\gamma_{M1i}=\gamma_{1\zeta_{S1i}{}^*\varXi}$ by \eqref{knot9/2}
trivially, $h_i:\gamma_{0\zeta_{S0i}{}^*\varXi}\Rightarrow 
\gamma_1{}^{-1_\circ}\circ\gamma_{1\zeta_{S1i}{}^*\varXi}\circ \gamma_1$.

The homotopy $h_i$ is the composition of the standard thin homotopy 
$\gamma_{0\zeta_{S0i}{}^*\varXi}\Rightarrow 
\iota_{p_{M0}}{}^{-1_\circ}\circ\gamma_{0\zeta_{S0i}{}^*\varXi}\circ \iota_{p_{M0}}$ and 
the homotopy $\widehat{h}_i$ defined 
in \eqref{knot37} in terms of the family of curves $\widehat{\gamma}_{Myi}$.
$\widehat{\gamma}_{Myi}$ in turn is expressed according to \eqref{knot36} in terms
of $\gamma_y$, $\gamma_{Myi}$. The image of $\gamma_y$ lies in that of $\varXi$ as shown in the proof
of prop. \ref{prop:s0knot1}. Since $\gamma_{Myi}=W_{\alpha(y)}\circ\gamma_{M0i}=
W_{\alpha(y)}\circ\varXi\circ\gamma_{S0i}=\varXi\circ k_{\alpha(y)}\circ \gamma_{S0i}$,
image of $\gamma_{Myi}$ is also contained in that of $\varXi$. The same holds thus true also
for the image of $\widehat{h}_i$ and consequently $h_i$. \hfill $\Box$





\subsection{\normalsize \textcolor{blue}{$2$--knots and source autodiffeomorphism action}}\label{sec:diffeo}

\hspace{.5cm}
In this subsection, 
we consider the role source autodiffeomorphisms on knots.
We begin by reviewing this matter for ordinary knots and the we show 
how the analysis extends to higher knots. 

We let again $C$ be an oriented differentiable closed curve with a choice of pointing
(cf. subsect. \ref{sec:knot}). We let further $M$ be a manifold, but we assume no
$C$--pointing in it.

Autodiffeomorphisms of $C$ act on free $C$--knots by right composition.

\begin{lemma} \label{lemma:diffeo1}
Let $\xi$ be a free $C$--knot and let $f$ be an orientation preserving  
autodiffeomorphism of $C$. Then, $\xi\circ f$ is a free $C$--knot too. 
\end{lemma}

\noindent
{\it Proof}. The statement is obvious. \hfill $\Box$

\begin{defi} \label{def:diffeo1}
For a free $C$--knot $\xi$ and an orientation preserving autodiffeomorphism $f$ 
of $C$, we set \hphantom{xxxxxxxxxxxxxxxxxx}
\begin{equation}
f^*\xi=\xi\circ f.
\label{diffeo3/0}
\end{equation}
$f^*\xi$ is called the pull--back of $\xi$ by $f$. 
\end{defi}

Free $C$--knot pull-back is compatible with ambient isotopy. 

\begin{prop} \label{prop:cdiffeo1}
If $\xi_0$, $\xi_1$ are ambient isotopic free $C$--knots, so are their pull--backs 
$f^*\xi_0$, $f^*\xi_1$ by any orientation preserving autodiffeomorphism $f$.
\end{prop}

\noindent
{\it Proof}. Let $F$ be an 
ambient isotopy relating $\xi_0$, $\xi_1$. Then, relation \eqref{knot9/1} holds. 
From \eqref{diffeo3/0}, it follows that 
$f^*\xi_1   
=F_1\circ f^*\xi_0$.
Consequently, $f^*\xi_0$, $f^*\xi_1$ are also ambient isotopic. 
\hfill $\Box$

We denote by $\Diff(C)$ the group of orientation preserving autodiffeomorphisms of $C$.
It follows that the pull--back action descends to a right action of $\Diff(C)$ on $\Knot_{\mathrm{F}C}(M)$, 
called source autodiffeomorphism action. 

The fact that every orientation preserving autodiffeomorphism of $C$
is isotopic to $\id_C$ entails that the source autodiffeomorphism action of $\Diff(C)$ on $\Knot_{\mathrm{F}C}(M)$
is completely trivial. 

\begin{prop} \label{prop:cdiffeo2} 
For any free $C$--knot $\xi$ and any orientation preserving autodiffeomorphism 
$f$ of $C$, the free $C$--knot $f^*\xi$ is ambient isotopic to $\xi$. 
\end{prop}

\noindent
{\it Proof}. Since all orientation preserving autodiffeomorphisms of $C$ are 
isotopic to $\id_C$, there is an orientation preserving
isotopy $k_y$, $y\in\mathbb{R}$, transforming $\id_C$ into $f$.
The composition $\xi\circ k_y$ is then an isotopy of the $C$--knots $\xi$, 
$f^*\xi$, by \eqref{diffeo3/0}. By the isotopy extension theorem, there is an ambient isotopy 
$W_y$, $y\in\mathbb{R}$ of $\xi$, $f^*\xi$ such that $W_y\circ \xi=\xi\circ k_y$.
\hfill $\Box$

Next, we shall study the properties of the curve $\gamma_\xi:\xi(p_C)\rightarrow \xi(p_C)$ 
associated with a free $C$--knot $\xi$ under the source autodiffeomorphism action $\xi\mapsto f^*\xi$. 
The following proposition is expected from prop. \ref{prop:cdiffeo2}. 

\begin{prop} \label{prop:diffeo2}
Let $\xi$ be a free $C$--knot and let $f$ be an orientation 
preserving autodiffeomorphisms of $C$. 
Then, there are a curve $\gamma_f:\xi(p_C)\rightarrow f^*\xi(p_C)$ of $M$ and a thin homotopy
$h_f:\gamma_{\xi}\Rightarrow \gamma_f{}^{-1_\circ}\circ \gamma_{f^*\xi}$ $\circ\,\gamma_f$, 
both with image contained in that of $\xi$. 
\end{prop}

\noindent
In this subsection, we continue to use the ``compose the rightmost first'' rule for the composition of curves
introduced in subsect. \ref{sec:knot}.

\noindent
{\it Proof}. 
$\xi$, $f^*\xi$ are $C$--knots belonging to the $C$--pointings $\xi(p_C)$, $f^*\xi(p_C)$ of $M$,
respectively. As such, they are freely ambient isotopic by prop. \ref{prop:cdiffeo2}.
By prop. \ref{prop:knot4} with $p_{M0}=\xi(p_C)$, $p_{M1}=f^*\xi(p_C)$ and 
$\xi_0=\xi$ and $\xi_1$ $=f^*\xi$, there is then a homotopy 
$h:\gamma_\xi\Rightarrow \gamma_1{}^{-1_\circ}\circ\gamma_{f^*\xi}\circ \gamma_1$,
where $\gamma_1:\xi(p_C)\rightarrow f^*\xi(p_C)$ is a curve of $M$.
$h$ is the composition of the standard thin homotopy $\gamma_\xi\Rightarrow
\iota_{\xi(p_C)}{}^{-1_\circ}\circ\,\gamma_\xi\circ\iota_{\xi(p_C)}$ and the homotopy 
$\widehat{h}:\iota_{\xi(p_C)}{}^{-1_\circ}\circ\gamma_\xi\circ\iota_{\xi(p_C)}
\Rightarrow \gamma_1{}^{-1_\circ}\circ\gamma_{f^*\xi}\circ \gamma_1$ defined 
by \eqref{knot36/0}, \eqref{knot37/0}. 
The curves $\gamma_y$ and $\gamma_{\xi_y}$ entering in \eqref{knot36/0} 
are built from \eqref{knot39/0}, \eqref{knot35/0}, 
where $W$ is the free ambient isotopy of $\xi$, $f^*\xi$ satisfying
$W_y\circ \xi=\xi\circ k_y$, as shown in the proof of prop. \ref{prop:cdiffeo2}.
Reasoning as in the proof of prop.
\ref{prop:c0knot}, one shows that the images of $\gamma_y$ and $\gamma_{\xi_y}$ are contained in that of 
$\xi$, which is a differentiable curve, so that $\widehat{h}$ satisfies \eqref{path11} for dimensional reasons. 
$\widehat{h}$ is therefore a thin homotopy. It follows that $h$ is a thin homotopy.
By the above property of the curves $\gamma_y$, $\gamma_{\xi_y}$, 
the images of the curve $\gamma_1$ and homotopy $h$ lie in that of the knot $\xi$. 
The proposition now follows setting $\gamma_f=\gamma_1$ and $h_f=h$. 
\hfill $\Box$



We now switch to surface knot theory. 
We let again $S$ be an oriented differentiable closed surface with a choice of marking
(cf. subsect. \ref{sec:knot}). We let further $M$ be a manifold, but we assume no
$S$--marking in it. 

Autodiffeomorphisms of $S$ act on free $S$--surface knots through right composition.
\begin{lemma} \label{lemma:sdiffeo1}
Let $\varXi$ be a free $S$--surface knot and let $f$ be an orientation preserving  
autodiffeomorphism of $S$. Then, $\varXi\circ f$ is a free $S$--surface knot too. 
\end{lemma}

\noindent
{\it Proof}. The statement is obvious. \hfill $\Box$

\begin{defi} \label{def:sdiffeo1}
For a free $S$--surface knot $\varXi$ and an orientation preserving autodiffeomorphism $f$ 
of $S$, we set \hphantom{xxxxxxxxxxxxxxxxxx}
\begin{equation}
f^*\varXi=\varXi\circ f.
\label{sdiffeo3/0}
\end{equation}
$f^*\varXi$ is called the pull--back of $\varXi$ by $f$. 
\end{defi}

Free $S$--surface knot pull-back is compatible with ambient isotopy. 

\begin{prop} \label{prop:sdiffeo1}
If $\varXi_0$, $\varXi_1$ are ambient isotopic free $S$--surface knots, so are their pull--backs 
$f^*\varXi_0$, $f^*\varXi_1$ by any orientation preserving autodiffeomorphism $f$.
\end{prop}

\noindent
{\it Proof}. Let $F$ be an 
ambient isotopy relating $\varXi_0$, $\varXi_1$. Then, relation \eqref{knot9} holds. 
From \eqref{sdiffeo3/0}, it follows that 
$f^*\varXi_1   
=F_1\circ f^*\varXi_0$.
Consequently, $f^*\varXi_0$, $f^*\varXi_1$ are also ambient isotopic. 
\hfill $\Box$

\noindent
We denote by $\Diff(S)$ the group of orientation preserving autodiffeomorphisms of $S$.
It follows that the pull--back action descends to a right action of $\Diff(S)$ on $\Knot_{\mathrm{F}S}(M)$, 
we shall call source autodiffeomorphism action. 

Unlike $C$, not every orientation preserving autodiffeomorphism of $S$
is isotopic to $\id_C$. Prop. \ref{prop:cdiffeo2}, so, extends to surface knots 
in the following weaker form. 

\begin{prop} \label{prop:sdiffeo2}
For any free $S$--surface knot $\varXi$ and any two isotopic orientation preserving autodiffeomorphisms 
$f_0,f_1$ of $S$, the free $S$--surfaces knots $f_0{}^*\varXi$, $f_1{}^*\varXi$ are ambient isotopic.
\end{prop}

\noindent 
{\it Proof}. Since the orientation preserving autodiffeomorphisms $f_0$, $f_1$ of $S$ are 
isotopic, there is an orientation preserving
isotopy $k_z$, $z\in\mathbb{R}$, transforming $f_0$ into $f_1$.
The composition $\varXi\circ h_z$ is then an isotopy of the $S$--surface knots 
$f_0{}^*\varXi$, $f_1{}^*\varXi$, by \eqref{sdiffeo3/0}. 
By the isotopy extension theorem, there is an ambient isotopy 
$W_z$, $z\in\mathbb{R}$ of $f_0{}^*\varXi$ and $f_1{}^*\varXi$ such that $W_z\circ f_0{}^*\varXi=\varXi\circ k_z$.
\hfill $\Box$

Let $\Diff_0(S)$ denote the normal subgroup of $\Diff(S)$ formed by the $S$--auto\-diffeomorphisms
isotopic to $\id_S$. The classic mapping class group 
\begin{equation}
\MCG(S)=\Diff(S)/\Diff_0(S)
\end{equation}
is then defined. By prop. \ref{prop:sdiffeo2}, the source autodiffeomorphism action of 
$\Diff(S)$ on $\Knot_{\mathrm{F}S}(M)$ descends to one of $\MCG(S)$ on $\Knot_{\mathrm{F}S}(M)$. 

Next, we shall study the properties of the surface $\varSigma^\sharp{}_\varXi:
\iota_{\varXi(p_S)}\Rightarrow\iota_{\varXi(p_S)}$ 
associated with a free $S$--surface knot $\varXi$ under the source autodiffeomorphism action
$\varXi\mapsto f^*\varXi$. 
These properties will not be anything simple, unless the the concomitant reference $S$--surface knot 
$\varXi_{M\varXi}$ (cf. def. \ref{defi:concom}) behave naturally under the same action. 
Again, since we cannot prove this in general, we assume it as a
working hypothesis. 
The following couple of propositions are expected from prop. \ref{prop:sdiffeo2}.

\vspace{1.5truemm}\eject

\begin{prop} \label{prop:sdiffeo3} $\vphantom{\dot{\dot{\dot{\dot{x}}}}}$
Let $\varXi$ be a free $S$--surface  knot and let $f_0$, $f_1$ be isotopic orientation 
preserving autodiffeomorphisms of $S$. 
Suppose that the concomitant reference $S$--surface knots 
$\varXi_{Mf_0{}^*\varXi}$, $\varXi_{Mf_1{}^*\varXi}$ 
of $f_0{}^*\varXi$, $f_1{}^*\varXi$ are $f_0{}^*\varXi_{M\varXi}$, $f_1{}^*\varXi_{M\varXi}$, 
respectively, where $\varXi_{M\varXi}$ is the concomitant 
reference $S$--surface knot of $\varXi$. Suppose further that the free ambient isotopies 
relating $f_0{}^*\varXi$, $f_1{}^*\varXi$ and $f_0{}^*\varXi_{M\varXi}$, $f_1{}^*\varXi_{M\varXi}$ 
according to prop. \ref{prop:sdiffeo2} are concordant 
on the $S$--marking $\{f_0{}^*\varXi(p_S)$, $\zeta_{Si}{}^*f_0{}^*\varXi\}$ of $M$.
Then, there are a curve $\gamma_{f_0f_1}:f_0{}^*\varXi(p_S)\rightarrow f_1{}^*\varXi(p_S)$ 
of $M$ and a thin homotopy $H_{f_0f_1}:\varSigma^\sharp{}_{f_0{}^*\varXi}\Rrightarrow 
I_{\gamma_{f_0f_1}}{}^{-1_\circ}\circ \varSigma^\sharp{}_{f_1{}^*\varXi} 
\circ I_{\gamma_{f_0f_1}}$, both with image lying in that of $\varXi$. 
\end{prop}

\noindent
Concordance is defined in \ref{def:concoknot}.
As usual the ``compose the rightmost first'' rule for the composition of surfaces is used. 

\noindent 
{\it Proof}. $f_0{}^*\varXi$, $f_1{}^*\varXi$ are $S$--surface knots belonging to the $S$--markings
$\{f_0{}^*\varXi(p_S)$, $\zeta_{Si}{}^*f_0{}^*\varXi\}$, $\{f_1{}^*\varXi(p_S),\zeta_{Si}{}^*f_1{}^*\varXi\}$,
respectively. As such, they are freely ambient isotopic by prop. \ref{prop:sdiffeo2}. 
The concomitant reference $S$--surface knots $f_0{}^*\varXi_{M\varXi}$, $f_1{}^*\varXi_{M\varXi}$ of 
$f_0{}^*\varXi$, $f_1{}^*\varXi$ are freely ambient isotopic by the same reason. 
By hypothesis, the ambient isotopy of $f_0{}^*\varXi_{M\varXi}$, $f_1{}^*\varXi_{M\varXi}$ is
 concordant with that of $f_0{}^*\varXi$, $f_1{}^*\varXi$
on the $S$--marking $\{f_0{}^*\varXi(p_S)$, $\zeta_{Si}{}^*f_0{}^*\varXi\}$ of $M$.
By prop. \ref{prop:knot9} using $\{p_{M0},\zeta_{M0i}\}=\{f_0{}^*\varXi(p_S),\zeta_{Si}{}^*f_0{}^*\varXi\}$, 
$\{p_{M1},\zeta_{M1i}\}=\{f_1{}^*\varXi(p_S),\zeta_{Si}{}^*f_1{}^*\varXi\}$ and 
$\varXi_0=f_0{}^*\varXi$, $\varXi_1=f_1{}^*\varXi$, 
$\varXi_{M0}=f_0{}^*\varXi_{M\varXi}$, $\varXi_{M1}=f_1{}^*\varXi_{M\varXi}$, there is a homotopy 
$H:\varSigma^\sharp{}_{f_0{}^*\varXi}\Rightarrow I_{\gamma_1}{}^{-1_\circ}\circ\varSigma_{f_1{}^*\varXi}\circ I_{\gamma_1}$,
where $\gamma_1:f_0{}^*\varXi(p_S)\rightarrow f_1{}^*\varXi(p_S)$ is a curve of $M$.
$H$ is the composition of the standard thin homotopy 
$\varSigma^\sharp{}_{f_0{}^*\varXi(p_S)}\Rrightarrow 
I_{\iota_{f_0{}^*\varXi(p_S)}}{}^{-1_\circ}\circ\varSigma^\sharp{}_{f_0{}^*\varXi}\circ I_{\iota_{f_0{}^*\varXi(p_S)}}$
and the homotopy $\widehat{H}:I_{\iota_{f_0{}^*\varXi(p_S)}}{}^{-1_\circ}\circ\varSigma^\sharp{}_{f_0{}^*\varXi}\circ I_{\iota_{f_0{}^*\varXi(p_S)}}
\Rrightarrow I_{\gamma_1}{}^{-1_\circ}\circ\varSigma^\sharp{}_{f_1{}^*\varXi}\circ I_{\gamma_1}$
defined by \eqref{knot41}--\eqref{knot43}. 
The curve $\gamma_z$ and the surface $\varSigma^\sharp{}_{\varXi_z}$ entering in \eqref{knot42} 
are built from \eqref{knot35/0} and \eqref{knot39} and its reference analog via \eqref{knot40} 
with $p_{M0}=f_0{}^*\varXi(p_S)$, $\varXi_0=f_0{}^*\varXi$, $\varXi_{M0}=f_0{}^*\varXi_{M\varXi}$
and $W$, $W_M$ the free ambient isotopies of $f_0{}^*\varXi$, $f_1{}^*\varXi$ 
and $f_0{}^*\varXi_{M\varXi}$, $f_1{}^*\varXi_{M\varXi}$, respectively, 
satisfying $W_z\circ \varXi=\varXi\circ h_z$, $W_{Mz}\circ \varXi_{M\varXi}=\varXi_{M\varXi}\circ h_z$
as shown in  the proof of prop. \ref{prop:sdiffeo2}.
Reasoning as in the proof of prop. \ref{prop:s0knot1}, one shows that the images of 
$\gamma_z$, $\varSigma^\sharp{}_{(f_0{}^*\varXi)_z}$ are contained in that $f_0{}^*\varXi$
and hence $\varXi$, which is a differentiable surface, so that $\widehat{H}$
satisfies \eqref{path30} by dimensional reasons. 
$\widehat{H}$ is therefore a thin homotopy. It follows that $H$ is a thin homotopy.
By the above property of the curve $\gamma_z$ and surface $\varSigma^\sharp{}_{(f_0{}^*\varXi)_z}$
the images of the curve $\gamma_1$ and homotopy $H$ lie in that of the knot $\varXi$. 
The proposition now follows setting $\gamma_{f_0f_1}=\gamma_1$ and $H_{f_0f_1}=H$. 
\hfill $\Box$

\begin{prop} \label{prop:sdiffeo4}
Let $\varXi$ be a free $S$--surface knot and let $f_0$, $f_1$ be isotopic orientation 
preserving autodiffeomorphisms of $S$. 
Then, there are a curve $\gamma_{f_0f_1}:f_0{}^*\varXi(p_S)\rightarrow f_1{}^*\varXi(p_S)$ 
of $M$ and $2\ell_S$ homotopies 
$h_{f_0f_1i}:\gamma_{\zeta_{Si}{}^*f_0{}^*\varXi}\Rightarrow \gamma_{f_0f_1}{}^{-1_\circ}\circ\gamma_{\zeta_{Si}{}^*f_1{}^*\varXi}\circ \gamma_{f_0f_1}$,
all with image is contained in that of $\varXi$. Furthermore, $\gamma_{f_0f_1}$ can be taken to be the same curve as
in prop. \ref{prop:sdiffeo3} under the hypotheses of this latter. 
\end{prop}

\noindent 
{\it Proof}. $f_0{}^*\varXi$, $f_1{}^*\varXi$ are $S$--surface knots belonging to the $S$--markings
$\{f_0{}^*\varXi(p_S)$, $\zeta_{Si}{}^*f_0{}^*\varXi\}$, $\{f_1{}^*\varXi(p_S),\zeta_{Si}{}^*f_1{}^*\varXi\}$,
respectively, which are freely ambient isotopic by prop. \ref{prop:sdiffeo2}. Then, by prop.
\ref{prop:knot9/1} with $\{p_{M0},\zeta_{M0i}\}=\{f_0{}^*\varXi(p_S),\zeta_{Si}{}^*f_0{}^*\varXi\}$, 
$\{p_{M1},\zeta_{M1i}\}=\{f_1{}^*\varXi(p_S),\zeta_{Si}{}^*f_1{}^*\varXi\}$ 
and $\varXi_0=f_0{}^*\varXi$, $\varXi_1=f_1{}^*\varXi$, there are a curve 
$\gamma_1:f_0{}^*\varXi(p_S)\rightarrow f_1{}^*\varXi(p_S)$ of $M$ and $2\ell_S$ homotopies 
$h_i:\gamma_{\zeta_{Si}{}^*f_0{}^*\varXi}\Rightarrow 
\gamma_1{}^{-1_\circ}\circ\gamma_{\zeta_{Si}{}^*f_1{}^*\varXi}\circ \gamma_1$.  
For each $i$, 
$h_i$ is the composition of the standard thin homotopy $\gamma_{\zeta_{Si}{}^*f_0{}^*\varXi}\Rightarrow
\iota_{f_0{}^*\varXi(p_S)}{}^{-1_\circ}\circ\,\gamma_{\zeta_{Si}{}^*f_0{}^*\varXi}\circ\iota_{f_0{}^*\varXi(p_S)}$ and the homotopy 
$\widehat{h}_i:\iota_{f_0{}^*\varXi(p_S)}{}^{-1_\circ}\circ\gamma_{\zeta_{Si}{}^*f_0{}^*\varXi}\circ\iota_{f_0{}^*\varXi(p_S)}
\Rightarrow \gamma_1{}^{-1_\circ}\circ\gamma_{\zeta_{Si}{}^*f_1{}^*\varXi}\circ \gamma_1$ defined according to 
\eqref{knot36}, \eqref{knot37}. The curves $\gamma_z$ and $\gamma_{Mzi}$ entering in \eqref{knot36}
are built from \eqref{knot35/0} and \eqref{ccknot2} via \eqref{xknot3}
with $p_{M0}=f_0{}^*\varXi(p_S)$, $\varXi_{M0}=f_0{}^*\varXi_{M\varXi}$
and $W$, the free ambient isotopies of $f_0{}^*\varXi$, $f_1{}^*\varXi$ 
satisfying $W_z\circ \varXi=\varXi\circ h_z$, 
as shown in  the proof of prop. \ref{prop:sdiffeo2}. Reasoning as in the proof of prop.
\ref{prop:s0knot2}, one shows that the images of $\gamma_z$ and $\gamma_{Mzi}$ are contained 
in that $f_0{}^*\varXi$ and hence $\varXi$. The same thus holds for the images of 
$\gamma_1$ and the $h_i$. The proposition now follows setting $\gamma_{f_0f_1}=\gamma_1$ and $h_{f_0f_1i}=h_i$. 
\hfill $\Box$

\subsection{\normalsize \textcolor{blue}{Sample computations}}\label{sec:sample}

\hspace{.5cm} In this subsection, we present a few simple sample computations
to illustrate the abstract constructions of the preceding subsections. 

To describe knots in terms of curves and surfaces with sitting instant, it is useful to introduce
a function $\alpha:\mathbb{R}\rightarrow [0,1]$ such that
\begin{subequations}
\begin{align}
&d_x\alpha(x)\geq 0,
\vphantom{\Big]}
\label{sample1}
\\
&\alpha(x)=0 \quad\text{for $x<\epsilon$},\qquad \alpha(x)=1 \quad\text{for $x>1-\epsilon$}.
\vphantom{\Big]}
\label{sample2}
\end{align}
\end{subequations}
for some number $\epsilon>0$ with $\epsilon<1/2$. For instance, 
\begin{equation}
\alpha(u)=g_\beta\bigg(\frac{1-2u}{(u-\epsilon)(1-\epsilon-u)}\bigg) 
\vphantom{\Big]}
\label{sample2/12}
\end{equation}
for $\epsilon<u<1-\epsilon$, $\alpha(u)=0$ for $u\leq\epsilon$ and $\alpha(u)=1$ for $u\geq 1-\epsilon$, where 
\begin{equation}
g_\beta(w)=\frac{1}{\exp(\beta w)+1}
\vphantom{\Big]}
\label{sample2/3}
\end{equation}
with $\beta>0$ is the Fermi--Dirac function.

For ordinary knots, the base manifold is $C=S^1$, the circle. To aid geometrical intuition, it is 
convenient to view $S^1$ as an embedded submanifold of $\mathbb{R}^2$ with radius $1$.
A standard embedding of this kind is   
\begin{equation}
s_{S^1}(\vartheta)=(\cos\vartheta,\sin\vartheta),
\label{sample3}
\end{equation}
with $\vartheta\in [0,2\pi)$. A natural choice of pointing is $p_{S^1}=(1,0)$. 
A compatible curve $\gamma_{S^1}:\mathbb{R}\rightarrow S^1$ is now given by 
\begin{equation}
\gamma_{S^1}(x)=s_{S^1}(2\pi\alpha(x)). 
\label{sample3/1}
\end{equation}

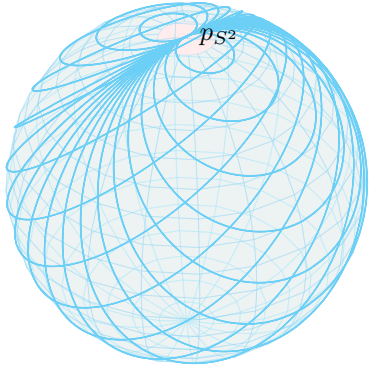
\begin{figure}[!t]
\begin{pspicture}(-7,-2.6)(5,3.)%
\psset{viewpoint= 20 -40 38  rtp2xyz,Decran=48} 

\defFunction[algebraic]%
{sphere}(u,v)
{sin(u)*cos(v)}{sin(u)*sin(v)}{cos(u)}

\psSolid[
object=surfaceparametree,
base= 0 pi 0 pi 2 mul,
function=sphere, 
fillcolor=cyan!20,incolor=red!20, 
opacity=.35,
linewidth=0.3\pslinewidth,
linecolor=cyan!40,
ngrid=20 20 
]%

\multido{\n=.0+.1571}{21}{
\defFunction[algebraic]{zero}(t)
{cos(\n)*sin(\n)*(1+cos(t))}{sin(\n)*sin(t)}{1-sin(\n)^2*(1+cos(t))}
\psSolid[
object=courbe, r=0.002,
range= 0 pi 2 mul,
function=zero, 
linewidth=0.2pt, 
linecolor=cyan!50,
]%
}

\composeSolid

\psPoint(0,0,1){PP}

\uput[r](PP){\footnotesize $p_{S^2}$}

\end{pspicture}
\caption{the surface $\varSigma_{S^2}$. \label{fig:sphere}}
\end{figure}

For genus $0$ surface knots, the base manifolds is $S=S^2$, the usual sphere.
To allow for a pictorial representation, we
view $S^2$ as an embedded submanifold of $\mathbb{R}^3$ with radius $1$.
A standard embedding of this kind is   
\begin{equation}
S_{S^2}(\vartheta,\varphi)=(\cos\vartheta\sin\vartheta(1-\cos\varphi),
-\sin\vartheta\sin\varphi,1-\sin^2\vartheta(1-\cos\varphi)),
\label{sample4}
\end{equation} 
where $\vartheta\in(0,\pi)$ and $\varphi\in[0,2\pi)$.
A customary choice of marking of $S^2$ consists of the point $p_{S^2}=(0,0,1)$, the sphere's north pole. 
A compatible surface $\varSigma_{S^2}:\mathbb{R}^2\rightarrow S^2$ is now given by 
\begin{equation}
\varSigma_{S^2}(x,y)=S_{S^2}(\pi\alpha(y),2\pi\alpha(x)).
\label{sample4/1}
\end{equation}
In fig. \ref{fig:sphere}, a sequence of constant $y$ curves sweeping $S^2$ is drawn in cyan 
\cite{Valdivia:2014zla}. 

For genus $1$ surface knots, $S=T^2$, the familiar torus. For a graphic representation again, 
we view $T^2$ as an embedded submanifold of $\mathbb{R}^3$ with toroidal radius $1$ and 
poloidal radius $r<1$. A standard embedding of this kind is 
\begin{equation}
S_{T^2}(\vartheta_1,\vartheta_2)
=(\cos\vartheta_1(1+r\cos\vartheta_2),
\sin\vartheta_1(1+r\cos\vartheta_2),r\sin\vartheta_2),
\label{sample7}
\end{equation} 
where $\vartheta_1,\vartheta_2\in[0,2\pi)$. A standard marking of $T^2$ consists of the point 
$p_{T^2}=(1+r,0,0)$ and the curves 
\begin{subequations}
\begin{align}
\xi_{T^2}(\vartheta)&=((1+r)\cos\vartheta,(1+r)\sin\vartheta,0),
\vphantom{\Big]}
\label{sample5}
\\
\eta_{T^2}(\vartheta)&=(1+r\cos\vartheta,0,r\sin\vartheta),
\vphantom{\Big]}
\label{sample6}
\end{align}
\end{subequations} %
where $\vartheta\in[0,2\pi)$. 
A compatible surface $\varSigma_{T^2}:\mathbb{R}^2\rightarrow T^2$ 
is given by  
\begin{equation}
\varSigma_{T^2}(x,y)=S_{T^2}(2\pi c_1(x,y)),2\pi c_2(x,y)))
\label{sample7/0}
\end{equation}
where the functions $c_1,c_2:\mathbb{R}^2\rightarrow [0,1]$ are given by 
\begin{subequations}
\label{sample8,9}
\begin{align}
c_1(x,y)&=\varrho(4\alpha(x),\alpha(y))-\varrho(4\alpha(x)-2,\alpha(y)),
\vphantom{\Big]}
\label{sample8}
\\
c_2(x,y)&=\varrho(4\alpha(x)-1,\alpha(y))-\varrho(4\alpha(x)-3,\alpha(y)),
\vphantom{\Big]}
\label{sample9}
\end{align}
\end{subequations}
\begin{figure}[!t]
\begin{pspicture}(-7,-2.65)(5,2.65)     

\psset{viewpoint= 20 20 32  rtp2xyz,Decran=21} %

\defFunction[algebraic]{aa}(s,t){t/(1+2.7183^(-(s-1/2)/(1-t)^(1/2)))}{}{}

\defFunction[algebraic]{xx}(s,t){aa(4*s,t)-aa(4*(s-1/2),t)}{}{}

\defFunction[algebraic]{yy}(s,t){aa(4*(s-1/4),t)-aa(4*(s-3/4),t)}{}{}

\defFunction[algebraic]%
{torus}(u,v)
{cos(2*\psPi*
u
)*(2+1*cos(2*\psPi*
v
))}
{sin(2*\psPi*
u
)*(2+1*cos(2*\psPi*
v
))}
{1*sin(2*\psPi*
v
)}

\psSolid[
object=surfaceparametree,
base= 0 1 0 1,    
function=torus, 
fillcolor=cyan!20, incolor=red!20, opacity=.35,
linewidth=0.3\pslinewidth,
linecolor=gray!40,
ngrid=30 30 
]%

\defFunction[algebraic]%
{circle1}(u)
{cos(2*\psPi*
u
)*(2+1*cos(2*\psPi*
1
))}
{sin(2*\psPi*
u
)*(2+1*cos(2*\psPi*
1
))}
{1*sin(2*\psPi*
1
)}

\defFunction[algebraic]%
{circle2}(v)
{cos(2*\psPi*
1
)*(2+1*cos(2*\psPi*
v
))}
{sin(2*\psPi*
1
)*(2+1*cos(2*\psPi*
v
))}
{1*sin(2*\psPi*
v
)}

\psSolid[
object=courbe, r=0.002,
range= 0 1,
function=circle1, 
linewidth=0.2pt, 
linecolor=blue!25,plotpoints=100 
]%

\psSolid[
object=courbe, r=0.002,
range= 0 1,
function=circle2, 
linewidth=0.2pt, 
linecolor=blue!25,plotpoints=100 
]%

\multido{\n=.01+.05}{20}{
\defFunction[algebraic]{square}(t)
{cos(2*\psPi*
xx(t,\n)
)*(2+1*cos(2*\psPi*
yy(t,\n)
))}
{sin(2*\psPi*
xx(t,\n)
)*(2+1*cos(2*\psPi*
yy(t,\n)
))}
{1*sin(2*\psPi*
yy(t,\n)
)}
\psSolid[
object=courbe, r=0.002,
range= .5 neg 1.5,  
function=square, 
linewidth=0.2pt, 
linecolor=cyan!50, plotpoints=100 
]%
}

\composeSolid

\psPoint(3,0,0){PP}
\psPoint(3 2 pi mul .1 mul Cos mul,3 2 pi mul .1 mul Sin mul,0){C1}
\psPoint(2 2 pi mul .1 mul Cos add,0,2 pi mul .1 mul Sin){C2}

\uput[l](PP){\footnotesize $p_{T^2}$}
\uput[d](C1){\footnotesize $\xi_{T^2}\!\!\!\!\!\!$}
\uput[r](C2){\footnotesize $\eta_{T^2}$}

\end{pspicture}
\caption{the surface $\varSigma_{T^2}$.
\label{fig:torus}}
\vfill
\end{figure}
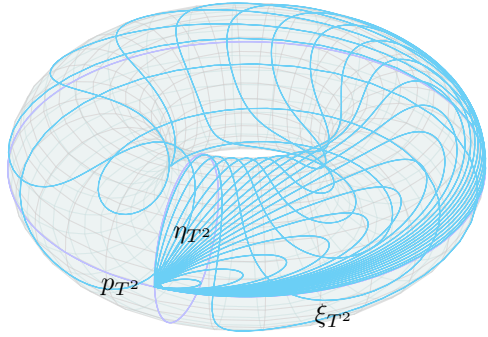
\begin{figure}[!b]
\begin{pspicture}(-5,-5.2)(7,1.) 

\psset{viewpoint= 10  180  90 rtp2xyz,Decran=45} 

\psPoint(0,0,0){O}
\psPoint(1,0,0){X}
\psPoint(0,1,0){Y}
\psPoint(1,1,0){D}
\psPoint(.5,0,0){A}
\psPoint(1,.5,0){B}
\psPoint(.5,1,0){A1}
\psPoint(0,.5,0){B1}

\defFunction[algebraic]%
{a}(u)
{u}{0}{0}

\defFunction[algebraic]%
{b}(u)
{1}{u}{0}

\defFunction[algebraic]%
{a1}(u)
{1-u}{1}{0}

\defFunction[algebraic]%
{b1}(u)
{0}{1-u}{0}

\psSolid[
object=courbe, r=0.002,
range= 0 1,
function=a, 
linewidth=0.2pt, 
linecolor=cyan!100,plotpoints=100 
]%

\psSolid[
object=courbe, r=0.002,
range= 0 1,
function=b, 
linewidth=0.2pt, 
linecolor=cyan!100,plotpoints=100 
]%

\psSolid[
object=courbe, r=0.002,
range= 0 1,
function=a1, 
linewidth=0.2pt, 
linecolor=cyan!100,plotpoints=100 
]%

\psSolid[
object=courbe, r=0.002,
range= 0 1,
function=b1, 
linewidth=0.2pt, 
linecolor=cyan!100,plotpoints=100 
]%

\defFunction[algebraic]{box}(u,v)
{u}
{v}
{0}

\psSolid[
object=surfaceparametree,
base= 0 1 0 1,    
function=box, 
fillcolor=cyan!20, incolor=red!20, opacity=.35,
linewidth=0.3\pslinewidth,
linecolor=gray!40,
ngrid=30 30 
]%

\defFunction[algebraic]{aa}(s,t){t/(1+2.7183^(-(s-1/2)/(1-t)^(1/2)))}{}{}

\defFunction[algebraic]{xx}(s,t){aa(4*s,t)-aa(4*(s-1/2),t)}{}{}

\defFunction[algebraic]{yy}(s,t){aa(4*(s-1/4),t)-aa(4*(s-3/4),t)}{}{}

\multido{\n=.01+.05}{20}{\defFunction[algebraic]{square}(x)
{xx(x,\n)}
{yy(x,\n)}
{0}
\psSolid[
object=courbe, r=0.002,
range= .5 neg 1.5, 
function=square, 
linewidth=0.2pt, 
linecolor=cyan!50,plotpoints=100 
]%
}

\composeSolid

\uput[l](A){\footnotesize $\xi_{T^2}$}
\uput[d](B){\footnotesize $\eta_{T^2}$}
\uput[r](A1){\footnotesize $\xi_{T^2}{}^{-1}$}
\uput[u](B1){\footnotesize $\eta_{T^2}{}^{-1}$}

\uput[ul](O){\footnotesize $p_{T^2}\!$}
\uput[dl](X){\footnotesize $p_{T^2}\!$}
\uput[ur](Y){\footnotesize $p_{T^2}$}
\uput[dr](D){\footnotesize $p_{T^2}$}

\end{pspicture}
\caption{the surface $\varSigma_{T^2}$ when $T^2$ is unfolded.
\label{fig:torusopen}}
\end{figure}
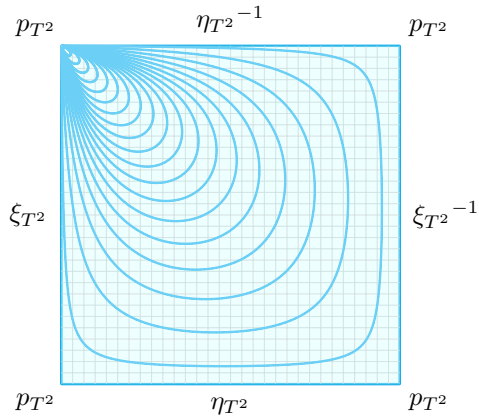
$\varrho:\mathbb{R}\times[0,1]\rightarrow [0,1]$ is defined by 
\begin{equation}
\varrho(s,t)=tg_\beta\bigg(\frac{1-2s}{(1+s-t)(2-s-t)}\bigg).
\label{sample10}
\end{equation}
for $t-1<s<2-t$, $\varrho(s,t)=0$ for \pagebreak $s\leq t-1$ and $\varrho(s,t)=1$ for $s\geq 2-t$ and
the function $g_\beta$ is defined in \eqref{sample2/3}.
In fig. \ref{fig:torus}, a sequence of constant $y$ curves sweeping $T^2$ are drawn in cyan.
In fig. \ref{fig:torusopen}, the sequence of curves is shown when the torus is unfolded for
a clearer visualization. 
The curve $\tau_{T^2}:\mathbb{R}\rightarrow T^2$ as 
\begin{equation}
\tau_{T^2}(x)=S_{T^2}(2\pi c_1(x,y)),2\pi c_2(x,y))),
\label{sample11}
\end{equation}
where $y>1-\epsilon$ whose choice is immaterial.

\vfil\eject

\section{\normalsize \textcolor{blue}{Higher parallel transport and holonomy}}\label{sec:hiholo}

\hspace{.5cm} In this section, we first review the theory of parallel transport both in ordinary and 
in strict higher gauge theory. In our presentation, we follow ref. \cite{SZ:2015}, to which the reader is 
referred to for further details. We then expound our theory of surface holonomy with the main results 
of this paper.

The relevant algebraic and differential geometric structures on which the following analysis rests are 
those of Lie group, Lie algebra, Lie crossed module and differential Lie crossed module. These are 
reviewed in some detail in the appendix of ref. \cite{Soncini:2014ara}, whose notational conventions we adopt. 
We now recall briefly these notions and their mail properties for the sake of completeness and to the reader's benefit. 
Below, unless otherwise stated, we work in the category of smooth manifolds and maps. 

\begin{defi}
A {\it Lie crossed module} consists of 
a pair of Lie groups $G$, $H$ together with two Lie group morphisms $t:H\rightarrow G$ and 
$m:G\rightarrow\Aut(H)$, where $\Aut(H)$ is the automorphism group of $H$,   
obeying the relations 
\begin{subequations}
\label{twogr3}
\begin{align}
&t(m(a)(A))=a t(A)a^{-1},
\vphantom{\Big]}
\label{twogr3a}
\\
&m(t(A))(B)=ABA^{-1}
\vphantom{\Big]}
\label{twogr3b}
\end{align}
\end{subequations}
with $a\in G$, $A,B\in H$.
\end{defi}

\noindent
Just as the notion of Lie group and Lie algebra are intimately related, so are those of Lie crossed module 
and differential Lie crossed module.

\begin{defi}
A {\it differential Lie crossed module} consists of 
a pair of Lie algebras $\mathfrak{g}$, $\mathfrak{h}$ together with 
two Lie algebra morphism $\tau:\mathfrak{h}\rightarrow\mathfrak{g}$.
and $\mu:\mathfrak{g}\rightarrow\mathfrak{der}(\mathfrak{h})$, where
$\mathfrak{der}(\mathfrak{h})$ is the Lie algebra of 
derivations of $\mathfrak{h}$, obeying the relations 
\begin{subequations}
\label{liecross}
\begin{align}
&\tau(\mu(x)(X))=[x,\tau(X)]_{\mathfrak{g}},
\vphantom{\Big]}
\label{liecrossa}
\end{align}
\begin{align}
&\mu(\tau(X))(Y)=[X,Y]_{\mathfrak{h}}
\vphantom{\Big]}
\label{liecrossb}
\end{align}
\end{subequations}
with $x\in\mathfrak{g}$, $X,Y\in\mathfrak{h}$.
\end{defi}

In the same way as there is a Lie algebra canonically derived from any Lie group, there is a differential Lie crossed module 
canonically derived from any Lie crossed module. 

\begin{prop}
With each Lie crossed module $(G,H,t,m)$ there is associated a differential Lie crossed module
$(\mathfrak{g},\mathfrak{h},\dot t,\widehat{m})$ with 
\begin{subequations}
\begin{align}
&\dot t(X)=\frac{dt(C(v))}{dv}\Big|_{v=0},
\vphantom{\Big]}
\label{hgholo1}
\\
&\widehat{m}(x,X)
=\frac{\partial}{\partial u}\Big(\frac{\partial m(c(u))(C(v))}{\partial v}\Big|_{v=0}\Big)\Big|_{u=0}
\vphantom{\Big]}
\label{hgholo2}
\end{align}
\end{subequations}
for $x\in\mathfrak{g}$, $X\in\mathfrak{h}$, where $c(u)$ is any curve in $G$ 
such that $c(u)\big|_{u=0}=1_G$ and $dc(u)/du\big|_{u=0}=x$ and  $C(v)$ is any curve 
in $H$ such that $C(v)\big|_{v=0}=1_H$ and $dC(v)/dv\big|_{v=0}=X$. 
\end{prop}

\noindent
In this paper, we shall need two more mappings arising in any Lie crossed module
defined below. 

\begin{defi}
Each Lie crossed module $(G,H,t,m)$ is characterized by two canonical mappings
$\dot m:G\times\mathfrak{h}\rightarrow \mathfrak{h}$ and 
$Q:\mathfrak{g}\times H\rightarrow \mathfrak{h}$ defined by
\begin{subequations}
\begin{align}
&\dot m(a)(X)=\frac{d}{dv}m(a)(C(v))\Big|_{v=0},
\vphantom{\Big]}
\label{hgholo3}
\\
&Q(x,A)=\frac{d}{du}m(c(u))(A)A^{-1}\Big|_{u=0} 
\vphantom{\Big]}
\label{hgholo4}
\end{align}
\end{subequations}
for $a\in G$, $X\in\mathfrak{h}$, $x\in\mathfrak{g}$, $A\in H$, where $c(u)$ is a curve in $G$ 
such that $c(u)\big|_{u=0}$ $=1_G$ and $dc(u)/du\big|_{u=0}=x$ and $C(v)$ is a curve 
in $H$ such that $C(v)\big|_{v=0}=1_H$ and $dC(v)/dv\big|_{v=0}=X$. 
\end{defi}


\vfil\eject

\subsection{\normalsize \textcolor{blue}{$2$--parallel transport}}\label{sec:twopara}

\hspace{.5cm} 
In this subsection, we review the construction of higher 
parallel transport in strict higher gauge theory. 
This subject has been analyzed  in depth in refs.
\cite{Baez:2004in,Baez:2005qu,Schrei:2009,Schrei:2011,Schrei:2008,
Martins:2007,Martins:2008,Chatterjee:2009ne,Chatterjee:2014pna}. We follow mainly
ref. \cite{Soncini:2014zra}. 
This material is not original and is presented mainly to make this 
paper self--contained and for later reference. For this reason, we give no proof of the basic results.

As an introduction to the topic, we review first parallel transport in ordinary gauge theory. 
Let $M$ be a manifold and $G$ be a Lie group. The trivial principal $G$--bundle on $M$
is assumed as differentiable background.

We begin by reviewing the notion of $G$--connection. 
\begin{defi} \label{def:gconn}
A $G$--connection on $M$, or simply a $G$--connection, is a form $\theta\in\Omega^1(M,\mathfrak{g})$. 
A $G$--connection $\theta$ is said flat if 
\begin{equation}
d\theta+\frac{1}{2}[\theta,\theta]=0. 
\label{twoholo7}
\end{equation}
\end{defi}

\noindent
We now proceed to the definition of parallel transport on ordinary gauge theory.

\begin{defi} \label{def:twofholo1}
Let $\theta$ be a $G$--connection on $M$. 
For any curve $\gamma$ of $M$, 
the element $F_\theta(\gamma)\in G$ defined by \hphantom{xxxxxxxxxxxxxxxxxxx}
\begin{equation}
F_\theta(\gamma)=u(1),
\label{twofholo1}
\end{equation}
where $u:\mathbb{R}\rightarrow G$ is the unique solution of the differential problem 
\begin{equation}
d_xu(x)u(x)^{-1}=-\gamma^*\theta_x(x), \qquad u(0)=1_G,
\label{twofholo2}
\end{equation}
is called the parallel transport along $\gamma$ induced by $\theta$. 
\end{defi}
The basic datum required by the construction of parallel transport is thus a $G$--connection 
$\theta$ on $M$, which we fix once and for all.

Parallel transport is compatible with the operations on curves of def. \ref{def:homotp2}.

\begin{prop} \label{prop:twoholo1} For any point $p$ of $M$, 
\begin{equation}
F_\theta(\iota_p)=1_G.
\label{twoholo2}
\end{equation}
For any curve $\gamma$ of $M$, \hphantom{xxxxxxxxxxxxx}
\begin{equation}
F_\theta(\gamma^{-1_\circ})=F_\theta(\gamma)^{-1}.
\label{twoholo3}
\end{equation}
For any two composable curves $\gamma_1,\gamma_2$ of $M$, 
\begin{equation}
F_\theta(\gamma_2\circ\gamma_1)=F_\theta(\gamma_2)F_\theta(\gamma_1).
\label{twoholo4}
\end{equation} 
\end{prop}

$F_\theta$ has the fundamental property of homotopy invariance as detailed  by the following proposition.

\begin{prop} \label{prop:twoholo2}
Let $\gamma_y$, $y\in\mathbb{R}$, be a smooth $1$--parameter family of curves of $M$ with $y$ independent 
source and target points 
such that the mapping $h:\mathbb{R}^2\rightarrow M$ defined by $h(x,y)=\gamma_y(x)$
is a thin homotopy $h:\gamma_0\Rightarrow\gamma_1$ (cf. def. \ref{def:homotp3}). Then,
\begin{equation}
F_\theta(\gamma_1)=F_\theta(\gamma_0). 
\label{twoholo6}
\end{equation}
The same relation holds if $\theta$ is flat and $h$ is a homotopy
\end{prop}

Parallel transport defines a mapping $F_\theta:\Pi_1M\rightarrow G$
of the set $\Pi_1M$ of curves of $M$ into the group $G$.
Because of the properties stated by props. \ref{prop:twoholo1}, \ref{prop:twoholo2}, 
this mapping descends to a functor $\bar F_\theta:(M,P_1M)\rightarrow BG$
from the path groupoid $(M,P_1M)$ of $M$ into the delooping $BG$ of $G$. 
When $\theta$ is flat, we have further a functor $\bar F^0{}_\theta:(M,P^0{}_1M)\rightarrow BG$
from the fundamental groupoid $(M,P^0{}_1M)$ of $M$ into $BG$. This is an important categorical 
characterization of parallel transport, but it will play no role in the following. 

We consider next parallel transport in strict higher gauge theory. 
Let $M$ be a manifold and $(G,H)$ be a Lie crossed module. 
The trivial principal $(G,H)$--$2$--bundle on $M$
is assumed as differentiable background.

\begin{defi} \label{def:ghconn}
A $(G,H)$--$2$--connection doublet on $M$, or simply a $(G,H)$--con\-nection doublet,
is a pair $(\theta,\varUpsilon)$ of a $1$--form $\theta\in\Omega^1(M,\mathfrak{g})$ and a $2$--form 
$\varUpsilon\in\Omega^2(M,\mathfrak{h})$ 
satisfying the zero fake curvature condition
\begin{equation}
d\theta+\frac{1}{2}[\theta,\theta]-\dot t(\varUpsilon)=0. 
\label{twoholo8}
\end{equation}
A $(G,H)$--connection $(\theta,\varUpsilon)$ is said flat if 
\begin{equation}
d\varUpsilon+\widehat{m}(\theta,\varUpsilon)=0.
\label{twoholo27}
\end{equation}
\end{defi}

Parallel transport 
in higher gauge theory is defined in analogy to ordinary one as follows.

\begin{defi} \label{def:twofholo6} 
Let $(\theta,\varUpsilon)$ be a $(G,H)$--$2$--connection doublet on $M$. 

For any curve $\gamma$ of $M$, 
the element $F_\theta(\gamma)\in G$ defined as in \ref{def:twofholo1} is the 
$1$--parallel transport along $\gamma$ induced by $(\theta,\varUpsilon)$. 

For any 
any surface $\varSigma$ of $M$ 
the element $F_{\theta,\varUpsilon}(\varSigma)\in H$ defined by 
\begin{equation}
F_{\theta,\varUpsilon}(\varSigma)=E(0,1), 
\label{twofholo3}
\end{equation}
where $E:\mathbb{R}^2\rightarrow H$ is the unique solution of the two step differential problem
\begin{subequations}
\label{cycle36,37,38}
\begin{align}
&\partial_xu(x,y)u(x,y)^{-1}=-\varSigma^*\theta_x(x,y),\qquad u(1,y)=1_G,
\vphantom{\Big]}
\label{cycle36}
\\
&\partial_yv(x,y)v(x,y)^{-1}=-\varSigma^*\theta_y(x,y),\,\,\qquad v(x,0)=1_G,
\vphantom{\Big]}
\label{cycle37}
\\
&\partial_x(\partial_yE(x,y)E(x,y)^{-1})
=-\dot m(v(1,y)^{-1}u(x,y)^{-1})(\varSigma^*\varUpsilon_{xy}(x,y)) 
\vphantom{\Big]}
\label{cycle38}
\\
\text{or}~~&\partial_y(E(x,y)^{-1}\partial_xE(x,y))
=-\dot m(u(x,0)^{-1}v(x,y)^{-1})(\varSigma^*\varUpsilon_{xy}(x,y))
\vphantom{\Big]}
\nonumber
\\
&\hspace{6cm}E(1,y)=E(x,0)=1_H.
\vphantom{\Big]}
\nonumber
\end{align}
\end{subequations}
with $u,v:\mathbb{R}^2\rightarrow G$, is called the $2$--parallel transport along $\varSigma$
induced by $(\theta,\varUpsilon)$. 
\end{defi}
The two forms of the differential problem \eqref{cycle38} are equivalent: 
any solution of one is automatically solution of the other. 
The basic datum required by the construction of $1$-- and $2$--parallel transport is a $(G,H)$--connection 
doublet $(\theta,\varUpsilon)$ on $M$, which we fix once and for all.


The $2$--parallel transport along a surface is simply related to the $1$--parallel
transport along its source and target curves. 

\begin{prop} \label{prop:holotar} Let 
$\gamma_0,\gamma_1$ be curves with equal source and target points 
and $\varSigma:\gamma_0\Rightarrow \gamma_1$ be a surface of $M$. Then, 
\begin{equation}
F_\theta(\gamma_1)=t(F_{\theta,\varUpsilon}(\varSigma))F_\theta(\gamma_0).
\label{twoholo14}
\end{equation}
\end{prop}

Parallel transport is compatible with the operations on curves and surfaces of def. \ref{def:homotp5}
analogously to the ordinary case. 

\begin{prop} \label{prop:twoholo6} For any point $p$ of $M$, \eqref{twoholo2} holds. 
For any 
curve $\gamma$ of $M$, \eqref{twoholo3}
also holds. For any two composable curves $\gamma_1,\gamma_2$ of $M$, 
\eqref{twoholo4} holds too. 

\noindent
For any 
curve $\gamma$ of $M$, one has \hphantom{xxxxxxxxxxxxxxx}
\begin{equation}
F_{\theta,\varUpsilon}(I_\gamma)=1_H.
\label{twoholo18}
\end{equation}
For any surface $\varSigma$ of $M$, one has \hphantom{xxxxxxxxxxxxxxx}
\begin{equation}
F_{\theta,\varUpsilon}(\varSigma^{-1\bullet})=F_{\theta,\varUpsilon}(\varSigma)^{-1}.
\label{twoholo19}
\end{equation}
For any two vertically composable surfaces $\varSigma_1,\varSigma_2$ of $M$, one has 
\begin{equation}
F_{\theta,\varUpsilon}(\varSigma_2\bullet\varSigma_1)
=F_{\theta,\varUpsilon}(\varSigma_2)F_{\theta,\varUpsilon}(\varSigma_1).
\label{twoholo20}
\end{equation}
For any surface $\varSigma$ of $M$, one has \hphantom{xxxxxxxxxxxxxxx}
\begin{equation}
F_{\theta,\varUpsilon}(\varSigma^{-1\circ})
=m(F_\theta(\gamma_0)^{-1})(F_{\theta,\varUpsilon}(\varSigma)^{-1}),
\label{twoholo21}
\end{equation}
where $\gamma_0$ is the source curve of $\varSigma$. 
For any two horizontally composable surfaces $\varSigma_1,\varSigma_2$ of $M$, one has 
\begin{equation}
F_{\theta,\varUpsilon}(\varSigma_2\circ\varSigma_1)
=F_{\theta,\varUpsilon}(\varSigma_2)m(F_\theta(\gamma_2))(F_{\theta,\varUpsilon}(\varSigma_1)),
\label{twoholo22}
\end{equation}
\eject\noindent
where $\gamma_2$ is the source curve of $\varSigma_2$. 
\end{prop}

Analogously to the ordinary case, $F_\theta$ and $F_{\theta,\varUpsilon}$ enjoy the property of homotopy invariance 
as established by the following proposition.

\begin{prop} \label{theor:twoholo3}
Let $\gamma_y$, $y\in\mathbb{R}$, be a smooth $1$--parameter family of curves of $M$ with $y$ independent 
source and target points 
such that the mapping $h:\mathbb{R}^2\rightarrow M$ defined by $h(x,y)=\gamma_y(x)$
is a thin homotopy $h:\gamma_0\Rightarrow\gamma_1$ (cf. def. \ref{def:homotp6}). Then, \eqref{twoholo6}
holds


Let  $\gamma_{0z},\gamma_{1z}$ and $\varSigma_z:\gamma_{0z}\Rightarrow\gamma_{1z}$, $z\in\mathbb{R}$,
be respectively two $1$--parameter families of curves with shared $z$ independent source and 
target points and a $1$--parameter family of surfaces such that the mapping $H:\mathbb{R}^3\rightarrow M$ 
defined by $H(x,y,z)=\varSigma_z(x,y)$ is a thin homotopy $H:\varSigma_0\Rrightarrow\varSigma_1$.  
Then, 
\begin{subequations}
\label{twoholo28,29,30}
\begin{align}
&F_\theta(\gamma_{01})=F_\theta(\gamma_{00}),
\vphantom{\Big]}
\label{twoholo28}
\\
&F_\theta(\gamma_{11})=F_\theta(\gamma_{10}),
\vphantom{\Big]}
\label{twoholo29}
\\
&F_{\theta,\varUpsilon}(\varSigma_1)=F_{\theta,\varUpsilon}(\varSigma_0).
\vphantom{\Big]}
\label{twoholo30}
\end{align}
\end{subequations}
The same relations hold if $(\theta,\varUpsilon)$ is flat and $H$ is a homotopy.
\end{prop}

$1$-- and $2$--parallel transport define mappings $F_\theta:\Pi_1M\rightarrow G$,
$F_{\theta,\varUpsilon}:\Pi_2M\rightarrow H$
of the sets $\Pi_1M$, $\Pi_2M$ of curves and surfaces of $M$ into the groups $G$, $H$, respectively.
Because of the properties stated by props. \ref{prop:twoholo6}, \ref{theor:twoholo3}, 
these mappings descend to a strict $2$--functor $\bar F_{\theta,\varUpsilon}:(M,P_1M,P_2M)\rightarrow B(G,H)$
from the path $2$--groupoid $(M,P_1M,P_2M)$ of $M$ into the delooping $B(G,H)$ of the Lie crossed module
$(G,H)$ seen as a strict $2$--group. 
When $(\theta,\varUpsilon)$ is flat, we have further a strict $2$--functor $\bar F^0{}_{\theta,\varUpsilon}:
(M,P_1M,P^0{}_2M)\rightarrow B(G,H)$
from the fundamental $2$--groupoid $(M,P_1M,P^0{}_2M)$ of $M$ into $B(G,H)$. Again, this is an important categorical 
characterization of higher parallel transport, but it will have little import in the following.

\vfil\eject

\subsection{\normalsize \textcolor{blue}{$2$--parallel transport and $1$-- and $2$--gauge transformation}}
\label{sec:gaupara}

\hspace{.5cm}
In this subsection, we review gauge transformation of higher parallel transport in strict higher gauge theory
for self--containedness and later reference. Again the material is not original and no proofs are provided
\cite{Baez:2004in,Baez:2005qu,Schrei:2009,Schrei:2011,Schrei:2008,
Martins:2007,Martins:2008,Chatterjee:2009ne,Chatterjee:2014pna}. We follow mainly ref. \cite{Soncini:2014zra}. 

We begin by reviewing gauge transformation of parallel transport in ordinary gauge theory. 
Let $M$ be a manifold and $G$ be a Lie group. As before, we assume 
the trivial principal $G$ bundle on $M$ as differentiable background.

\begin{defi} A $G$--gauge transformation is a map $g\in\Map(M,G)$. 
\end{defi}

$G$--gauge transformations act on $G$--connections (cf. def. \ref{def:gconn}).

\begin{defi} \label{def:gautransf}
Let $\theta$ be a $G$--connection and $g$ be a $G$--gauge transformation.
The gauge transformed $G$--connection ${}^g\theta$ is 
\begin{equation}
{}^g\theta=\Ad g (\theta)-dgg^{-1}.
\vphantom{\Big]}
\label{gauholo1}
\end{equation}
\end{defi}

\noindent
Gauge transformations preserves flatness of connections.

\begin{prop} If $\theta$ is a flat $G$--connection, then, for any $G$--gauge transformation $g$,
${}^g\theta$ is also a flat $G$--connection (cf. def. \ref{def:gconn}). 
\end{prop}

Let us now fix a $G$--connection $\theta$ and a $G$--gauge transformation $g$. 
The following proposition is a classic result.

\begin{prop} \label{theor:pargau1} 
For any curve $\gamma:p_0\rightarrow p_1$ of $M$, the parallel transports $F_\theta(\gamma)$ 
and $F_{{}^g\theta}(\gamma)$ along $\gamma$ are related as 
\begin{equation}
F_{{}^g\theta}(\gamma)=g(p_1)F_\theta(\gamma)g(p_0)^{-1}.
\vphantom{\Big]}
\label{gauholo2}
\end{equation}
\end{prop}

From a categorical perspective, \pagebreak it can be shown that a $G$--gauge transformation $g$ encodes 
a natural transformation $\bar F_\theta\Rightarrow \bar F_{{}^g\theta}$ of parallel transport functors
(cf. subsect. \ref{sec:hiholo}). 
Furthermore, when $\theta$ is flat, $g$ yields a natural transformation 
$\bar F^0{}_\theta$ $\Rightarrow \bar F^0{}_{{}^g\theta}$ of flat parallel transport functors.

We now shift to higher gauge theory and review gauge transformation of higher 
parallel transport. 
Let $M$ be a manifold and $(G,H)$ be a Lie crossed module. We assume again 
the trivial principal $(G,H)$--bundle  on $M$ as differentiable background.

\begin{defi} A $(G,H)$--$1$--gauge transformation is a pair $(g,J)$ 
of a map $g\in\Map(M,G)$ and a $1$--form $J\in\Omega^1(M,\mathfrak{h})$. 
\end{defi}

$(G,H)$--$1$--gauge transformations  act on $(G,H)$--$2$--connection doublets 
(cf. def. \ref{def:ghconn}).

\begin{defi}  \label{def:gau1transf}
Let $(\theta,\varUpsilon)$ be a $(G,H)$--$2$--connection doublet and $(g,J)$  be a 
$(G$, $H)$--$1$--gauge transformation.
The gauge transformed $(G,H)$--$2$--connection doublet $({}^{g,J}\theta,{}^{g,J}\varUpsilon)$ is 
\begin{subequations}
\label{gauholo3,4}
\begin{align}
&{}^{g,J}\theta=\Ad g(\theta)-dgg^{-1}-\dot t(J),
\vphantom{\Big]}
\label{gauholo3}
\\
&{}^{g,J}\varUpsilon=\dot m(g)(\varUpsilon)-dJ
-\frac{1}{2}[J,J]-\widehat{m}(\Ad g(\theta) -dgg^{-1}-\dot t(J),J).
\vphantom{\Big]}
\label{gauholo4}
\end{align}
\end{subequations}
\end{defi}
It can be checked that this gauge transformation is compatible with the zero fake
curvature condition \eqref{twoholo8}. It also preserves flatness of connection doublets. 

\begin{prop} If $(\theta,\varUpsilon)$ is a flat $(G,H)$--$2$--connection doublet, then, 
for every $(G,H)$--$1$--gauge transformation 
$(g,J)$, $({}^{g,J}\theta,{}^{g,J}\varUpsilon)$ is also a flat $(G,H)$--$2$--connection doublet
(cf. def. \ref{def:ghconn}). 
\end{prop}

The $1$--gauge transformation of $1$-- and $2$--parallel transport
is more involved than its ordinary counterpart. 
Let us now fix a $(G,H)$--$2$--connection doublet $(\theta,\varUpsilon)$
and a $(G,H)$--$1$--gauge transformation $(g,J)$. 

\begin{defi} \label{def:parfgau1}
For any curve $\gamma$ of $M$, the element $G_{g,J;\theta}(\gamma)\in H$ defined by
\begin{equation}
G_{g,J;\theta}(\gamma)=\varLambda(0),
\vphantom{\Big]}
\label{parfgau1}
\end{equation}
where $\varLambda:\mathbb{R}\rightarrow H$ is the unique solution of the differential problem
\begin{equation}
\varLambda(x)^{-1}d_x\varLambda(x)=-\dot m(u(x)^{-1}\gamma^*g(x)^{-1})(\gamma^*J_x(x)), \qquad \varLambda(1)=1_H, 
\vphantom{\Big]}
\label{parfgau2}
\end{equation}
with $u(x)$ being the solution of the differential problem \eqref{twofholo2} but 
with initial condition $u(1)=1_G$, is called 
$1$--gauge parallel transport along $\gamma$. 
\end{defi}

\noindent
The following proposition generalizes prop. \ref{theor:pargau1}.

\begin{prop} \label{theor:pargau2} 
For any curve $\gamma:p_0\rightarrow p_1$ of $M$, one has 
\begin{equation}
F_{{}^{g,J}\theta}(\gamma)=g(p_1)
t(G_{g,J;\theta}(\gamma))F_\theta(\gamma)g(p_0)^{-1}.
\label{gauholo16/1}
\end{equation} 
For any two curves $\gamma_0,\gamma_1:p_0\rightarrow p_1$ and surface 
$\varSigma:\gamma_0\Rightarrow\gamma_1$ of $M$, one has 
\begin{equation}
F_{{}^{g,J}\theta,{}^{g,J}\varUpsilon}(\varSigma)
=m(g(p_1))\big(G_{g,J;\theta}(\gamma_1)
F_{\theta,\varUpsilon}(\varSigma)G_{g,J;\theta}(\gamma_0)^{-1}\big).
\label{gauholo7}
\end{equation}
\end{prop}

\noindent
$1$--gauge parallel transport is compatible with the curve operations of def. \ref{def:homotp5}.

\begin{prop} \label{prop:gauholo8} For any point $p$ of $M$, one has 
\begin{equation}
G_{g,J;\theta}(\iota_p)=1_H.
\label{gauholo15}
\end{equation}
For any curve $\gamma$ of $M$, one has 
\begin{equation}
G_{g,J;\theta}(\gamma^{-1_\circ})=m(F_\theta(\gamma)^{-1})(G_{g,J;\theta}(\gamma)^{-1}).
\label{gauholo16}
\end{equation}
For any two composable curves $\gamma_1,\gamma_2$ of $M$, one has 
\begin{equation}
G_{g,J;\theta}(\gamma_2\circ\gamma_1)=G_{g,J;\theta}(\gamma_2)
m(F_\theta(\gamma_2))(G_{g,J;\theta}(\gamma_1)).
\label{gauholo17}
\end{equation}
\end{prop}

\noindent
$1$--gauge parallel transport is thin homotopy invariant. 

\begin{prop} \label{theor:gauholo3/1}
Let $\gamma_y$, $y\in\mathbb{R}$, be a smooth $1$--parameter family of curves with $y$ independent 
source and target points such that the mapping $h:\mathbb{R}^2\rightarrow M$ defined by $h(x,y)=\gamma_y(x)$
is a thin homotopy of $\gamma_0$, $\gamma_1$. Then,
\begin{equation}
G_{g,J;\theta}(\gamma_1)=G_{g,J;\theta}(\gamma_0). 
\label{gauholo6/2}
\end{equation}
\end{prop}

By prop. \ref{theor:gauholo3/1}, the map 
$G_{g;J;\theta}:\Pi_1M\rightarrow H$ of the set $\Pi_1M$ of curves of $M$ into the group $H$ 
induces one $\bar G_{g;J;\theta}:P_1M\rightarrow H$ of the set $P_1M$ of thin homotopy classes
of curves of $M$ into $H$. From a categorical perspective, props. \ref{theor:pargau2} and 
\ref{prop:gauholo8} entail that a $(G,H)$--$1$--gauge transformation $(g,J)$ encodes a pseudonatural transformation 
$\bar G_{g,J;\theta}:\bar F_{\theta,\varUpsilon}\Rightarrow \bar F_{{}^{g,J}\theta,{}^{g,J}\varUpsilon}$
of parallel transport $2$--functors. 
If $(\theta,\varUpsilon)$ is flat, then $(g,J)$ yields a pseudonatural transformation 
$\bar G^0{}_{g,J;\theta}:\bar F^0{}_{\theta,\varUpsilon}\Rightarrow \bar F^0{}_{{}^{g,J}\theta,{}^{g,J}\varUpsilon}$
of flat parallel transport $2$--functors.



Higher gauge theory enjoys also gauge for gauge symmetry. This plays no role
in the theory of surface knot invariants. We touch it briefly only for the sake of completeness.   

\begin{defi}
A $(G,H)$--$2$--gauge transformation is a map $\varOmega\in \Map(M, H)$.
\end{defi}

$(G,H)$--$2$--gauge transformations act on $(G,H)$--$1$--gauge transformations, the action depending on 
an assigned $(G,H)$--$2$--connection doublet.

\begin{defi} Let $(\theta,\varUpsilon)$ be a $(G,H)$--$2$--connection doublet, 
$(g,J)$ be a $(G,H)$--$1$--gauge transformation
and $\varOmega$ a $(G,H)$--$2$--gauge transformation. The $2$--gauge transformed
$1$--gauge transformation $({}^{\varOmega} g_{|\theta},{}^{\varOmega}J_{|\theta})$ is given by 
\begin{subequations}
\label{twogauholo1,2}
\begin{align}
&{}^{\varOmega}g_{|\theta}=t(\varOmega )g,
\vphantom{\Big]}
\label{twogauholo1}
\\
&{}^{\varOmega}J_{|\theta}=\Ad\varOmega(J) -d\varOmega \varOmega ^{-1}-
Q({}^{g,J}\theta,\varOmega ).
\vphantom{\Big]}
\label{twogauholo2}
\end{align}
\end{subequations}
where ${}^{g,J}\theta$ is given in \eqref{gauholo3}.
\end{defi}

Let us now fix a $(G,H)$--connection \pagebreak doublet $(\theta,\varUpsilon)$ and a $(G,H)$--$1$--gauge
transformation $(g,J)$. 
$2$--gauge equivalent $1$--gauge  transformations yield the the same gauge transformed 
connection doublet. 

\begin{prop}
Let $\varOmega$ be a $(G,H)$--$2$--gauge transformation. Then, 
\begin{subequations}
\label{twogauholo4,5}
\begin{align}
&{}^{{}^{\varOmega}g_{|\theta},{}^{\varOmega}J_{|\theta}}\theta={}^{g,J}\theta,
\vphantom{\Big]}
\label{twogauholo4}
\\
&{}^{{}^{\varOmega}g_{|\theta},{}^{\varOmega}J_{|\theta}}\varUpsilon={}^{g,J}\varUpsilon.
\vphantom{\Big]}
\label{twogauholo5}
\end{align}
\end{subequations}
\end{prop}

\noindent
The action of $2$--gauge transformations on $1$--gauge transformations
translates into one on the map $\gamma\rightarrow G_{g,J;\theta}(\gamma)$. 

\begin{prop}
Let $\varOmega$ be a $(G,H)$--$2$--gauge transformation. Then, 
for any curve $\gamma:p_0\rightarrow p_1$ of $M$, one has 
\begin{equation}
G_{{}^{\varOmega} g_{|\theta},{}^{\varOmega}J_{|\theta};\theta}(\gamma)
=\tilde\varOmega(p_1)^{-1}G_{g,J;\theta}(\gamma)m(F_\theta(\gamma))(\tilde\varOmega(p_0))
\label{twogauholo6}
\end{equation}
where $\tilde\varOmega$ is related to $\varOmega$ by the relation 
\begin{equation}
\tilde\varOmega=m(g^{-1})(\varOmega). 
\label{twogauholo7}
\end{equation}
\end{prop}

From a categorical point of view, it can be shown that a $(G,H)$--$2$--gauge transformation
$\varOmega$ encodes a modification $\bar G_{g,J;\theta}\Rrightarrow
\bar G_{{}^{\tilde \varOmega} g_{|\theta},{}^{\tilde \varOmega}J_{|\theta};\theta}$ of gauge 
pseudonatural transformations of parallel transport functors.


\subsection{\normalsize \textcolor{blue}{$2$--holonomy}}\label{sec:twoholo}

\hspace{.5cm} In this subsection containing the main results of this paper, 
we shall illustrate the theory of higher holonomy based on  the framework of sect. \ref{sec:hiknot}. 

We consider first ordinary knot holonomy in a gauge theory with gauge Lie group $G$ on a manifold $M$.

Let $C$ be a differentiable compact closed curve with a choice of pointing $p_C$
(cf. def. \ref{def:knot3}).  
Recall that with any free $C$--knot $\xi$ of $M$ (cf. def. \ref{def:cfree})
there is associated a curve 
$\gamma_\xi:\xi(p_c)\rightarrow \xi(p_c)$ via eq. \eqref{knot21/0}.  

\begin{defi} \label{def:2holo1} Let $\theta$ be a flat $G$--connection on $M$. For any free 
$C$--knot $\xi$ of $M$, the holonomy of $\xi$ is the element 
$F_\theta(\xi)\in G$ defined by 
\begin{equation}
F_\theta(\xi)=F_\theta(\gamma_\xi).
\label{hiholo1}
\end{equation}
\end{defi}

\noindent
The datum required by the construction of knot holonomies is thus a flat $G$--connection 
$\theta$ on $M$ (cf. def. \ref{def:gconn}, eq. \eqref{twoholo7}), 
which we fix once and for all.

The construction of the curve $\gamma_\xi$ of a free $C$--knot $\xi$ according to \eqref{knot21/0}
involves a choice of a compatible curve $\gamma_C:p_C\rightarrow p_C$ of $C$ (cf. def. \ref{def:knot10}). 
The value $F_\theta(\xi)$ of the holonomy of $\xi$ is however independent from the choice made.

\begin{prop} \label{prop:cknotind} 
Knot holonomy is independent from the choice of the compatible curve $\gamma_C$. 
\end{prop}

\noindent
{\it Proof}. 
By virtue of prop. \ref{prop:knot2}, the curve $\gamma_\xi$ of a free $C$--knot $\xi$ 
does not depend on the choice of $\gamma_C$ up to thin homotopy. Since the map $\gamma\mapsto F_\theta(\gamma)$ is thin 
homotopy invariant for a $G$--connection $\theta$ (cf. prop. \ref{prop:twoholo2}), $F_\theta(\gamma_\xi)$
does not depend on the choice of $\gamma_C$. The statement follows now readily from \eqref{hiholo1}. 
\hfill $\Box$

In order the definition of knot holonomy that we have given to be topologically meaningful, 
the holonomies of ambient isotopic free  $C$--knots (cf. def. \eqref{def:cfreeamb})
should be equivalent in a suitable sense. As it turns out, they are equivalent modulo $G$--conjugation. 

To study this matter, we first assign a $C$--pointing $p_M$ to $M$ (cf. def. \ref{def:knot4}) and consider 
$C$--knots of $M$ (cf. def. \ref{def:knot8}, eq. \eqref{knot5/0}). 
Recall that $C$--knot ambient isotopy is $C$--pointing preserving (cf. def. \ref{def:knot9}).

\begin{prop} \label{prop:2holo1} 
If $\xi_0$, $\xi_1$ are $C$--knots of $M$ and $\xi_0$, $\xi_1$ are ambient isotopic, 
\begin{equation}
F_\theta(\xi_1)=F_\theta(\xi_0). 
\label{xiholo0}
\end{equation}
\end{prop}

\noindent
{\it Proof}. 
By prop. \ref{prop:knot3}, under the stated hypotheses, \pagebreak there is a homotopy 
$\gamma_{\xi_0}\Rightarrow \gamma_{\xi_1}$.
Since the map $\gamma\mapsto F_\theta(\gamma)$ is homotopy invariant for a flat $G$--connection $\theta$
(cf. prop. \ref{prop:twoholo2}), $F_\theta(\gamma_{\xi_0})=F_\theta(\gamma_{\xi_1})$. By \eqref{hiholo1}, 
\eqref{xiholo0} then holds. \hfill $\Box$ 

Suppose now that two distinct $C$--pointings $p_{M0}$, $p_{M1}$ of $M$ are given. 
The holonomies of two $C$--knot $\xi_0$, $\xi_1$ of respectively $M_0$ and $M_1$ 
which are freely ambient isotopic (cf. 
def. \ref{def:cknotfreeamb}), though not strictly equal, are related by a simple $G$--conjugation.

\begin{prop} \label{prop:cfaieqcj}
If $\xi_0$, $\xi_1$ are $C$--knots of $M_0$ and $M_1$, respectively, and $\xi_0$, $\xi_1$ are freely ambient isotopic, 
then
\begin{equation}
F_\theta(\xi_1)=F_\theta(\gamma_1)F_\theta(\xi_0)F_\theta(\gamma_1)^{-1},
\label{xiholo1}
\end{equation}
where $\gamma_1:p_{M0}\rightarrow p_{M1}$ is a curve of $M$.
\end{prop}

\noindent
{\it Proof}. 
By prop. \ref{prop:knot4}, under the stated hypotheses, there exist a curve $\gamma_1:p_{M0}\rightarrow p_{M1}$ of $M$
and a homotopy $\gamma_{\xi_0}\Rightarrow \gamma_1{}^{-1_\circ}\circ\gamma_{\xi_1}\circ \gamma_1$.
Again, as the map $\gamma\mapsto F_\theta(\gamma)$ is homotopy invariant for a flat $G$--connection $\theta$
(cf. prop. \ref{prop:twoholo2}), $F_\theta(\gamma_{\xi_0})=F_\theta(\gamma_1{}^{-1_\circ}\circ\gamma_{\xi_1}\circ \gamma_1)
=F_\theta(\gamma_1)^{-1}F_\theta(\gamma_{\xi_1})F_\theta(\gamma_1)$, where \eqref{twoholo3}, \eqref{twoholo4} have been used. 
By \eqref{hiholo1}, 
\eqref{xiholo1} then holds. \hfill $\Box$ 

Next, we allow the pointing of $C$ to be changed. 
Suppose that two pointings $p_{C0}$, $p_{C1}$ of $C$ are given
and that $M$ is $C$--pointed by $p_{M0}$, $p_{M1}$. 
We denote by $F_{\theta 0}$, $F_{\theta 1}$ the holonomy maps associated with the two pointings of $C$, that is 
$F_{\theta 0}(\xi)=F_\theta(\gamma_{0\xi})$, $F_{\theta 1}(\xi)=F_\theta(\gamma_{1\xi})$ with $\xi$
a free $C_0$-- and $C_1$--knot, respectively, in accordance with \eqref{hiholo1} (cf. subsect. \ref{sec:knot}).

\begin{prop}
If $\xi$ is both a $C_0$--knot of $M_0$ and a $C_1$--knot of $M_1$, then 
\begin{equation}
F_{\theta 1}(\xi)=F_\theta(\gamma_1)F_{\theta 0}(\xi)F_\theta(\gamma_1)^{-1},
\label{xiholo2}
\end{equation}
where $\gamma_1:p_{M0}\rightarrow p_{M1}$ is a curve of $M$ with image lying in that of $\xi$.
\end{prop}

\noindent
{\it Proof}. \pagebreak 
By prop. \ref{prop:c0knot}, under the hypotheses, there are a curve $\gamma_1:p_{M0}\rightarrow p_{M1}$ of $M$and 
a thin homotopy $\gamma_{0\xi}\Rightarrow 
\gamma_1{}^{-1_\circ}\circ\gamma_{1\xi}\circ \gamma_1$, both with image contained in that of $\xi$.
Again, since the map $\gamma\mapsto F_\theta(\gamma)$ is homotopy invariant for a flat $G$--connection $\theta$, 
$F_\theta(\gamma_{0\xi})
=F_\theta(\gamma_1)^{-1}F_\theta(\gamma_{\xi_1})F_\theta(\gamma_1)$, by a calculation similar to that done to prove 
\eqref{xiholo1}. By \eqref{hiholo1}, \eqref{xiholo2} then follows readily. \hfill $\Box$ 

Knot holonomy $\xi\mapsto F_\theta(\xi)$ depends on the flat $G$--connection $\theta$ used in a gauge 
covariant manner (cf. def. \ref{def:gautransf}, eq. \eqref{gauholo1}). Let again $M$ be $C$--pointed by $p_M$. 

\begin{prop} \label{prop:2holo2} 
Let $\xi$ be a $C$--knot of $M$. Then, for any $G$--gauge transformation $g$, then one has 
\begin{equation}
F_{{}^g\theta}(\xi)=g(p_M)F_\theta(\xi)g(p_M)^{-1}.
\vphantom{\Big]}
\label{hiholo2}
\end{equation}
\end{prop}

\noindent
{\it Proof}. If $p$ is a point and $\gamma:p\rightarrow p$ is a closed curve based at $p$,
then one has $F_{{}^g\theta}(\gamma)=g(p)F_\theta(\gamma)g(p)^{-1}$, by \eqref{gauholo2}. 
Then, on account of \eqref{hiholo1} and the fact that $\gamma_\xi:p_M\rightarrow p_M$, we have 
$F_{{}^g\theta}(\gamma_\xi)=g(p_M)F_\theta(\gamma_\xi)g(p_M)^{-1}$. 
\eqref{hiholo2} follows immediately from this relation and \eqref{hiholo1}. \hfill $\Box$ 


We consider next surface knot holonomy in a strict higher gauge theory with gauge Lie crossed module $(G,H)$ on a manifold $M$. 
Let $S$ be a compact closed differentiable surface with a choice of marking (cf. def. \ref{def:knot17}). 
Recall that with any $S$--surface knot $\varXi$ and its $2\ell_S$ characteristic $C$--knots $\zeta_{Si}{}^*\varXi$
(cf. defs. \ref{def:sfree}, \ref{def:knot22/0}), 
there are associated the normalized surface $\varSigma^\sharp{}_\varXi:\iota_{p_M}\Rightarrow \iota_{p_M}$ 
and the curves $\gamma_{\zeta_{Si}{}^*\varXi}:p_M\rightarrow p_M$ via eqs. \eqref{knot21}, \eqref{knot40} and 
\eqref{knot9/2}. Recall also that the definition of $\varSigma^\sharp{}_\varXi$ involves the fixing of
a concomitant reference $S$--surface knot $\varXi_{M\varXi}$ (cf. def. \ref{defi:concom}).

\begin{defi} \label{def:2holo2} Let $(\theta,\varUpsilon)$ be a flat $(G,H)$--$2$--connection doublet on $M$. 
For any free $S$--surface knot $\varXi$ of $M$, the $2$--holonomy 
of $\varXi$ is the element $F_{\theta,\varUpsilon}(\varXi)$ $\in H$ defined by \hphantom{xxxxxxxxxxxxxxxxxx}
\begin{equation}
F_{\theta,\varUpsilon}(\varXi)=F_{\theta,\varUpsilon}(\varSigma^\sharp{}_\varXi)
\label{hiholo4}
\end{equation}
and the $1$--holonomies of the characteristic $C$--knots $\zeta_{Si}{}^*\varXi$ of $\varXi$
are the $2\ell_S$ elements $F_\theta(\zeta_{Si}{}^*\varXi)\in G$ defined by 
\begin{equation}
F_\theta(\zeta_{Si}{}^*\varXi)=F_\theta(\gamma_{\zeta_{Si}{}^*\varXi}).
\label{hiholo4/0}
\end{equation}
\end{defi}
We note that, by \eqref{knot40} and \eqref{twoholo19}, \eqref{twoholo20},
\begin{equation}
F_{\theta,\varUpsilon}(\varXi)=F_{\theta,\varUpsilon}(\varSigma_{M\varXi})^{-1}F_{\theta,\varUpsilon}(\varSigma_\varXi),
\label{hiholo5}
\end{equation}
where $\varSigma_{M\varXi}=\varSigma_{\varXi_{M\varXi}}$. 
The datum required by the construction of the above knot holonomies is thus a flat $(G,H)$--$2$--connection 
doublet $(\theta,\varUpsilon)$ on $M$ (cf. def. \ref{def:ghconn}, eq. \eqref{twoholo27}), 
which we fix once and for all.

\begin{prop} \label{prop:2holo4} For any $S$--surface knot $\varXi$ of $M$, we have 
\begin{equation}
t(F_{\theta,\varUpsilon}(\varXi))=1_H.
\label{cholo1}
\end{equation}
\end{prop}

\noindent
Non trivial $2$--holonomy is thus possible only if the kernel of $t$ is non trivial.

\noindent{\it Proof}. 
If $p$ is a point, $\gamma:p\rightarrow p$ is a closed curve based at $p$ and 
and $\varSigma:\gamma\Rightarrow\gamma$ is a closed surface based at $\gamma$, then, 
by \eqref{twoholo14}, $F_\theta(\gamma)
=t(F_{\theta,\varUpsilon}(\varSigma))F_\theta(\gamma)$.
It follows that $t(F_{\theta,\varUpsilon}(\varSigma))=1_H$.
By \eqref{hiholo4}, applying this result to the closed 
 surface $\varSigma^\sharp{}_\varXi:\iota_{p_M}\Rightarrow \iota_{p_M}$, we find 
$t(F_{\theta,\varUpsilon}(\varSigma^\sharp{}_\varXi))=1_H$. \eqref{cholo1} follows now immediately from 
\eqref{hiholo4}. \hfill$\Box$

The construction of the normalized surface surface $\varSigma^\sharp{}_\varXi$ of a free $S$--surface knot $\varXi$ 
according to \eqref{knot21}, \eqref{knot40} and the curves $\gamma_{\zeta_{Si}{}^*\varXi}$ according to \eqref{knot9/2}
involves a choice of a compatible curve $\gamma_C$ of $C$ and a compatible surface 
$\varSigma_S:\iota_{p_S}\Rightarrow \tau_S$ of $S$ (cf. def. \ref{def:knot24}) congruent with it (cf. def. \ref{def:consch}). 
The values $F_{\theta,\varUpsilon}(\varXi)$ and $F_\theta(\zeta_{Si}{}^*\varXi)$ of the $2$--holonomy of $\varXi$ 
and the $1$--holonomies of the $\zeta_{Si}{}^*\varXi$ are however independent from the choices made.

\begin{prop} \label{prop:sknotind} 
Surface knot holonomy is independent from the choice of the compatible curve $\gamma_C$ and surface $\varSigma_S$
congruent with it. 
\end{prop}

\noindent
{\it Proof}. 
By virtue of prop. \ref{prop:knot6}, the normalized surface $\varSigma^\sharp{}_\varXi$ of a free $S$--surface knot $\varXi$ 
does not depend on the congruent choice of $\gamma_C$ and $\varSigma_S$ up to thin homotopy. Since the map 
$\varSigma\mapsto F_{\theta,\varUpsilon}(\varSigma)$ is thin
homotopy invariant for a $(G,H)$--$2$--connection doublet $(\theta,\varUpsilon)$ 
(cf. prop. \ref{theor:twoholo3}), $F_{\theta,\varUpsilon}(\varSigma^\sharp{}_\varXi)$
does not depend on the congruent choice of $\gamma_C$ and $\varSigma_S$. 
Reasoning as in the proof of prop. \ref{prop:cknotind}, one finds that 
$F_\theta(\gamma_{\zeta_{Si}{}^*\varXi})$ does not depend on the choice of $\gamma_C$
The statement follows now readily from \eqref{hiholo4} and \eqref{hiholo4/0}. 
\hfill $\Box$

As for ordinary knots, in order the definition of surface knot holonomy that we have given to be topologically meaningful, 
the $2$--holonomies of ambient isotopic free $S$--surface knots (cf. def. \eqref{def:sfreeamb})
as well as the $1$--holonomies of their attached characteristic $C$--knots should be equivalent in a suitable sense.
Further, the $2$--holonomy of a given $S$--surface knot corresponding to ambient isotopic choices of the concomitant 
reference $S$--surface knots (cf. def. \ref{defi:concom}) should also be equivalent in the same sense. 
As we are going to show below, all these holonomies are equivalent modulo the appropriate form of
$(G,H)$--crossed module conjugation. 

To study this matter, we first assign a $S$--marking $\{p_M,\zeta_{Mi}\}$ to $M$ 
(cf. def. \ref{def:knot18}) and restrict ourselves to $S$--surface knots of $M$
(cf. def. \ref{def:knot22}, eqs. \eqref{knot5,6}). 
Recall that $S$--surface knot ambient isotopy is $S$--marking preserving (cf. def. \ref{def:knot23}). 


Suppose that we are given two choices $\varXi_{M0}$, $\varXi_{M1}$ of the reference 
$S$--surface knot. For any $S$--surface knot $\varXi$, we let 
$F_{\theta,\varUpsilon}(\varXi_{|0})=F_{\theta,\varUpsilon}(\varSigma^\sharp{}_{\varXi|0})$, 
$F_{\theta,\varUpsilon}(\varXi_{|1})$ $=F_{\theta,\varUpsilon}(\varSigma^\sharp{}_{\varXi|1})$ 
be the corresponding $2$--holonomies of $\varXi$ constructed 
according to \eqref{hiholo4} (cf. subsect. \ref{sec:knot}). 

\begin{prop} \label{prop:2holo5/00} 
If the reference $S$--surface knots $\varXi_{M0}$, $\varXi_{M1}$ are ambient isotopic, then
for any $S$--surface knot $\varXi$
\begin{equation}
F_{\theta,\varUpsilon}(\varXi_{|1})=F_{\theta,\varUpsilon}(\varXi_{|0}).
\label{mhiholo0}
\end{equation}
\end{prop}

\noindent
{\it Proof}. Under the above hypothesis, for every $S$--surface knot $\varXi$ 
there is a homotopy $H_M{}^\sharp:\varSigma^\sharp{}_{\varXi|0}\Rrightarrow \varSigma^\sharp{}_{\varXi|1}$ 
(cf. prop. \ref{prop:refind}). Since the map
$\varSigma\rightarrow F_{\theta,\varUpsilon}(\varSigma)$ is homotopy invariant 
for a flat $(G,H)$--$2$--connection doublet $(\theta,\varUpsilon)$ (cf. prop. \ref{theor:twoholo3}), 
$F_{\theta,\varUpsilon}(\varSigma^\sharp{}_{\varXi|0})=
F_{\theta,\varUpsilon}(\varSigma^\sharp{}_{\varXi|1})$.
From \eqref{hiholo4}, the statement then follows. 
\hfill $\Box$


\begin{prop} \label{prop:2holo5} 
If $\varXi_0$, $\varXi_1$ are $S$--surface knots of $M$ and $\varXi_0$, $\varXi_1$ are ambient isotopic, 
then one has \hphantom{xxxxxxxxxxx}
\begin{equation}
F_{{\theta,\varUpsilon}}(\varXi_1)=F_{\theta,\varUpsilon}(\varXi_0)
\vphantom{\Big]}
\label{ahiholo1}
\end{equation}
and likewise \hphantom{xxxxxxxxxxx}
\begin{equation}
F_\theta(\zeta_{Si}{}^*\varXi_1)=F_\theta(\zeta_{Si}{}^*\varXi_0). 
\label{ahiholo2}
\end{equation}
\end{prop}

\noindent
{\it Proof}. 
By  prop. \ref{prop:knot7}, under the stated hypotheses,
there is a homotopy $\varSigma^\sharp{}_{\varXi_0}\Rrightarrow \varSigma^\sharp{}_{\varXi_1}$.
Again, as the map $\varSigma\rightarrow F_{\theta,\varUpsilon}(\varSigma)$ is homotopy invariant 
for a flat $(G,H)$--$2$--connection doublet $(\theta,\varUpsilon)$ (cf. prop. \ref{theor:twoholo3}), 
$F_{\theta,\varUpsilon}(\varSigma^\sharp{}_{\varXi_0})=F_{\theta,\varUpsilon}(\varSigma^\sharp{}_{\varXi_1})$. 
From \eqref{hiholo4}, \eqref{ahiholo1} follows. 

By prop. \ref{prop:knot4/1}, under the same hypotheses, $\gamma_{\zeta_{Si}{}^*\varXi_1}=\gamma_{\zeta_{Si}{}^*\varXi_0}$,
so that $F_\theta(\gamma_{\zeta_{Si}{}^*\varXi_1})=F_\theta(\gamma_{\zeta_{Si}{}^*\varXi_0})$. 
From \eqref{hiholo4/0}, \eqref{ahiholo2} also follows. \hfill $\Box$ 

Suppose now that two distinct $S$--markings $\{p_{M0},\zeta_{M0i}\}$, $\{p_{M1},\zeta_{M1i}\}$ of $M$ are given. 
The $2$ holonomies of two $S$--surface knots $\varXi_0$, $\varXi_1$ of respectively $M_0$ and $M_1$
which are freely ambient isotopic (cf. 
def. \ref{def:cknotfreeamb})
as well as the $1$--holonomies of their attached characteristic $C$--knots $\zeta_{Si}{}^*\varXi_0$, 
$\zeta_{Si}{}^*\varXi_1$, though different in general, are related by a form of $(G,H)$--conjugation.
We remind the reader the notion of concordance introduced in def. \ref{def:concoknot}. 



\begin{prop} \label{prop:2holo7} 
Assume that the reference $S$--surface knots $\varXi_{M0}$, $\varXi_{M1}$ 
of $M_0$. $M_1$, respectively, are freely ambient isotopic. 
If $\varXi_0$, $\varXi_1$ are $S$--surface knots of $M_0$. $M_1$, respectively, 
and $\varXi_0$, $\varXi_1$ are freely ambient isotopic through an isotopy 
concordant with that of $\varXi_{M0}$, $\varXi_{M1}$ on 
$\{p_{M0},\zeta_{M0i}\}$, \pagebreak then there are a curve $\gamma_1:p_{M0}\rightarrow p_{M1}$ and $2\ell_S$ surfaces 
$\varSigma_i:\gamma_{\zeta_{Si}{}^*\varXi_0}\Rightarrow \gamma_1{}^{-1_\circ}\circ\gamma_{\zeta_{Si}{}^*\varXi_1}\circ \gamma_1$,
of $M$ such that 
\begin{equation}
F_{{\theta,\varUpsilon}}(\varXi_1)=m(F_\theta(\gamma_1))(F_{\theta,\varUpsilon}(\varXi_0))
\vphantom{\Big]}
\label{hiholo8}
\end{equation}
and similarly that 
\begin{equation}
F_\theta(\zeta_{Si}{}^*\varXi_1)=F_\theta(\gamma_1)t(F_{\theta,\varUpsilon}(\varSigma_i))
F_\theta(\zeta_{Si}{}^*\varXi_0)F_\theta(\gamma_1)^{-1}.
\label{hiholo7}
\end{equation}
\end{prop}

\noindent
{\it Proof}. According to prop. \ref{prop:knot9}, 
under the above hypotheses, there exists a curve $\gamma_1:p_{M0}\rightarrow p_{M1}$ and a homotopy 
$\varSigma^\sharp{}_{\varXi_0}\Rrightarrow I_{\gamma_1}{}^{-1_\circ}\circ\varSigma^\sharp{}_{\varXi_1}\circ I_{\gamma_1}$.
Again, by the homotopy invariance of the map $\varSigma\rightarrow F_{\theta,\varUpsilon}(\varSigma)$ 
for a flat $(G,H)$--$2$--connection doublet $(\theta,\varUpsilon)$ (cf. prop. \ref{theor:twoholo3}), 
we have 
\begin{align}
F_{\theta,\varUpsilon}(\varSigma^\sharp{}_{\varXi_0})
&=F_{\theta,\varUpsilon}(I_{\gamma_1}{}^{-1_\circ}\circ\varSigma^\sharp{}_{\varXi_1}\circ I_{\gamma_1})
\vphantom{\Big]}
\nonumber 
\\
&=m(F_\theta(\gamma_1)^{-1})\big(F_{\theta,\varUpsilon}(I_{\gamma_1})^{-1}
F_{\theta,\varUpsilon}(\varSigma^\sharp{}_{\varXi_1})m(F_\theta(\iota_{p_{M1}}))(F_{\theta,\varUpsilon}(I_{\gamma_1}))\big)
\vphantom{\Big]}
\nonumber 
\\
&=m(F_\theta(\gamma_1)^{-1})(F_{\theta,\varUpsilon}(\varSigma^\sharp{}_{\varXi_1})),
\vphantom{\Big]}
\nonumber 
\end{align}
where \eqref{twoholo2}, \eqref{twoholo3}, \eqref{twoholo18}, 
\eqref{twoholo21}, \eqref{twoholo22} have been used. 
By \eqref{hiholo4}, this relation is equivalent to \eqref{hiholo8}. 

According to prop. \ref{prop:knot9/1}, under the same hypotheses, for each $i$ there exists a homotopy 
$\gamma_{\zeta_{Si}{}^*\varXi_0}\Rightarrow \gamma_1{}^{-1_\circ}\circ\gamma_{\zeta_{Si}{}^*\varXi_1}\circ \gamma_1$, 
which we presently view as a surface $\varSigma_i$ with the same  source and target curves,
$\gamma_1$ being the same curve as the one featured above. 
Again, by the homotopy invariance of the map $\gamma \rightarrow F_\theta(\gamma)$ 
for a flat $(G,H)$--$2$--connection doublet $(\theta,\varUpsilon)$ (cf. prop. \ref{theor:twoholo3}), 
we have 
\begin{align}
t(F_{\theta,\varUpsilon}(\varSigma_i))F_\theta(\gamma_{\zeta_{Si}{}^*\varXi_0})
&=F_\theta(\gamma_1{}^{-1_\circ}\circ\gamma_{\zeta_{Si}{}^*\varXi_1}\circ \gamma_1)
\vphantom{\Big]}
\nonumber 
\\
&=F_\theta(\gamma_1)^{-1}F_\theta(\gamma_{\zeta_{Si}{}^*\varXi_1})F_\theta(\gamma_1),
\vphantom{\Big]}
\nonumber 
\end{align}
where \eqref{twoholo3}, \eqref{twoholo4} have been used. 
Using \eqref{hiholo4/0}, this relation immediately reduces to \eqref{hiholo7}.  \hfill $\Box$ 

Next, we allow the marking of $S$ to be changed. \pagebreak 
Suppose that two markings $\{p_{S0},\zeta_{S0i}\}$, $\{p_{S1},\zeta_{S1i}\}$ of $S$ 
with the same underlying $C$--pointing $p_C$ are given
and that $M$ is $S$--marked by $\{p_{M0},\zeta_{M0i}\}$, $\{p_{M1},\zeta_{M1i}\}$. 
We denote by $F_{\theta,\varUpsilon 0}$, $F_{\theta,\varUpsilon 1}$ and $F_{\theta 0}$, $F_{\theta 1}$ 
the $2$-- and $1$--holonomy maps associated with the 
two markings of $S$, that is $F_{\theta ,\varUpsilon 0}(\varXi)=F_\theta(\varSigma^\sharp{}_{0\varXi})$, 
$F_{\theta,\varUpsilon 1}(\varXi)=F_\theta(\varSigma^\sharp{}_{1\varXi})$ 
and $F_{\theta 0}(\zeta_{S0i}{}^*\varXi)
=F_\theta(\gamma_{0\zeta_{S0i}{}^*\varXi})$, $F_{\theta 1}(\zeta_{S1i}{}^*\varXi)=F_\theta(\gamma_{1\zeta_{S1i}{}^*\varXi})$ 
with $\varXi$ a free $S_0$-- and $S_1$--surface knot, respectively, in accordance with \eqref{hiholo4},
\eqref{hiholo4/0}  (cf. subsect. \ref{sec:knot}).

\begin{prop} \label{prop:2holo7/01} 
Suppose that the reference $S_0$--surface knot $\varXi_{M0}$ of $M_0$ and $S_1$--surface knot $\varXi_{M1}$
of $M_1$ are both equal to a simultaneous $S_0$-- and $S_1$ surface knot $\varXi_M$  of $M_0$ 
and $M_1$ perhaps up to $S_0$-- and $S_1$--marking preserving ambient isotopy, respectively.
Let $\varXi$ be a simultaneous $S_0$-- and $S_1$--surface knot of $M_0$ and $M_1$, respectively, 
having the following properties: $\varXi\circ k_z(p_{S0})=\varXi_M\circ k_z(p_{S0})$ and 
$\varXi\circ k_z\circ \zeta_{S0i}=\varXi_M\circ k_z\circ \zeta_{S0i}$ for all 
$z$, where $k_z$ is an orientation preserving isotopy with the properties stated 
in lemma \ref{lemma:s0knot}. Then, there are a curve $\gamma_1:p_{M0}\rightarrow p_{M1}$ 
and $2\ell_S$ surfaces 
$\varSigma_i:\gamma_{0\zeta_{S0i}{}^*\varXi}\Rightarrow \gamma_1{}^{-1_\circ}\circ\gamma_{1\zeta_{S1i}{}^*\varXi_1}\circ \gamma_1$,
of $M$ all with image lying in that of $\varXi$ such that \hphantom{xxxxxxxxxxxxxx}
\begin{equation}
F_{\theta,\varUpsilon 1}(\varXi)=m(F_\theta(\gamma_1))(F_{\theta,\varUpsilon 0}(\varXi))
\vphantom{\Big]}
\label{dhiholo8}
\end{equation}
and similarly that 
\begin{equation}
F_{\theta 1}(\zeta_{S1i}{}^*\varXi)=F_\theta(\gamma_1)t(F_{\theta,\varUpsilon}(\varSigma_i))
F_{\theta 0}(\zeta_{S0i}{}^*\varXi)F_\theta(\gamma_1)^{-1}.
\label{dhiholo7}
\end{equation}
\end{prop}

\noindent
{\it Proof}. By prop. \ref{prop:s0knot1}, under the above hypotheses, 
there exist a curve $\gamma_1:p_{M0}\rightarrow p_{M1}$ and a thin 
homotopy $\varSigma^\sharp{}_{0\varXi}\Rrightarrow 
I_{\gamma_1}{}^{-1_\circ}\circ\varSigma^\sharp{}_{1\varXi}\circ I_{\gamma_1}$,
both with image contained in that of $\varXi$.
Using once more the homotopy invariance of the map $\varSigma\rightarrow F_{\theta,\varUpsilon}(\varSigma)$ 
for a flat $(G,H)$--$2$--connection doublet $(\theta,\varUpsilon)$ (cf. prop. \ref{theor:twoholo3}), 
by a calculation similar to that carried out to prove \eqref{hiholo8}, we find that 
$F_{\theta,\varUpsilon}(\varSigma^\sharp{}_{0\varXi})=m(F_\theta(\gamma_1)^{-1})(F_{\theta,\varUpsilon}(\varSigma^\sharp{}_{1\varXi}))$. 
By \eqref{hiholo4}, this relation is equivalent to \eqref{dhiholo8}. 

According to prop. \ref{prop:s0knot2}, \pagebreak under the same hypotheses,  there exist $2\ell_S$ homotopies 
$\gamma_{0\zeta_{S0i}{}^*\varXi}\Rightarrow \gamma_1{}^{-1_\circ}\circ\gamma_{1\zeta_{S1i}{}^*\varXi_1}\circ \gamma_1$
with image contained in that of $\varXi$,
which we now view as surfaces $\varSigma_i$ with the same  source and target curves,
$\gamma_1$ being the same curve as the one described above. 
Exploiting again the homotopy invariance of the map $\gamma \rightarrow F_\theta(\gamma)$ 
for a flat $(G,H)$--$2$--connection doublet $(\theta,\varUpsilon)$ (cf. prop. \ref{theor:twoholo3}), 
by a calculation similar to that worked out to show \eqref{hiholo7}, we find that 
$t(F_{\theta,\varUpsilon}(\varSigma_i))F_\theta(\gamma_{0\zeta_{S0i}{}^*\varXi})
=F_\theta(\gamma_1)^{-1}F_\theta(\gamma_{1\zeta_{S1i}{}^*\varXi_1})F_\theta(\gamma_1)$.
Using \eqref{hiholo4/0}, this relation immediately reduces to \eqref{dhiholo7}. 
\hfill $\Box$

As in the case of ordinary knots, surface knot holonomy $\xi\mapsto F_\theta(\xi)$ depends on the flat 
$(G,H)$--$2$--connection doublet $(\theta,\varUpsilon)$ used in a gauge 
covariant manner (cf. def. \ref{def:gau1transf}, eqs. \eqref{gauholo3,4}). 
Let again $M$ be $S$--marked by $\{p_M,\zeta_{Mi}\}$.

\begin{prop} \label{prop:2holo6} 
Let $\varXi$ be an $S$--surface knot of $M$. Then, for any $(G,H)$--$1$--gauge transformation $(g,J)$, 
one has 
\begin{equation}
F_{{}^{g,J}\theta,{}^{g,J}\varUpsilon}(\varXi)=m(g(p_M))(F_{\theta,\varUpsilon}(\varXi))
\vphantom{\Big]}
\label{hiholo6}
\end{equation}
and correspondingly that \hphantom{xxxxxxxxxxxxxxxxxxxxxxxxxxx}
\begin{equation}
F_{{}^{g,J}\theta}(\zeta_{Si}{}^*\varXi)=g(p_M)t(G_{g,J;\theta}(\zeta_{Si}{}^*\varXi))
F_\theta(\zeta_{Si}{}^*\varXi)g(p_M)^{-1},
\vphantom{\Big]}
\label{hiholo6/0}
\end{equation}
where the gauge $1$--holonomies $G_{g,J;\theta}(\zeta_{Si}{}^*\varXi))$ are given by 
\begin{equation}
G_{g,J;\theta}(\zeta_{Si}{}^*\varXi)=G_{g,J;\theta}(\gamma_{\zeta_{Si}{}^*\varXi}).
\vphantom{\Big]}
\label{hiholo6/1}
\end{equation}
\end{prop}

\noindent
{\it Proof}. If $p$ is a point,  $\gamma:p\rightarrow p$ is a closed curve based at $p$ and 
$\varSigma:\gamma\Rightarrow\gamma$ is a closed surface based at $\gamma$, then 
$F_{{}^{g,J}\theta,{}^{g,J}\varUpsilon}(\varSigma)
=m\big(g(p)t(G_{g,J;\theta}(\gamma))\big)(F_{\theta,\varUpsilon}(\varSigma))$, by
\eqref{gauholo7}.
Then, by virtue the fact that $\iota_{p_M}:p_M\rightarrow p_M$ and 
$\varSigma^\sharp{}_\varXi:$ $\iota_{p_M}\Rightarrow \iota_{p_M}$, \linebreak 
we obtain that $F_{{}^{g,J}\theta,{}^{g,J}\varUpsilon}(\varSigma^\sharp{}_\varXi)
=m\big(g(p_M)t(G_{g,J;\theta}(\iota_{p_M}))\big)(F_{\theta,\varUpsilon}(\varSigma^\sharp{}_\varXi))$. 
Using the property that $G_{g,J;\theta}(\iota_{p_M})=1_H$ following 
immediately from def. \ref{def:parfgau1}, eqs. \eqref{parfgau1}, \eqref{parfgau2}, 
and the fact that $\rank(d\iota_{p_M}(x))=0$, we find so
$F_{{}^{g,J}\theta,{}^{g,J}\varUpsilon}(\varSigma^\sharp{}_\varXi)
=m(g(p_M))(F_{\theta,\varUpsilon}(\varSigma^\sharp{}_\varXi))$. 
Recalling \eqref{hiholo4}, we reach \eqref{hiholo6}. 

If $p$ is a point and $\gamma:p\rightarrow p$ is a closed curve based at $p$, then
$F_{{}^{g,J}\theta}(\gamma)=g(p)t(G_{g,J;\theta}(\gamma))
F_\theta(\gamma)g(p)^{-1}$, by \eqref{gauholo16/1}. 
Then, by \eqref{hiholo4/0} and the fact that $\gamma_{\zeta_{Si}{}^*\varXi}:p_M\rightarrow p_M$, 
we have $F_{{}^{g,J}\theta}(\gamma_{\zeta_{Si}{}^*\varXi})=
g(p_M)t(G_{g,J;\theta}(\gamma_{\zeta_{Si}{}^*\varXi}))F_\theta(\gamma_{\zeta_{Si}{}^*\varXi})g(p_M)^{-1}$. 
Recalling \eqref{hiholo4/0} and using the notation \eqref{hiholo6/1}, we reach \eqref{hiholo6/0}. 
\hfill $\Box$

The gauge $1$--holonomies of the characteristic $C$--knots of two 
freely ambient isotopic $S$--surface knots are related in a computable way. 

\begin{prop} \label{prop:2holo8} 
Under the same hypotheses and with the same specifications 
of prop. \ref{prop:2holo7}, the gauge $1$--holonomies 
$G_{g,J;\theta}(\zeta_{Si}{}^*\varXi_0)$, $G_{g,J;\theta}(\zeta_{Si}{}^*\varXi_1)$ are related as 
follows,
\begin{align}
G_{g,J;\theta}(\zeta_{Si}{}^*\varXi_1)
&=G_{g,J;\theta}(\gamma_1)m(F_\theta(\gamma_1))\big(
m(g(p_{M0})^{-1})(F_{{}^{g,J}\theta,{}^{g,J}\varUpsilon}(\varSigma_i))
\vphantom{\Big]}
\label{hiholo9}
\\
&\hspace{-.5cm}\times G_{g,J;\theta}(\zeta_{Si}{}^*\varXi_0)
m\big(F_\theta(\zeta_{Si}{}^*\varXi_0)F_\theta(\gamma_1)^{-1}\big)(G_{g,J;\theta}(\gamma_1)^{-1})
F_{\theta,\varUpsilon}(\varSigma_i)^{-1}\big).
\vphantom{\Big]}
\nonumber
\end{align}
\end{prop}

\noindent
{\it Proof}. 
As shown in the proof of prop. \ref{prop:2holo7} based on prop. \ref{prop:knot9/1}, for each $i$ 
there exists a surface $\varSigma_i:\gamma_{\zeta_{Si}{}^*\varXi_0}\Rightarrow 
\gamma_1{}^{-1_\circ}\circ\gamma_{\zeta_{Si}{}^*\varXi_1}\circ \gamma_1$. 
Using \eqref{gauholo7}, we have 
\begin{align}
G_{g,J;\theta}&(\gamma_1{}^{-1_\circ}\circ\gamma_{\zeta_{Si}{}^*\varXi_1}\circ \gamma_1)
\label{hiholoc1}
\\
&=m(g(p_{M0})^{-1})(F_{{}^{g,J}\theta,{}^{g,J}\varUpsilon}(\varSigma_i))
G_{g,J;\theta}(\gamma_{\zeta_{Si}{}^*\varXi_0})F_{\theta,\varUpsilon}(\varSigma_i)^{-1}.
\nonumber
\end{align}
From \eqref{twoholo3} and \eqref{gauholo16}, \eqref{gauholo17}, we find
\begin{align}
G_{g,J;\theta}&(\gamma_1{}^{-1_\circ}\circ\gamma_{\zeta_{Si}{}^*\varXi_1}\circ \gamma_1)
\vphantom{\Big]}
\label{hiholoc3}
\\
&
=m(F_\theta(\gamma_1)^{-1})\big(G_{g,J;\theta}(\gamma_1)^{-1}
G_{g,J;\theta}(\gamma_{\zeta_{Si}{}^*\varXi_1})
m(F_\theta(\gamma_{\zeta_{Si}{}^*\varXi_1}))(G_{g,J;\theta}(\gamma_1))\big).
\vphantom{\Big]}
\nonumber
\end{align}
Inserting \eqref{hiholoc3} into \eqref{hiholoc1}, we get 
\begin{align}
G_{g,J;\theta}(\gamma_{\zeta_{Si}{}^*\varXi_1})
&=G_{g,J;\theta}(\gamma_1)m(F_\theta(\gamma_1))\big(
m(g(p_{M0})^{-1})(F_{{}^{g,J}\theta,{}^{g,J}\varUpsilon}(\varSigma_i))
\vphantom{\Big]}
\label{hiholoc4}
\\
&\hspace{.5cm}\times G_{g,J;\theta}(\gamma_{\zeta_{Si}{}^*\varXi_0})F_{\theta,\varUpsilon}(\varSigma_i)^{-1}\big)
m(F_\theta(\gamma_{\zeta_{Si}{}^*\varXi_1}))(G_{g,J;\theta}(\gamma_1)^{-1}).
\vphantom{\Big]}
\nonumber
\end{align}
The last factor in the right hand side of \eqref{hiholoc4} can be reshaped using 
the calculation  of the proof of prop. \ref{prop:2holo7}, 
\begin{align}
&m(F_\theta(\gamma_{\zeta_{Si}{}^*\varXi_1}))(G_{g,J;\theta}(\gamma_1)^{-1}) \hspace{6cm}
\vphantom{\Big]}
\label{hiholoc5}
\end{align}
\begin{align}
&\hspace{1cm}
=m\big(F_\theta(\gamma_1)t(F_{\theta,\varUpsilon}(\varSigma_i))
F_\theta(\gamma_{\zeta_{Si}{}^*\varXi_0})F_\theta(\gamma_1)^{-1}\big)(G_{g,J;\theta}(\gamma_1)^{-1})
\vphantom{\Big]}
\nonumber
\\
&\hspace{1cm}
=m(F_\theta(\gamma_1))\big(F_{\theta,\varUpsilon}(\varSigma_i)
m\big(F_\theta(\gamma_{\zeta_{Si}{}^*\varXi_0})F_\theta(\gamma_1)^{-1}\big)(G_{g,J;\theta}(\gamma_1)^{-1})
F_{\theta,\varUpsilon}(\varSigma_i)^{-1}\big).
\vphantom{\Big]}
\nonumber
\end{align}
Substituting \eqref{hiholoc5} into \eqref{hiholoc4} and using
\eqref{hiholo4/0}  and \eqref{hiholo6/1}, we reach immediately \eqref{hiholo9}. \hfill $\Box$

The gauge $1$--holonomies of the characteristic $C$--knots of an
$S$--surface knot with respect two marking of $S$ are also related in a computable way
Below, we denote by $G_{g,J;\theta 0}$, $G_{g,J;\theta 1}$ 
the gauge $1$--holonomy maps associated with the 
two markings, that is $G_{g,J;\theta 0}(\zeta_{S0i}{}^*\varXi)
=G_{g,J;\theta}(\gamma_{0\zeta_{S0i}{}^*\varXi})$, $G_{g,J;\theta 1}(\zeta_{S1i}{}^*\varXi)=G_{g,J;\theta}(\gamma_{1\zeta_{S1i}{}^*\varXi})$ 
with $\varXi$ a free $S_0$-- and $S_1$--surface knot, respectively, in accordance with \eqref{hiholo6/1}.

\begin{prop} \label{prop:2holo9} 
Under the same hypotheses and with the same specifications 
of prop. \ref{prop:2holo7/01}, the gauge $1$--holonomies 
$G_{g,J;\theta 1}(\zeta_{S0i}{}^*\varXi)$, $G_{g,J;\theta 1}(\zeta_{S1i}{}^*\varXi)$ are related as 
\begin{align}
G_{g,J;\theta 1}(\zeta_{S1i}{}^*\varXi)
&=G_{g,J;\theta}(\gamma_1)m(F_\theta(\gamma_1))\big(
m(g(p_{M0})^{-1})(F_{{}^{g,J}\theta,{}^{g,J}\varUpsilon}(\varSigma_i))
\vphantom{\Big]}
\label{hiholo10}
\\
&\hspace{-.8cm}\times G_{g,J;\theta 0}(\zeta_{S0i}{}^*\varXi)
m\big(F_{\theta 0}(\zeta_{S0i}{}^*\varXi)F_\theta(\gamma_1)^{-1}\big)(G_{g,J;\theta}(\gamma_1)^{-1})
F_{\theta,\varUpsilon}(\varSigma_i)^{-1}\big).
\vphantom{\Big]}
\nonumber
\end{align}
\end{prop}

\noindent
{\it Proof}. The proof proceeds by a reasoning and calculations essentially identical to those of prop. \ref{prop:2holo8}.
\hfill $\Box$

We now analyse the effect of source autodiffeomorphism action on holonomy. 

We consider first an ordinary  gauge theory with gauge Lie group $G$ on a manifold $M$
with a flat $G$--connection $\theta$. 

\begin{prop} \label{prop:ffcfaieqcj}
Let $\xi$ be a free $C$--knot and let $f$ be an orientation 
preserving autodiffeomorphisms of $C$. 
Then, 
\begin{equation}
F_\theta(f^*\xi)=F_\theta(\gamma_f)F_\theta(\xi)F_\theta(\gamma_f)^{-1},
\label{fxiholo1}
\end{equation}
where $\gamma_f:\xi(p_C)\rightarrow
f^*\xi(p_C)$ is a curve of $M$ whose image lies in that of $\xi$.
\end{prop}

\noindent
{\it Proof}. By prop. \ref{prop:diffeo2}, under the above hypotheses, 
there are a curve $\gamma_f:\xi(p_C)\rightarrow f^*\xi(p_C)$
and a thin homotopy $\gamma_{\xi}\Rightarrow \gamma_f{}^{-1_\circ}\circ \gamma_{f^*\xi}
\circ\,\gamma_f$, both with image contained in that of $\xi$. 
The statement then follows from a reasoning analogous to that of the proof
of prop. \ref{prop:cfaieqcj}. \hfill $\Box$

We consider next a strict higher gauge theory with gauge Lie crossed module $(G,H)$ on a manifold $M$
with a flat $(G,H)$--$2$--connection doublet $(\theta,\varUpsilon)$. 

\begin{prop} \label{prop:holosdiffeo3}
Let $\varXi$ be a free $S$--surface knot and let $f_0$, $f_1$ be isotopic orientation 
preserving autodiffeomorphisms of $S$. 
Suppose that the concomitant reference $S$--surface knots 
$\varXi_{Mf_0{}^*\varXi}$, $\varXi_{Mf_1{}^*\varXi}$ 
of $f_0{}^*\varXi$, $f_1{}^*\varXi$ are $f_0{}^*\varXi_{M\varXi}$, $f_1{}^*\varXi_{M\varXi}$, 
respectively, where $\varXi_{M\varXi}$ is the concomitant 
reference $S$--surface knot of $\varXi$. Suppose further that the free ambient isotopies 
relating $f_0{}^*\varXi$, $f_1{}^*\varXi$ and $f_0{}^*\varXi_{M\varXi}$, $f_1{}^*\varXi_{M\varXi}$ 
according to prop. \ref{prop:sdiffeo2} are concordant 
on the $S$--marking $\{f_0{}^*\varXi(p_S)$, $\zeta_{Si}{}^*f_0{}^*\varXi\}$ of $M$.
Then, there are a curve $\gamma_{f_0f_1}:f_0{}^*\varXi(p_S)\rightarrow f_1{}^*\varXi(p_S)$ 
of $M$ and $2\ell_S$ surfaces
$\varSigma_{f_0f_1i}:\gamma_{\zeta_{Si}{}^*f_0{}^*\varXi}
\Rightarrow \gamma_{f_0f_1}{}^{-1_\circ}\circ\gamma_{\zeta_{Si}{}^*f_1{}^*\varXi}\circ \gamma_{f_0f_1}$,
all with image is contained in that of $\varXi$ such that 
\begin{equation}
F_{{\theta,\varUpsilon}}(f_1{}^*\varXi)=m(F_\theta(\gamma_{f_0f_1}))(F_{\theta,\varUpsilon}(f_0{}^*\varXi))
\vphantom{\Big]}
\label{ffhiholo8}
\end{equation}
and similarly that 
\begin{equation}
F_\theta(\zeta_{Si}{}^*f_1{}^*\varXi)=F_\theta(\gamma_{f_0f_1})t(F_{\theta,\varUpsilon}(\varSigma_{f_0f_1i}))
F_\theta(\zeta_{Si}{}^*f_0{}^*\varXi)F_\theta(\gamma_{f_0f_1})^{-1}.
\label{ffhiholo7}
\end{equation}
\end{prop}

\noindent
{\it Proof}. By prop. \ref{prop:sdiffeo3}, under the above hypotheses, 
there are a curve $\gamma_{f_0f_1}:f_0{}^*\varXi(p_S)\rightarrow f_1{}^*\varXi(p_S)$ 
of $M$ and a thin homotopy $\varSigma^\sharp{}_{f_0{}^*\varXi}\Rrightarrow 
I_{\gamma_{f_0f_1}}{}^{-1_\circ}\circ \varSigma^\sharp{}_{f_1{}^*\varXi} 
\circ I_{\gamma_{f_0f_1}}$, both with image lying in that of $\varXi$. 
Using once more the homotopy invariance of the map $\varSigma\rightarrow F_{\theta,\varUpsilon}(\varSigma)$ 
for a flat $(G,H)$--$2$--connection doublet $(\theta,\varUpsilon)$ (cf. prop. \ref{theor:twoholo3}), 
by a calculation similar to that carried out to prove \eqref{hiholo8}, we find that 
$F_{\theta,\varUpsilon}(\varSigma^\sharp{}_{f_0{}^*\varXi})
=m(F_\theta(\gamma_{f_0f_1})^{-1})(F_{\theta,\varUpsilon}(\varSigma^\sharp{}_{f_1{}^*\varXi} ))$. 
By \eqref{hiholo4}, this relation is equivalent to \eqref{ffhiholo8}. 

According to prop. \ref{prop:sdiffeo4},  under the same hypotheses,  there exist $2\ell_S$ homotopies 
$\gamma_{\zeta_{Si}{}^*f_0{}^*\varXi}\Rightarrow \gamma_{f_0f_1}{}^{-1_\circ}\circ\gamma_{\zeta_{Si}{}^*f_1{}^*\varXi}\circ \gamma_{f_0f_1}$
with image contained in that of $\varXi$,
which we now view as surfaces $\varSigma_{f_0f_1i}$ with the same source and target curves,
$\gamma_{f_0f_1}$ being the same curve as the one described above. 
Exploiting again the homotopy invariance of the map $\gamma \rightarrow F_\theta(\gamma)$ 
for a flat $(G,H)$--$2$--connection doublet $(\theta,\varUpsilon)$ (cf. prop. \ref{theor:twoholo3}), 
by a calculation similar to that worked out to show \eqref{hiholo7}, we find that 
$t(F_{\theta,\varUpsilon}(\varSigma_{f_0f_1i}))F_\theta(\gamma_{\zeta_{Si}{}^*f_0{}^*\varXi})
=F_\theta(\gamma_{f_0f_1})^{-1}F_\theta(\gamma_{\zeta_{Si}{}^*f_1{}^*\varXi})F_\theta(\gamma_{f_0f_1})$.
Using \eqref{hiholo4/0}, this relation immediately reduces to \eqref{ffhiholo7}. 
\hfill $\Box$

\begin{prop} 
Under the same hypotheses of prop. \ref{prop:holosdiffeo3}, one has 
\begin{align}
&G_{g,J;\theta}(\zeta_{Si}{}^*f_1{}^*\varXi)=G_{g,J;\theta}(\gamma_{f_0f_1})m(F_\theta(\gamma_{f_0f_1}))\big(
\label{ffhiholo9}
\\
&\hspace{1cm}
m(g(f_0{}^*\varXi(p_S))^{-1})(F_{{}^{g,J}\theta,{}^{g,J}\varUpsilon}(\varSigma_{f_0f_1i}))G_{g,J;\theta}(\zeta_{Si}{}^*f_0{}^*\varXi)
\vphantom{\Big]}
\nonumber
\\
&\hspace{2cm} \times
m\big(F_\theta(\zeta_{Si}{}^*f_0{}^*\varXi)F_\theta(\gamma_{f_0f_1})^{-1}\big)(G_{g,J;\theta}(\gamma_{f_0f_1})^{-1})
F_{\theta,\varUpsilon}(\varSigma_{f_0f_1i})^{-1}\big).
\vphantom{\Big]}
\nonumber
\end{align}
\end{prop}

\noindent
{\it Proof}. 
This follows from a calculation analogous to that leading \eqref{hiholo9}
using the homotopies $\gamma_{\zeta_{Si}{}^*f_0{}^*\varXi}
\Rightarrow \gamma_{f_0f_1}{}^{-1_\circ}\circ\gamma_{\zeta_{Si}{}^*f_1{}^*\varXi}\circ \gamma_{f_0f_1}$
discussed in the proof of \eqref{ffhiholo7} above. \hfill $\Box$

\vfil\eject

\end{document}